\documentclass[11pt,epsf,psfig]{article}
\usepackage{graphicx}

\begin{document}
\baselineskip 0.6cm

\newcommand{\gsim}{ \mathop{}_{\textstyle \sim}^{\textstyle >} }
\newcommand{\lsim}{ \mathop{}_{\textstyle \sim}^{\textstyle <} }
\newcommand{\vev}[1]{ \langle {#1} \rangle }
\newcommand{\EV}{ ~{\rm eV} }
\newcommand{\KEV}{ ~{\rm keV} }
\newcommand{\MEV}{ ~{\rm MeV} }
\newcommand{\GEV}{ ~{\rm GeV} }
\newcommand{\TEV}{ ~{\rm TeV} }
\newcommand{\eps}{\varepsilon}
\newcommand{\barr}[1]{ \overline{{#1}} }
\newcommand{\del}{\partial}
\newcommand{\nn}{\nonumber}
\newcommand{\ra}{\rightarrow}
\newcommand{\bino}{\tilde{\chi}}
\def\tr{\mathop{\rm tr}\nolimits}
\def\Tr{\mathop{\rm Tr}\nolimits}
\def\Re{\mathop{\rm Re}\nolimits}
\def\Im{\mathop{\rm Im}\nolimits}
\setcounter{footnote}{1}

\begin{titlepage}

\begin{flushright}
UT-03-25\\
\end{flushright}

\vskip 3cm
\begin{center}
{\Large \bf  Neutralino Dark Matter from MSSM Flat Directions \\
in light of WMAP Result}

\vskip 2.4cm

\center{Masaaki~Fujii  and Masahiro~Ibe}\\
{\it{Department of Physics, University of Tokyo, Tokyo 113-0033,
Japan}}\\

\vskip 2cm

\abstract{The minimal supersymmetric standard model (MSSM) has a truly 
supersymmetric way to explain both the baryon asymmetry and cold dark
 matter in the present Universe, that is,
``{\it Affleck--Dine baryo/DM-genesis.}'' 
The associated late-time decay 
of Q-balls directly connects the origins of the baryon asymmetry and dark
matter, and also predicts a specific nature of the LSP.
In this paper, we investigate the prospects for indirect detection 
of these dark matter candidates observing 
high energy neutrino flux from the Sun, 
and hard positron flux from the halo.
We also update the previous analysis of the direct detection 
in hep-ph/0205044 by implementing the recent result from  WMAP satellite. 
}

\end{center}
\end{titlepage}

\renewcommand{\thefootnote}{\arabic{footnote}}
\setcounter{footnote}{0}

%
%       *** Main Part ***
%
%%%%%%%%%%%%%%%%%%%%%%%%%%%%%%%%%%%%%
\section{Introduction}%%%%%%%%%%%%%%%
%%%%%%%%%%%%%%%%%%%%%%%%%%%%%%%%%%%%%
The origin of both the baryon asymmetry and dark matter in the present Universe
is one of the most fundamental puzzles in particle physics and cosmology.
The recent data from the WMAP satellite~\cite{WMAP} provides 
us the cosmological
abundance of these quantities with surprising accuracy:
\begin{equation}
 \Omega_{B}h^2=0.0224\pm0.0009,\qquad \Omega_{DM}h^2=0.1126^{+0.0161}_{-0.0181}\;.
\label{abundance}
\end{equation}
Important things remaining to be done are constructing a realistic
model that explains both the quantities simultaneously and investigating
its implications in low-energy experiments.

The Minimal Supersymmetric  
Standard Model (MSSM) is the most motivated framework for constructing such a model.
The MSSM inevitably contains the lightest supersymmetry (SUSY) particle (LSP), which 
is absolutely stable under the R-parity conservation and its thermal 
relic density can fall in the observed quantity in some specific parameter space.
Many works have been carried out to scrutinize this point 
by assuming various boundary conditions for SUSY breaking parameters
with gradually increasing accuracy in calculation of the relic abundance.
In this point, we have nothing to add to those works.
However, there is an implicit but very important assumption here.

In order for those researches to have something to do with real nature,
the generation of the observed baryon asymmetry must be completed well
before the freeze-out time of the LSP. The most natural answer for this 
is provided by leptogenesis~\cite{Fukugita_Yanagida}, 
which supplies the required baryon
asymmetry by non-equilibrium decays of thermally~\cite{thermal_L} 
(or non-thermally~\cite{non-thermal_L})
produced right-handed Majorana neutrinos. In this case, the scenario might be
indirectly tested  by discoveries of the CP violation in the neutrino
sector, the neutrinoless double beta $(0\nu\beta\beta)$ decay 
and lepton-flavor violations in the future experiments.

On the other hand, once we introduce SUSY 
to the Standard Model, the thermal leptogenesis 
is not the only minimal scenario to generate the observed baryon
asymmetry. Affleck and Dine proposed another minimal scenario
of baryogenesis by utilizing a flat direction 
existing in the MSSM, which carries non-zero baryon number.
That is, what we call,  Affleck--Dine (AD) baryogenesis~\cite{Affleck_Dine_org,Affleck_Dine_review}.
The AD field, which is a linear combination of squark and/or slepton
fields along flat directions of the MSSM, can naturally acquire a large 
expectation value during inflation because of the 
flatness of the potential. After the end of
inflation, the AD field starts coherent oscillation around the origin,
which can produce the required baryon asymmetry very  efficiently.

In recent developments, however, it becomes clear that this is 
not the whole story.
The coherent oscillation of the AD field is not stable under the spatial 
perturbations, and fragments into a non-topological soliton, called
Q-ball~\cite{Q-ball_org} after dozens of  
oscillations~\cite{Q-ball_gauge,Q-ball_sugra}. 
The large expectation value of the
AD field inside the Q-ball protects it from being thermalized, and the 
decay temperature of the Q-ball is expected to be well below the 
freeze-out temperature of the LSP. This fact has a very important
implication on neutralino dark matter. Particularly, the Bino-like LSP, which otherwise explains
the required mass density of dark matter in the standard scenario at least
in some specific parameter space, inevitably leads to
the overclosure of the Universe.  This is a generic problem in 
Affleck--Dine baryogenesis~\cite{Q-ball-problem}.

One natural answer for this problem
proposed by K.~Hamaguchi and one of the authors (M.F.) is to adopt the
LSP with a larger annihilation cross section, such as Higgsino or
Wino~\cite{Fujii_Hamaguchi1}. 
This choice of the LSP allows us to explain both the required
baryon asymmetry and dark matter mass density simultaneously by a single
mechanism. In fact, it was pointed out that some class of AD baryogenesis 
directly connects the ratio of baryon to neutralino mass density 
$\Omega_{B}/\Omega_{\chi}$ in terms of low-energy parameters, irrespective
of inflation models and other details in the history of the 
Universe~\cite{Fujii_Yanagida}.
Furthermore, the late time decay of Q-balls opens up  new cosmologically relevant parameter regions,
where the standard scenario gives only a very small fraction of the required 
mass density, for instance, $\Omega^{\rm th}_{\chi}h^2\simeq 10^{-2}\sim
10^{-3}$ or even smaller.

In the previous paper~\cite{Hamaguchi_Fujii2}, K.H. and M.F. investigated 
implications of the Higgsino- and Wino-like non-thermal dark matter  
in the direct and indirect detection by assuming the 
gravity-mediated (mSUGRA) and the anomaly-mediated (mAMSB)
SUSY-breaking models~\cite{AMSB_models}. As for the indirect detection, 
we consider the mono-energetic photons caused by the direct annihilation 
of the neutralinos, $\chi\chi\rightarrow
\gamma\gamma$~\cite{photon-photon}.

In this paper, we investigate the prospects of detection possibility 
in another promising method of indirect dark-matter search 
observing high-energy neutrino flux from the center of the Sun.
We also add the calculation of the expected high-energy positron flux,
which may also serve as a ``smoking-gun'' signal of the non-thermal dark
matter in the  near future.
Furthermore, we update the analysis of the direct detection rates 
in the previous work, since 
the conditions adopted for the non-thermal dark matter
seem to be too conservative after the report from  the WMAP satellite.
We implement the result to constrain the allowed parameter space 
in the presence of the late-time Q-ball decays, which allows us to have 
much more definite predictions of the present scenario. 
For that purpose, we calculate the thermal relic density of the LSP 
by using {\it micrOMEGAs} computer code~\cite{micrOMEGAs}, which includes all the 
possible co-annihilation effects.
As a bonus, we can also clarify the differences in the dark-matter detection rates between
the non-thermal and  the standard thermal scenario by appropriately 
scaling the detection rates by 
a factor, $(\Omega_{\chi}^{\rm th}h^2/\Omega_{DM}h^2)$.

As we will see, these new indirect methods provide 
additional promising ways to find signals  of the non-thermal dark matter 
in our Universe.
If indeed the existence of Higgsino- or Wino-like dark matter is
confirmed in future experiments, it strongly suggests that the
whole matter in the present Universe has a truly supersymmetric origin,
``{\it Affleck--Dine baryo/DM-genesis}.''~\footnote{
In this paper, we do not discuss the AD leptogenesis. In this case, 
the following arguments on the neutralino dark matter cannot be applied.
See discussion in Refs.~\cite{AD-leptogenesis1,AD-leptogenesis2}.}
%%%%%%%%%%%%%%%%%%%%%%%%%%%%%%%%%%%%%%%%%%%%%%%%%%%%%%%%%%%%%%%%%%%%
\section{Late-time Q-ball decay in AD baryogenesis}%%%%%%%%%%%%%%%%%%
%%%%%%%%%%%%%%%%%%%%%%%%%%%%%%%%%%%%%%%%%%%%%%%%%%%%%%%%%%%%%%%%%%%%
In the next two sections, we review the non-thermal dark-matter generation 
from late-time decay of Q-balls, which generally appears in 
AD baryogenesis. The readers who are interested in much details, please see
Ref.~\cite{Hamaguchi_Fujii2}.

AD baryogenesis utilizes the AD field $\phi$, which is a linear
combination of squarks and/or slepton fields along flat directions in
the MSSM. Each flat direction is labeled by a monomial of chiral
superfields, such as $\bar{U}\bar{D}\bar{D}$, $Q\bar{D}L$ and $QQQL$. 
A complete list of the flat directions in the MSSM is available in 
Ref.~\cite{flat_directions}.
During inflation, the field $\phi$ can get a large negative mass
term of the order of the Hubble parameter, $-c_{H}H^2 |\phi|^2$, where 
$c_{H}$ is $\cal{O}$(1) and positive~\cite{Randall_Thomas1}. 
This occurs if the inflaton 
has a four-point coupling with the $\phi$ field in the K\"ahler
potential as
\begin{equation}
\delta K=\frac{b}{M_{*}^2}I^{\dag}I\phi^\dag \phi\;,
\end{equation}
with $b\gsim 1$. Here, $I$ denotes the inflaton superfield, 
and $M_{*}=2.4\times 10^{18}\GEV$ is the reduced 
Planck scale. Actually,
such four-point couplings with the SM fields serve as dominant decay
modes of inflatons in many inflationary models.

In this case, the $\phi$ field is driven far away from the origin
during  inflation by this negative mass term. We assume this is the
case in the following discussion.~\footnote{Quantum fluctuations
during inflation may serve as a driving force of the $\phi$ field in the
special case, where $|c_{H}|<<1$.} 
After the end of inflation, 
the $\phi$ field starts coherent oscillation around the origin when its 
mass $m_{\phi}$ exceeds the Hubble parameter. This is the stage where 
the net baryon asymmetry is generated. 

The relevant baryon-number-violating
operators come from the superpotential or from the K\"ahler potential.
In the case of the superpotential, the operator generally has the
following form:
\begin{equation}
 \delta W=\frac{1}{n M^{n-3}}\phi^n\;,
\label{superpotential}
\end{equation}
with $n\geq 4$. Here, we treat $M$ as the effective scale where the
operator appears.~\footnote{Note that $M$ can exceed $M_{*}$ because we
include possible suppression effects coming from coupling constants.} 
Some examples of these terms are given by
$ \delta W\propto QQQL,\;\bar{U}\bar{U}\bar{D}\bar{E}$
for $n=4$, and 
$ \delta W \propto (\bar{U}\bar{D}\bar{D} )^2,\;
\bar{U}\bar{D}\bar{D}Q\bar{D}L$ 
for $n=6$. 
Through SUSY-breaking effects, these operators induce the scalar potential that generates the
baryon asymmetry: 
\begin{equation}
 \delta V=\frac{a_{m}m_{3/2}}{nM^{n-3}}\phi^n+{\rm h.c.}\;,
\label{fromsuper}
\end{equation}
where $a_{m}$ 
is a coupling constant, and $m_{3/2}$ denotes the gravitino
mass.~\footnote{Here, we have assumed for simplicity 
that there is no A-term of the
order of the Hubble parameter, which is true when the three point
coupling, $\delta K\propto
I\phi^\dag\phi$, is absent in the K\"ahler potential.
 Even if such an A-term exists, the
conclusions in the following do not change much.}
The generation of the baryon asymmetry can easily be seen from the 
equation of motion of the baryon-number density:
\begin{equation}
 \dot{n_{B}}+3 H n_{B}=2 \beta {\rm Im}\left(\frac{\del \delta V}{\del \phi}\phi\right)\;,
\label{equationofbaryon}
\end{equation}
which can be rewritten in the following integration form:
\begin{equation}
\left[R^3 n_{B}\right](t)=2 \beta \int^{t}dt\; R^3{\rm Im}\left(
\frac{\del \delta V}{\del\phi}\phi\right)\;,
\label{equationofbaryon2}
\end{equation}
where we define $\beta$ as the baryon charge of the AD field such that
$n_{B}=\beta (\dot{\phi}^{\dag}\phi-\phi^\dag \dot{\phi})$, and $R$ as
the scale factor of the Universe.

The non-renormalizable operator given in Eq.~(\ref{superpotential}) also 
lifts the flat direction, and 
the AD field evolves slowly  as $|\phi|\simeq (HM^{n-3})^{1/n-2}$ until it
starts oscillations around the origin.
This is the balance point between the F-term of $\delta W$
and the negative Hubble mass term $-c_{H}H^2|\phi|^2$.
During this stage, the baryon asymmetry is generated through 
Eq.~(\ref{equationofbaryon2}).
This baryon-number generation is 
terminated as soon as the AD field starts coherent oscillations,
because the baryon-number-violating operators given in
Eq.~(\ref{fromsuper}) damp very quickly after the start of oscillation. 
The amplitude of the AD field at this time is very important information 
for the following discussion, since it determines the typical size of
produced Q-balls, which in turn  determines the typical Q-ball decay
temperature, $T_{d}$.
In the scenario we are now considering, 
the initial oscillation amplitude of the AD field is 
given by $|\phi|_{\rm osc}\simeq (m_{\phi}M^{n-3})^{1/n-2}$.
Here, 
the scale of $M$ is naturally expected to be $\gsim M_{*}$.  In the case of
$n=4$, however, it is not the case. This is because  these operators 
are responsible for the proton decay~\cite{proton-decay_org}, 
and at lease for the most relevant operators $M\gsim 10^{25}\GEV$ 
should be satisfied in order to avoid a too rapid 
decay~\cite{proton-decay_constraints}.
In both the $n=4$ and $n=6$ cases, required reheating temperature of
inflation to explain
the correct amount of baryon asymmetry is about $T_{R}\sim 10^2\GEV$.
For an explicit expression in each case, see Ref.~\cite{Hamaguchi_Fujii2}.

In the case of the K\"ahler potential, the most relevant operators 
are given by~\cite{Affleck_Dine_org}
\begin{equation}
 \delta K = \frac{\lambda}{M_{*}^2} Q\bar{U}^{\dag}\bar{D}^{\dag}L,\;
QQ\bar{U}^\dag\bar{E}^\dag \;,
\label{Kahlerpotential}
\end{equation}
where $|\lambda|={\cal{O}}(1)$ is a coupling constant.~\footnote{In
order for these operators to be dominant, we generally have to assume suppressions
of non-renormalizable operators in the superpotential, which can be
done, for instance, by imposing R-symmetry.} 
These operators
do not lift the AD field, and other higher order terms in the K\"ahler
potential, or the $U_{B-L}$ D-term determines 
the initial amplitude of the AD 
field~\cite{Q-ball-problem}~\footnote{In this case, the AD field is
stopped at the $B-L$ breaking scale.}. 
The potential that is responsible for the generation of baryon asymmetry 
has the following form:
\begin{equation}
\delta V=\left(a_{m}\frac{m_{3/2}^2}{4 M_{*}^2}\phi^4
+a_{H}\frac{H^2}{4 M_{*}^2}\phi^4\right)+{\rm h.c.}\;,
\label{fromKahler}
\end{equation}
where $a_{m}$ and $a_{H}$ are coupling constants.
The generation mechanism of the baryon asymmetry is the same as 
the former example.
Note that,  the resultant baryon
asymmetry is completely independent of $T_{R}$, since the AD field
dominates the energy density of the Universe when it decays.~\footnote{
The condition for this statement 
to hold is discussed in Ref.~\cite{Fujii_Yanagida} with
thermal effects taken into account.} 
This interesting feature allows us to directly calculate
$\Omega_{B}/\Omega_{\chi}$ with low-energy parameters, which is, in
particular, independent of the initial amplitude of the AD
field and $T_{R}$~\cite{Fujii_Yanagida}.~\footnote{
The relation is given as follows:
\begin{equation}
\frac{\Omega_{B}}{\Omega_{\chi}}\simeq 10^{3\mbox{--}4}\left(
\frac{m_{\phi}^2}{\vev{\sigma v}_{\chi}^{-1}}\right)\left(
\frac{m_{p}}{m_{\chi}}\right)\delta_{\rm eff}\;.
\nonumber
\end{equation}
For the derivation, please see the reference.}
To obtain the correct abundance of the baryon asymmetry in 
this model, we need $|\phi|_{\rm
osc}\sim 10^{16}\GEV$ and hence the $U(1)_{B-L}$ D-term is perfectly suitable
for this purpose.

After dozens of oscillations, the AD fields fragment into the Q-balls,
which absorb almost all of the produced baryon asymmetry.
Here we quote the expected size of Q-balls, ``$Q$'', produced in each
case~\cite{Hamaguchi_Fujii2}. 
If we use the superpotential to generate the baryon asymmetry, it is
written as 
\begin{eqnarray}
Q\sim \left\{
\begin{array}{lll}
3\times 10^{20}\times \beta \delta_{\rm eff}|a_{m}|\left(\displaystyle{\frac{1\TEV}{m_{\phi}}}\right)
\left(\displaystyle{\frac{M}{10^{26}\GEV}}\right)\qquad\qquad\quad \mbox{for} \quad n=4
\\
3\times 10^{20}\times \beta \delta_{\rm eff}|a_{m}|\left(\displaystyle{\frac{1\TEV}{m_{\phi}}}\right)
\left(\displaystyle{\frac{1\TEV}{m_{\phi}}}\right)^{3/2}
\left(\displaystyle{\frac{M}{M_{*}}}\right)^{3/2}
\quad \mbox{for}\quad n=6\;,
\end{array}
\right.
\label{Q_super}
\end{eqnarray}
where $\delta_{\rm eff}={\cal{O}}(0.1)$ is an effective CP-violating phase. 
In the case of the K\"ahler potential, 
\begin{equation}
 Q\sim 10^{26}\left(\frac{|\phi|_{\rm osc}}{M_{*}}\right)^2 \left(
\frac{1\TEV}{m_{\phi}}\right)^2\;.
\label{Q_Kahler} 
\end{equation} 
These expressions were first derived analytically~\cite{Q-ball_sugra}, 
which has been  found fairly consistent with recent detailed lattice 
simulations~\cite{Q-ball_in_lattice}.

Although they can also be used in the anomaly-mediation
models, there appears one complication in this case. 
Because of the large gravitino mass, there appears a global (or local)
minimum displaced from the origin along the flat direction
(See, Eqs (\ref{fromsuper}) and (\ref{fromKahler})). 
If the AD field is trapped by this
minimum, it leads to  a color breaking Universe. In order to avoid this
disaster, we have to restrict the initial amplitude of the AD field as 
\begin{equation}
 |\phi|_{\rm ini}\lsim \left(\frac{m_{\phi}^2}{m_{3/2}}M^{n-3}\right)^{1/(n-2)}
\end{equation}
in the case of the superpotential~\cite{Q-ball-problem}, and 
\begin{equation}
 |\phi|_{\rm ini}\lsim M_{*}\frac{m_{\phi}}{m_{3/2}}\;,
\end{equation}
in the case of the K\"ahler potential~\cite{Hamaguchi_Fujii2}. 
These conditions can be easily satisfied if
we make use of the $U(1)_{B-L}$ D-term to stop the AD field.

Now, let us estimate the decay temperature of a Q-ball.
It is known that the decay rate of a Q-ball can be 
written as~\cite{Q-ball-decay}
\begin{equation}
\Gamma_{Q}\equiv -\frac{dQ}{dt}\lsim \frac{\omega^3 {\cal{A}}}{192 \pi^2}\;,
\label{Qdecayrate}
\end{equation}
where $\omega\simeq m_{\phi}$, ${\cal{A}}=4\pi R_{Q}^2$ is the surface
area of the Q-ball, and $R_{Q}\simeq \sqrt{2}/(m_{\phi}\sqrt{|K|})$ is
its radius. 
Here, $K$ denotes the one loop correction of the mass term of the AD
field:
\begin{equation}
 V(\phi)=m_{\phi}^2\left(1+K{\rm{log}}\left(\frac{|\phi|^2}{M_{G}^2}\right)\right)\;,
\end{equation}
where $M_{G}$ is the renormalization scale at which $m_{\phi}$ is defined.
Note that the negativeness of $K$ is the necessary and sufficient condition
for the Q-ball formation.
{}From Eq.~(\ref{Qdecayrate}), we can calculate the decay temperature
of the Q-ball as follows:
\begin{equation}
 T_{d}\lsim 2\GEV \times \left(\frac{0.03}{|K|}\right)^{1/2}
\left(\frac{m_{\phi}}{1\TEV}\right)^{1/2}\left(\frac{10^{20}}{Q}\right)^{1/2}\;.
\label{decaytemp}
\end{equation}
We can see that the expected decay temperature of Q-balls is about
$T_{d}\sim {\cal{O}}(1)\GEV$ if we use the superpotential, and 
${\cal{O}}(10)\MEV\lsim T_{d}\lsim {\cal{O}}(1)\GEV$ for $10^{17}\GEV\gsim |\phi|_{\rm
osc}\gsim $ $10^{15}\GEV$ in the case of the K\"ahler potential.
There is no big difference also in the anomaly-mediated SUSY breaking models.

%%%%%%%%%%%%%%%%%%%%%%%%%%%%%%%%%%%%%%%%%%%%%%%%%%%%%%%%%%%%%%
\section{DM-genesis}%%%%%%%%%%%%%%%%%%%%%%%%%%%%%%%%%%%%%%%%%%
%%%%%%%%%%%%%%%%%%%%%%%%%%%%%%%%%%%%%%%%%%%%%%%%%%%%%%%%%%%%%%
Finally, we explain the subsequent consequences of late-time decay of
Q-balls. Although the full Boltzmann equations to calculate the 
LSP relic density during the decay of Q-balls are rather complicated, 
especially in the case of the Q-ball dominated Universe, 
the final abundance of the LSP can be approximately expressed by 
a simple analytical form~\cite{Fujii_Hamaguchi1,Hamaguchi_Fujii2}.  

Note that, in any case, the Boltzmann equations for the neutralino LSP are 
reduced to the single form for $\tau<\tau_{d}$:
\begin{equation}
\dot{n}_{\chi}+3 H n_{\chi}=-\vev{\sigma v}_{\chi}n_{\chi}^2,
\label{Boltzamann} 
\end{equation}
where $\tau_{d}=Q/\Gamma_{Q}$ denotes the lifetime of the Q-ball.  Here we have
assumed that $T_{d}$ is well below the freeze-out temperature of the LSP.
Introducing 
the yield $Y_{\chi}\equiv n_{\chi}/s$, where $s$ is the entropy density
of the Universe, the above equation can be written as
\begin{equation}
\frac{d Y_{\chi}}{d T}=\sqrt{\frac{8 \pi^2 g_{*}}{45}}\left(
1+\frac{T}{g_{*}}\frac{d g_{*}}{d T}\right) \vev{\sigma v}_{\chi}M_{*}Y_{\chi}^2\;.
\label{yield_version} 
\end{equation}   
Here, $T$ denotes the cosmic temperature. 
Because the LSPs become highly non-relativistic soon after they are 
produced by the Q-ball decays, we can expect that the $T$-dependent
component  of the annihilation cross section is likely to be subdominant.
This is particularly true when the LSP has a non-negligible component of 
Higgsino and/or Wino.
In this case we can write $\vev{\sigma v}_{\chi}\simeq {\rm
const}$, and in conjunction with an additional
approximation $g_{*}(T)\simeq g_{*}(T_{d})={\rm const}$, we can solve
Eq.~(\ref{yield_version}) analytically~\cite{Fujii_Hamaguchi1}:
\begin{equation}
Y_{\chi}(T)=\left[\frac{1}{Y_{\chi}(T_{d})}+
\sqrt{\frac{8\pi^2 g_{*}(T_{d})}{45}}\vev{\sigma v}_{\chi}M_{*}(T_{d}-T)
\right]^{-1}. 
\label{final_yield2}
\end{equation}

We can see that, if the initial abundance $Y_{\chi}(T_{d})$ is large enough, 
the final abundance $Y_{\chi 0}$ for $T\ll T_{d}$ is expressed
independently of $Y_{\chi}(T_{d})$ as
\begin{equation}
 Y_{\chi 0}\simeq Y_{\chi}^{\rm approx}
=\sqrt{\frac{45}{8 \pi^2 g_{*}(T_{d})}}\;
\frac{\vev{\sigma v}_{\chi}^{-1}}{M_{*}T_{d}}\;.
\label{final_yield}
\end{equation}
In terms of the density parameter, this is rewritten
as~\cite{Fujii_Hamaguchi1,Hamaguchi_Fujii2} 
\begin{equation}
 \Omega_{\chi}h^2\simeq 0.1 \left(\frac{10}{g_{*}(T_{d})}\right)^{1/2}
\left(\frac{m_{\chi}}{100\GEV}\right)\left(\frac{300\MEV}{T_{d}}\right)
\left(\frac{10^{-7}\GEV^{-2}}{\vev{\sigma v}_{\chi}}\right)\;.
\label{density_parameter}
\end{equation}               
This result clearly shows that we need a fairly large annihilation cross 
section $\vev{\sigma v}_{\chi}$ $\sim$ $10^{-(8\sim7)}$ $\GEV^{-2}$ in
order to explain the required mass density of dark matter with 
a typical range of $T_{d}$. Interestingly, the typical annihilation
cross section of Higgsino and Wino has also this size.
This opens up a new possibility to explain both the baryon asymmetry and 
dark matter at the same time with a single mechanism.
                                       
On the other hand, if the first term in Eq.~(\ref{final_yield2}) dominates, 
the final abundance of the LSP is given by
\begin{equation}
 Y_{\chi 0}\simeq Y_{\chi}(T_{d})\approx \left(\frac{n_{B}}{s}\right)_{0}\;
\left(\frac{n_{\phi}}{n_{B}}\right)\;,
\label{Bino_case}
\end{equation}
where $(n_{B}/s)_{0}\simeq 10^{-10}$ is the current value of the 
baryon asymmetry and $(n_{\phi}/n_{B})$ is fixed when the AD field
starts coherent oscillations and remains the same until it finally
decays.  Such a situation appears when the LSP is nearly pure Bino.  
One can easily understand this relation by noting 
that each decay of the $\phi$ field produces nearly one LSP.
In this case, the late-time decay of Q-balls 
causes a big difficulty. {}From Eq.~(\ref{Bino_case}),
we can see that it results in too large mass density of the LSP: 
\begin{equation}
 \Omega_{\chi}\approx \left(\frac{n_{\phi}}{n_{B}}\right)\left(\frac{m_{\chi}}{m_{p}}\right)
\Omega_{B}\;.
\end{equation}
Note that, because of the R-parity conservation, the relation $(n_{\phi}/n_{B})\geq3$
always  hold. In order not to over-produce the LSPs in the presence of
the late-time decay of Q-balls, we need an extremely light Bino:
\begin{equation}
 m_{\chi}\lsim 1.7\GEV \left(\frac{\Omega_{\chi}}{5\;\Omega_{B}}\right)\;,
\end{equation}           
which is clearly unrealistic. 

%%%%%%%%%%%%%%%%%%%%%%%%%%%%%%%%%%%%%%%%%%%%%%%%%%%%%%%%%%%%%%%
\section{Low-energy consequences in direct and indirect detections}
%%%%%%%%%%%%%%%%%%%%%%%%%%%%%%%%%%%%%%%%%%%%%%%%%%%%%
%%%%%%%%%%%%%%%%%%%%%%%%%%%%%%%%%%%%%%%%%%%%%%%%%%%%%%%%%%%%%%%
As we have seen in the previous section, the late-time decay of Q-balls
requires a quite large annihilation cross section of the LSP to provide 
the required mass density of dark matter.
In the rest of the paper, we consider low-energy consequences of this
result in several dark-matter searches by 
adopting the mSUGRA and mAMSB models.
First, we investigate the detection possibility of 
the non-thermal dark matter in the direct detection, 
and then calculate
the indirect detection rate observing the 
high energy neutrino flux from the center of the Sun. 
We also add the estimation of the hard positron flux, which 
is produced by the direct decays of gauge 
bosons produced by pair annihilations of the LSPs.

We have already had an estimation of the direct detection rate of the 
non-thermal dark matter in the 
previous work~\cite{Hamaguchi_Fujii2}. However, in this time,
we further restricts the allowed parameter space by implementing 
the recent WMAP result~\cite{WMAP}, which gives us more definitive predictions 
of AD baryogenesis.
Furthermore, we compare various detection rates of 
the non-thermal dark matter
with those in the standard thermal scenario. 
The required annihilation cross section of the LSP in  AD
baryogenesis would lead to only a very 
small fraction of the required dark matter density in the standard
scenario. Because the detection rates are  
proportional to the local neutralino density 
in the first two detection methods, we can obtain the 
corresponding detection rates in the standard scenario by 
rescaling them by a factor $(\Omega_{\chi}^{\rm th}h^2/\Omega_{DM}h^2)$.
In the case of the positron flux, this rescaling can be done 
by multiplying a factor $(\Omega_{\chi}^{\rm th}h^2/\Omega_{DM}h^2)^2$.
These procedures clarify the differences of the detection rates 
between the non-thermal  and the standard scenarios at the same SUSY-breaking 
parameters.~\footnote{
We have independently constructed all the required computer programs
for the above mentioned detection methods. We found quite good
agreements to the  results in other papers based on {\it DarkSUSY}
computer code.}

%%%%%%%%%%%%%%%%%%%%%%%%%%%%%%%%%%%%%%%%%%
\subsection{\Large Direct detection}
%%%%%%%%%%%%%%%%%%%%%%%%%%%%%%%%%%%%%%%%%
If the neutralino LSP is a dominant component of the halo dark matter,
we may observe the small energy deposit within a detector due to
LSP-nucleus scattering. 
This observation may provide the most promising way to confirm the
existence of neutralino dark matter.
The interactions of neutralinos with matter are usually dominated by
scalar couplings for relatively heavy nuclei 
$A\gsim 20$~\cite{scalar_domination,Jungman}.
These interactions are mediated by light $h$ and heavy $H$ exchanges or
sfermion $\tilde{f}$ exchanges.
The former diagrams contain $h\chi\chi$  and $H\chi\chi$ couplings,
which are suppressed for Bino-like LSPs.
On the other hand, if the LSP has a significant 
component of the Higgsino, these couplings
are strongly  enhanced.
In the case of Wino-like dark matter, they are also enhanced by a factor
$g_2/(g_1 \tan \theta_W)$. 
These facts give us much more promising possibility to 
find signals of the non-thermal LSP dark matter in the near future 
experiments~\cite{Hamaguchi_Fujii2}.
In this section, we investigate $\chi$-proton scalar 
cross section~\cite{Drees_Nojiri,Jungman} in
the mSUGRA and mAMSB models, for both the non-thermal and the
standard thermal freeze-out scenarios.

\subsubsection{Parameter space and direct detection in the mSUGRA}
First, let us discuss the allowed parameter space and corresponding
decay temperature of Q-balls to explain the required dark matter density.  
In the framework of the minimal supergravity (mSUGRA), there are four
continuous free parameters and one binary choice:
\begin{eqnarray}
 m_0,\,\, M_{1/2},\,\, A_0,\,\, \tan \beta,\,\,  {\rm sgn}(\mu),
\end{eqnarray}
where $m_0$, $M_{1/2}$, $A_0$ are the universal soft scalar mass,
gaugino mass, and trilinear scalar coupling given at the grand unified
theory (GUT) scale $M_{G} \simeq 2\times 10^{16}$ GeV, respectively. 
All the couplings and mass parameters at the weak scale are obtained
through the renormalization group (RG) evolution. 
We have used {\em SoftSusy 1.7} code~\cite{SoftSusy} for this purpose.
The code includes two-loop RG equations, one-loop self-energies for all
the particles and one-loop threshold corrections from SUSY particles to
the gauge and Yukawa coupling constants following the method of
Ref.~\cite{Pierce:1996zz}.

\begin{figure}[t]
 \begin{minipage}{0.49\linewidth}
\begin{center}(a)

 \includegraphics[width=0.8\linewidth]{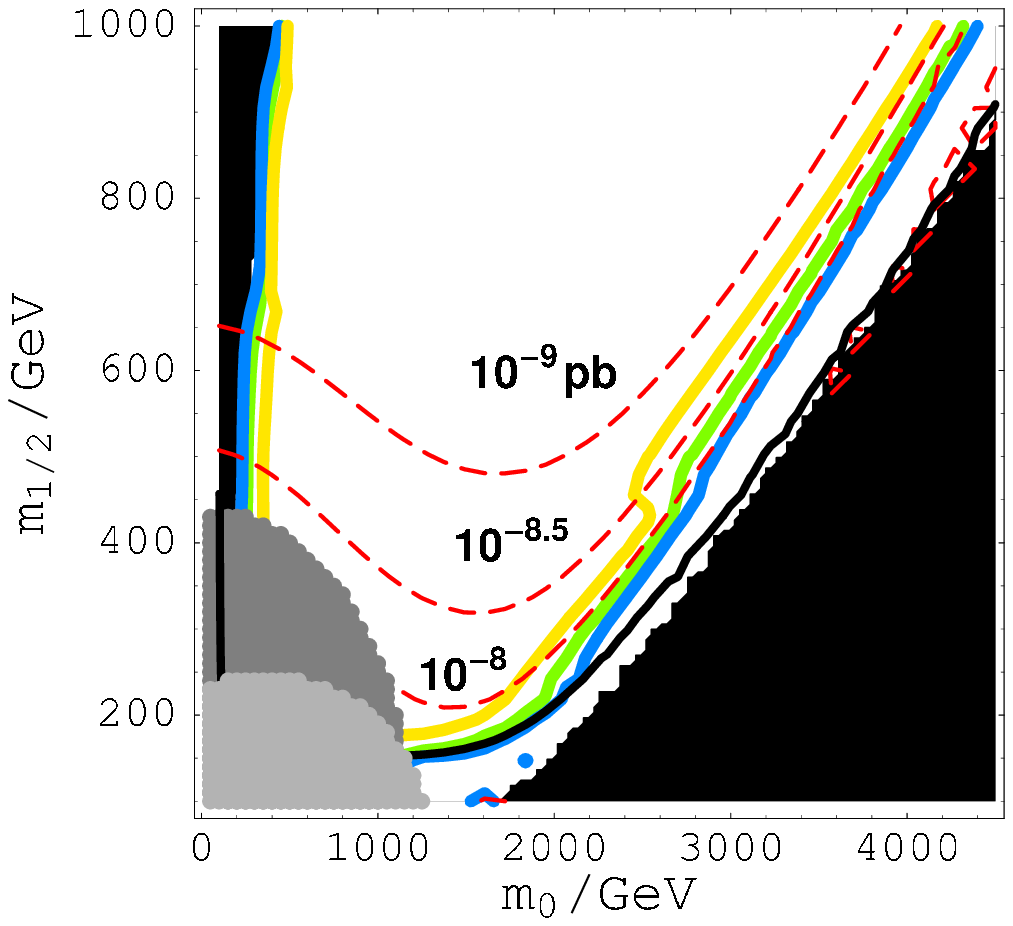} 
\end{center}
\end{minipage}
\begin{minipage}{0.49\linewidth}
\begin{center}(b)

  \includegraphics[width=0.8\linewidth]{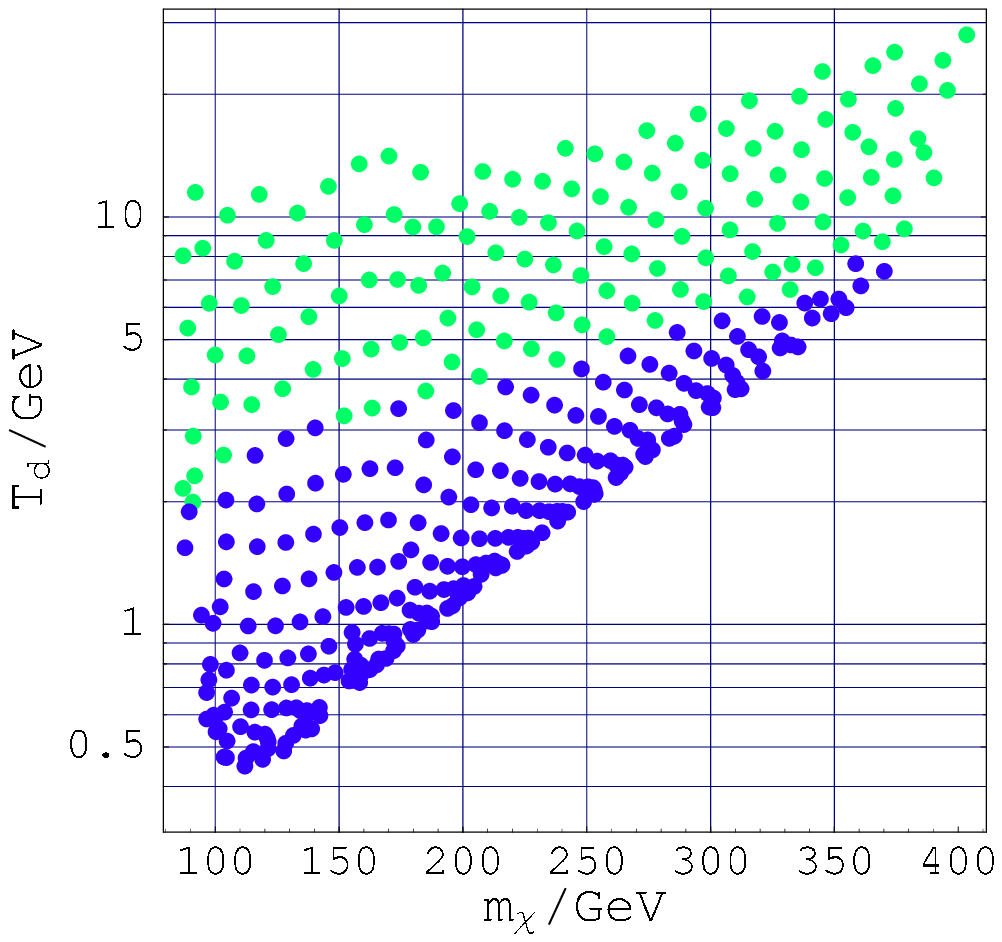} 
\end{center}
\end{minipage}
\caption{(a) The contours of $\Omega^{\rm th}_{\chi}h^2$ and
 $\sigma_{\chi-p}~[{\rm{pb}}]$. (b) The decay
 temperature of Q-ball which leads to the desired mass density of dark matter.
 ($\tan\beta=45$)}
\label{fig:sugra45}
\end{figure}

In Fig.~\ref{fig:sugra45}-(a), we show contours of the relic
neutralino density in the standard thermal freeze-out scenario,
$\Omega^{\rm th}_{\chi}h^2$, for $\tan\beta = 45$ in the ($m_0$, $M_{1/2}$)
plane. The figure also contains the contours of $\chi$-proton scalar cross
section $\sigma_{\chi-p}$, which will be explained later in this 
section.
We have used {\em micrOMEGAs} code~\cite{micrOMEGAs} to compute the relic density here.
The three thick lines are contours of $\Omega^{\rm th}_{\chi}h^2$,
corresponding to $\Omega^{\rm th}_{\chi}h^2 = 0.1$, $0.3$, and $1.0$
from the bottom up, respectively.

Here, we have taken $A_0=0$, and the sign of $\mu$ to be positive which
is desirable to avoid a large deviation of the branching ratio of the
$b\to  s\gamma$ from the observations.
We conservatively adopted the following constraint on the 
$b\to s\gamma$ branching ratio:
\begin{eqnarray}
 2\times 10^{-4} < B(B\to X_s \gamma) < 4\times 10^{-4}.
\end{eqnarray}
The dark shaded region denotes where the above constraint is 
violated.~\footnote{Even if we adopt the recent PDG average of CLEO
and Belle measurements, $B(B\to X_{s}\gamma)=(3.3\pm 0.4)\times
10^{-4}$~\cite{PDG}, the allowed parameter space in the focus 
point region is not affected at all.
On the other hand, 
the parameter space in the co-annihilation region is severely
constrained in the case of large $\tan \beta$.}
The region below the black solid line is excluded by the
chargino mass limit, $m_{\chi^\pm}\ge 104$ GeV~\cite{chargino-bound}.
The mass of the lightest Higgs boson is smaller than $114$ GeV in the
light shaded region, which is excluded by the CERN $e^+e^-$ collider LEP
II~\cite{Higgs-bound}.
The black shaded regions are excluded because the electroweak symmetry
breaking does not take place or the lightest stau becomes the LSP.

%%%%%%%%%%%%%%%%%%%%%%%%%%%%%%%%%%%%%%%%%%%%%%%%%%%%%%%%%%%%%%%%%%%%%
%%%%%%%%%%  Omega vs SUGRA parameter, T_d vs m_{\chi}  %%%%%%%%%%%%%%
%%%%%%%%%%%%%%%%%%%%%%%%%%%%%%%%%%%%%%%%%%%%%%%%%%%%%%%%%%%%%%%%%%%%%
%%%%%%%%%%%%%%%%%%%%%%%%%%%%%%%%%%%%%%%%%%%%%%%%%%%%%%%%%%%%%%%%%%%%%%%%%%%%

As we can see from Fig.~\ref{fig:sugra45}-(a),
$\Omega^{\rm th}_{\chi}h^2$ are too
large to be consistent with the WMAP result,
$\Omega_{DM}h^2=0.1126^{+0.0161}_{-0.0181}$, in most of the parameter space. 
However, as the parameter sets approach the ``focus
point''~\cite{focus-point} region ($m_0
\gsim$ TeV), the pair annihilation of neutralinos becomes more
efficient because of the increase of Higgsino component in the LSP, 
and then, there appears a very thin parameter region that gives the correct
abundance of the LSP~\cite{focus-point-DM}.~\footnote{  
At the left border of the figure, $\Omega^{\rm th}h^2$ can give the
required abundance of dark matter.
This is the so-called co-annihilation region where the Bino-like LSP is almost
degenerate with the lightest stau.}   
As we further increase $m_0$, $\Omega^{\rm th}h^2$ continues to decline,
and the late-time decay of Q-balls  
comes to be allowed to play an 
important role.

In Fig.~\ref{fig:sugra45}-(b), we show the decay temperature of Q-ball
in the ($m_{\chi}$, $T_d$) plane, which leads to the desired mass density
of dark matter (see Eq.~(\ref{density_parameter})).
The light shaded (green) points denote the required $T_d$'s for
$0.03\le \Omega^{\rm th}_{\chi}h^2\le 0.1$, and the dark shaded (blue)
points for $\Omega^{\rm th}_{\chi}h^2<0.03$. 
This result suggests that the desirable parameter sets for the present
scenario would give $\Omega^{\rm th}h^2 < 0.03$ in the thermal
freeze-out scenario, since $T_d\le {\cal O}(1)$ GeV is expected in the typical 
models of AD baryogenesis (Section 2).
We can also see that the 
anticipated 
Q-ball decay temperature prefers the existence of a relatively light 
neutralino, $m_{\chi}\lsim 300\GEV$.

In the above calculation of the decay temperature, we have included only the
$s$-wave contribution in the annihilation cross section of the 
neutralinos for the reasons explained before.
We have also neglected the possible co-annihilation effects with the
lightest charginos.
This procedure can be justified as long as the decay temperature of the
Q-ball is smaller than their mass difference $\delta m$.
Actually, this condition is satisfied in almost the entire parameter space, 
where we have confirmed $\delta m = {\cal O}(10)\GEV$.

%%%%%%%%%%%%%%%%%%%%%%%%%%%%%%%%%%%%%%%%%%%%%%%%%%%%%%%%%%%%%%%%%%%%

%%%%%%%%%%%%%%%%%%%%%%%%%%%%%%%%%%%%%%%%%%%%%%%%%%%%%%%%%%%%%%%%%%%%%
%%%%%%%%%%  The Direct detection rates                 %%%%%%%%%%%%%%
%%%%%%%%%%%%%%%%%%%%%%%%%%%%%%%%%%%%%%%%%%%%%%%%%%%%%%%%%%%%%%%%%%%%%
\begin{figure}[t]
 \begin{minipage}{0.49\linewidth}

\begin{center}(a)

 \includegraphics[width=0.8\linewidth]{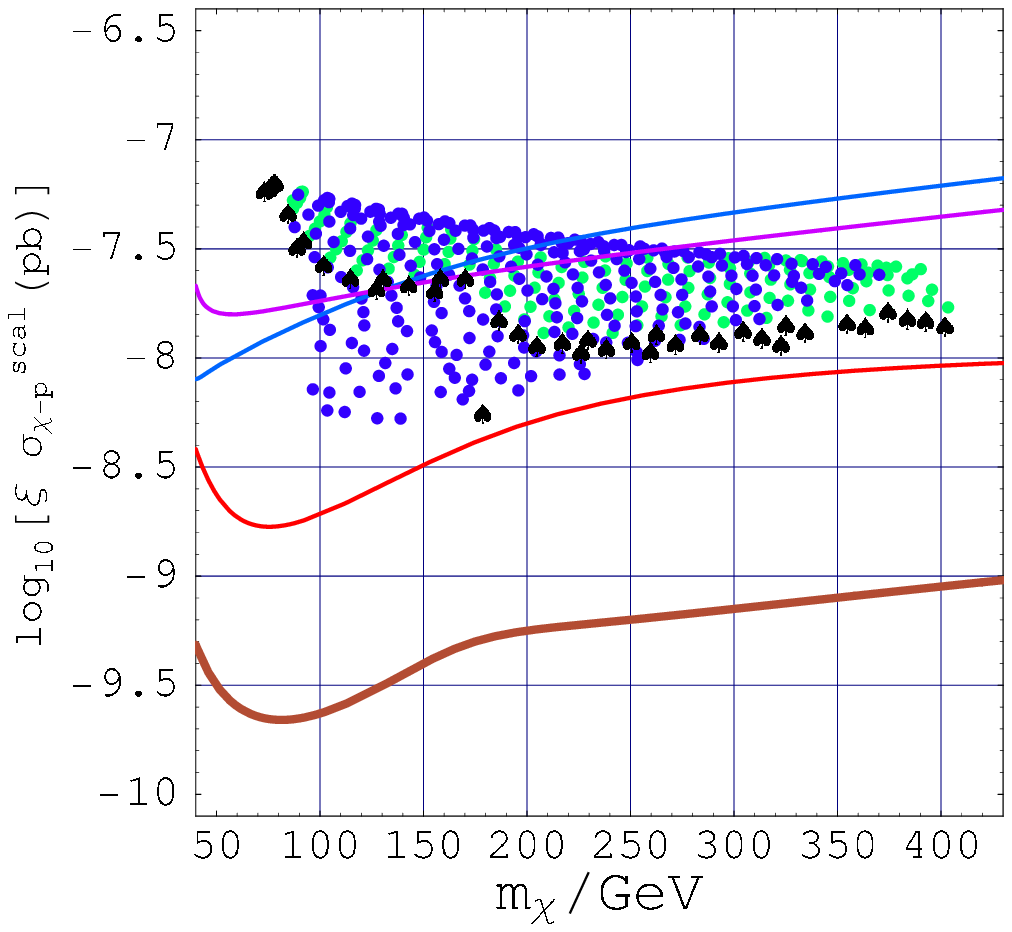} 
\end{center}
\begin{picture}(0,0)
\put(170,75){ZEPLIN}
\put(170,116){GENIUS}
\put(170,144){EDELWEISS}
\put(190,165){CDMS}
\end{picture}
\end{minipage}
\begin{minipage}{0.49\linewidth}
\vspace{-1cm}
\begin{center}(b)

  \includegraphics[width=0.8\linewidth]{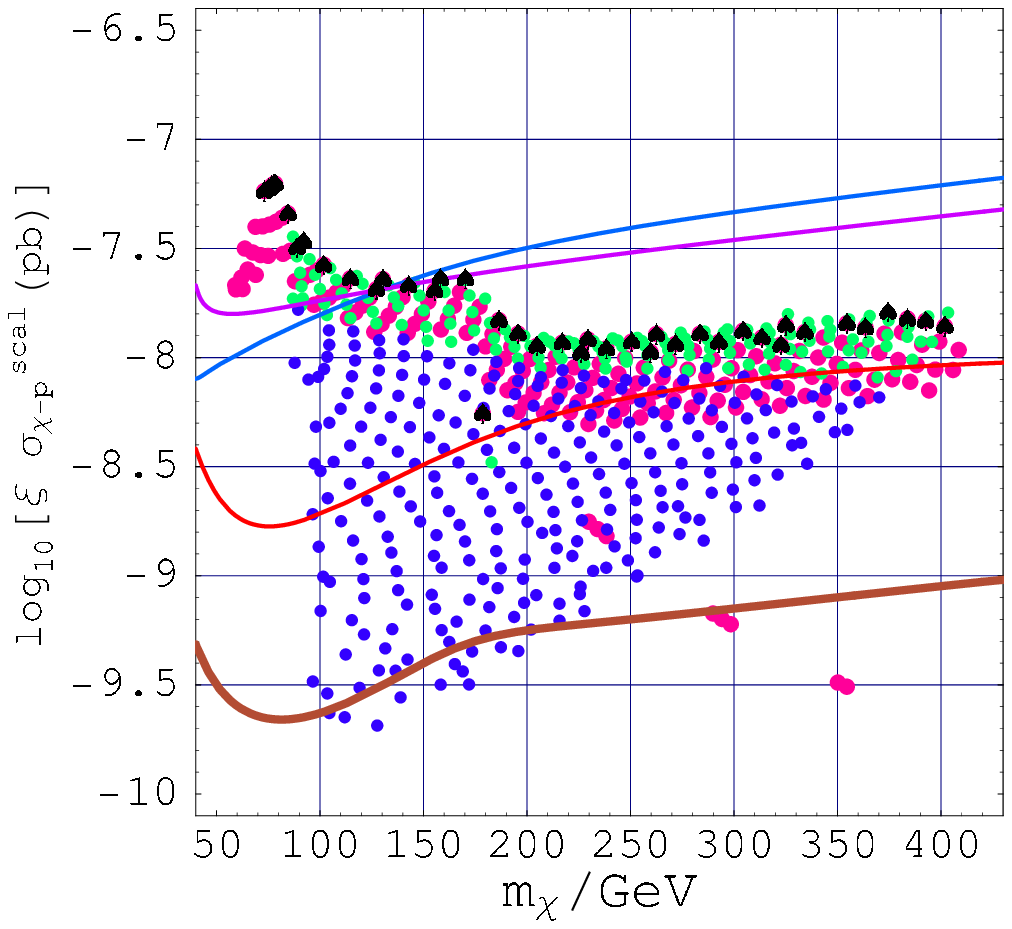} 
\end{center}
\end{minipage}
 
\caption{ The effective cross section of proton-$\chi$ scalar interaction in the
 mSUGRA scenario. 
(a) The non-thermal scenario.
(b) The standard thermal freeze-out scenario. ($\tan\beta=45$)
}
\label{fig:sugradirect45}
\end{figure}
%%%%%%%%%%%%%%%%%%%%%%%%%%%%%%%%%%%%%%%%%%%%%%%%%%%%%%%%%%%%%%%%%%%%%%%%%%%%
Now, let us turn our attention to the direct detection of the
non-thermal dark matter.
As one can see from  Fig.~\ref{fig:sugra45}-(a), the $\chi$-proton cross 
section $\sigma_{\chi-p}$ becomes larger as $m_{0}$ increases 
because of the increase of the Higgsino component in the LSP. 
Since the increase of Higgsino
fraction reduces the relic density, we can expect larger 
direct detection rate in the non-thermal scenario.
In Fig.~\ref{fig:sugradirect45}, we show the effective $\chi$-proton cross
section in the ($m_{\chi}$, $\xi\sigma_{\chi -p}$) plane in the mSUGRA scenario
with $\tan\beta = 45$.
Fig.~\ref{fig:sugradirect45}-(a) shows the cross section in the
non-thermal scenario $(\xi=1)$, and Fig.~\ref{fig:sugradirect45}-(b) shows it
in the standard thermal scenario.
In  Fig.~\ref{fig:sugradirect45}-(b) the cross section is rescaled by
multiplying a factor $\xi=(\Omega^{\rm th}_{\chi}h^2/\Omega_{DM}h^2)$ where
$\Omega^{\rm th}_{\chi}h^2$ is smaller than $\Omega_{DM}h^2$, since
the detection rate is proportional to the local neutralino 
density.~\footnote{When 
$\Omega_{\chi}h^2$ in the thermal scenario is smaller than the
$\Omega_{DM}h^2$, total dark matter should consist of several populations in
addition to the neutralino.}   

\begin{figure}[t]
 \begin{minipage}{0.49\linewidth}
\begin{center}(a)

 \includegraphics[width=0.8\linewidth]{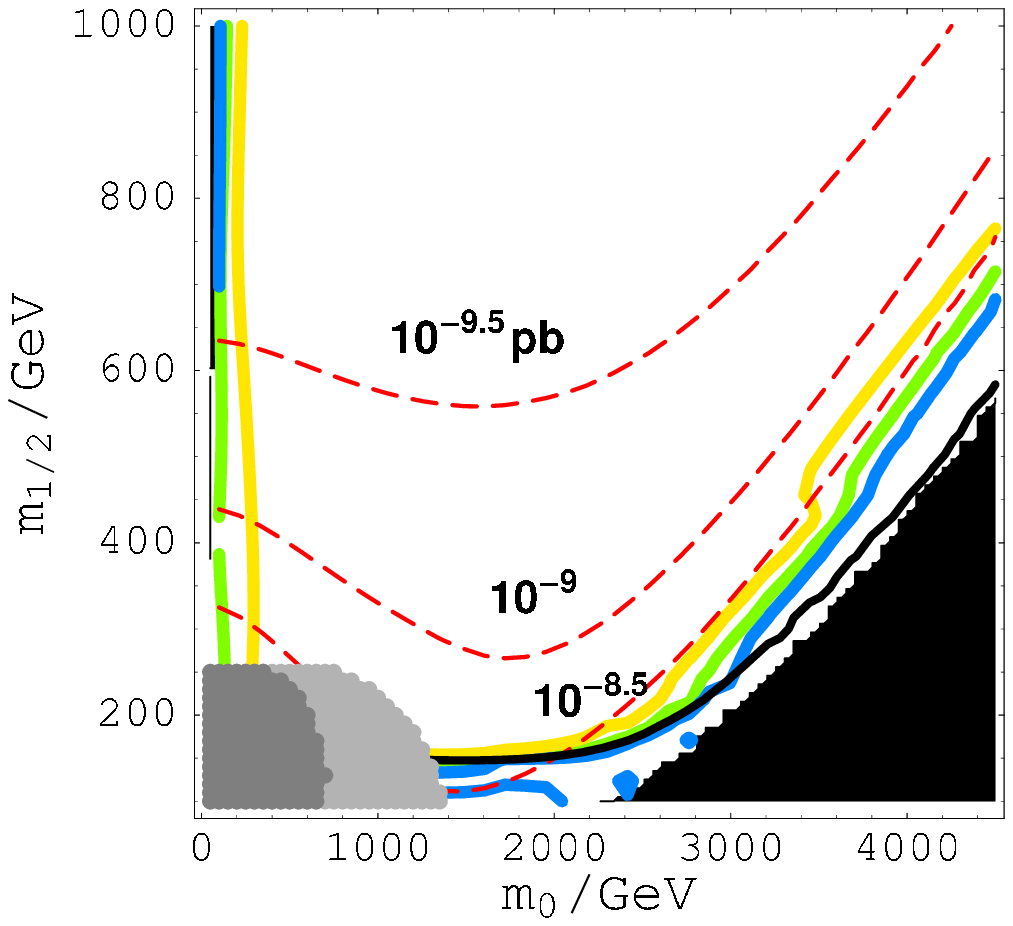} 
\end{center}
\end{minipage}
\begin{minipage}{0.49\linewidth}
\begin{center}(b)

  \includegraphics[width=0.8\linewidth]{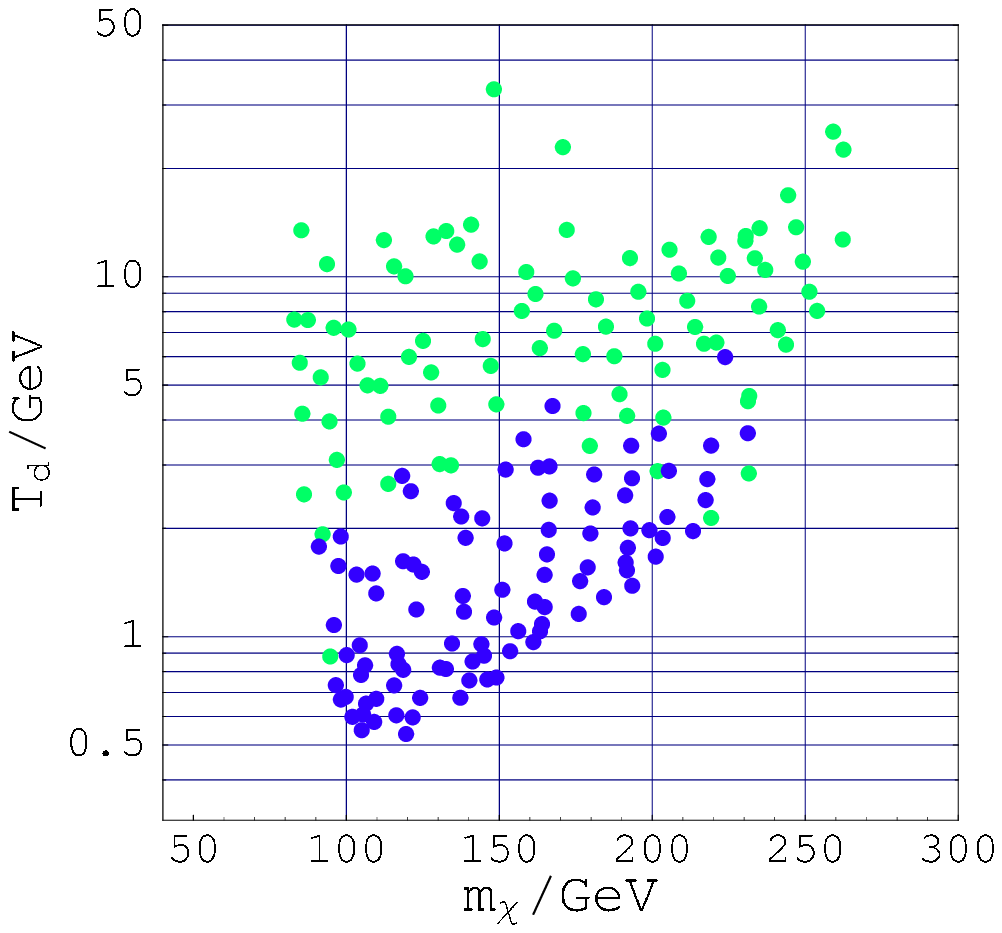} 
\end{center}
\end{minipage}
\caption{(a) The contours of $\Omega^{\rm th}_{\chi}h^2$
 and $\sigma_{\chi-p}~[\rm{pb}]$. (b) The decay
 temperature of Q-ball which leads to the desired mass density of dark
 matter.
($\tan\beta=10$)
}
\label{fig:sugra10}
\end{figure}

\begin{figure}[t!]
 \begin{minipage}{0.48\linewidth}
\begin{center}(a)

 \includegraphics[width=0.9\linewidth]{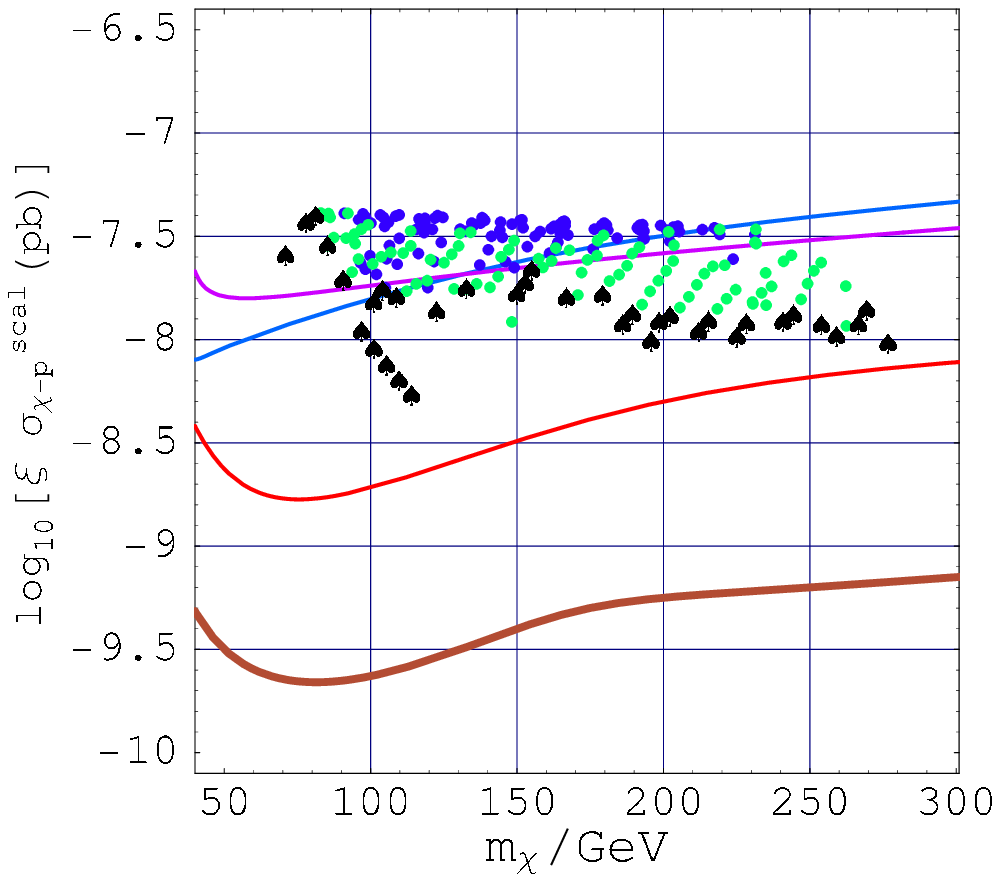} 
\end{center}
\end{minipage}
\begin{minipage}{0.48\linewidth}
\begin{center}(b)

  \includegraphics[width=0.9\linewidth]{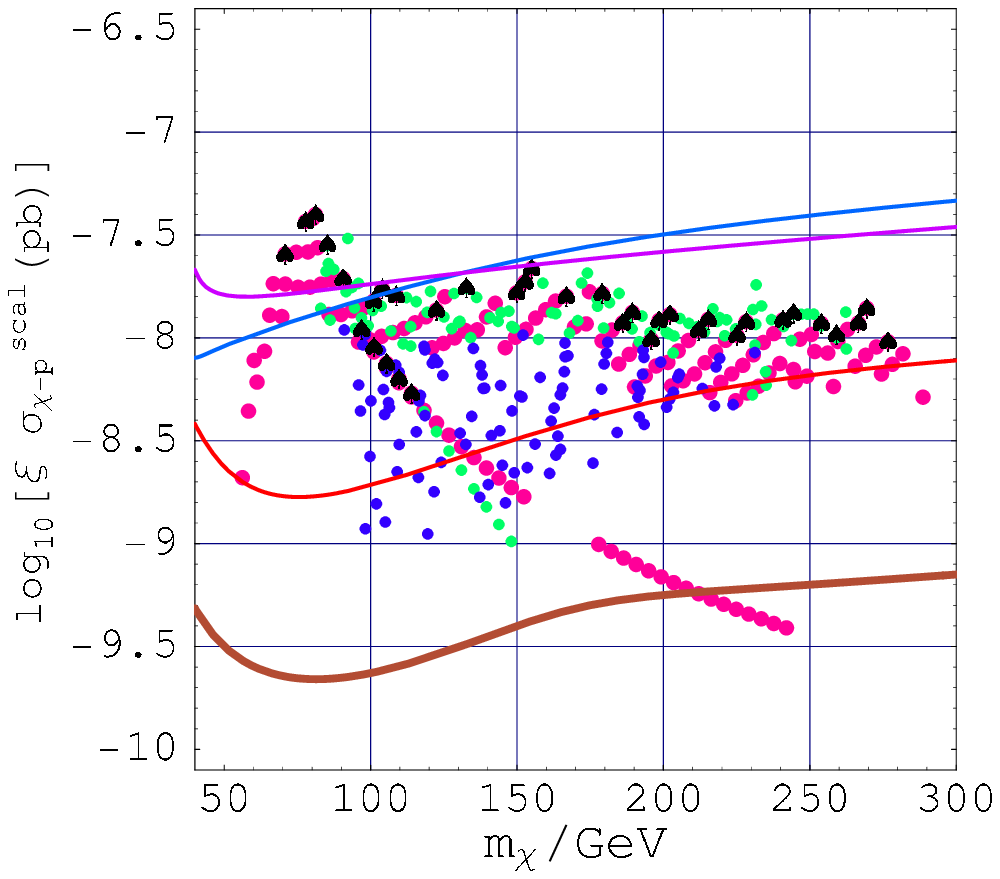} 
\end{center}\end{minipage}
 
\caption{The effective cross section of proton-$\chi$ interaction in the 
 mSUGRA scenario. 
(a) The non-thermal scenario.
(b) The standard thermal freeze-out scenario. ($\tan\beta=10$)
}
\label{fig:sugradirect10}
\end{figure}

The dark shaded (blue) points in the both figures correspond to the mSUGRA
parameters with $\Omega^{\rm th}_{\chi}h^2\le 0.03$, the light shaded
(green) points to those with $0.03\le\Omega^{\rm th}_{\chi}h^2\le 0.1$
and the medium shaded (purple) points to those with $0.1\le\Omega^{\rm
th}_{\chi}h^2\le 0.3$.    
%\footnote{The scarce points
%around the $\sigma_{\chi-p}^{\rm scal} \lsim 10^{-9}$pb correspond to
%the co-annihilation region at the right border of
%Fig. \ref{fig:sugra45}.  
%The scarcity of these points comes from the large intervals of the
%parameters we have swept, and with more small intervals this region can
%be expected to be plotted on a line of points (see
%Fig. \ref{fig:sugradirect10} also.). 
%}
We also plot the parameter sets which are consistent with the WMAP
experiment ($\Omega^{\rm th}_{\chi}h^2 = 0.1126^{+0.0161}_{-0.0181}$) in
the  standard thermal scenario as black ``$\spadesuit$''
points,~\footnote{Absence of $\spadesuit$ points in the co-annihilation
region is just because the small number of samplings in our calculations.} 
which are also plotted in  Fig.~\ref{fig:sugradirect45}-(a) 
just for convenience to comparison.
The four lines denote sensitivities of several direct detection
experiments: ZEPLIN MAX~\cite{ZEPLIN}, GENIUS~\cite{GENIUS}, EDELWEISS
II~\cite{EDELWEISS}, 
and CDMS (Soudan)~\cite{CDMS} from the bottom up, respectively.    
In Figs.~\ref{fig:sugra10} and \ref{fig:sugradirect10}, we also show
the corresponding figures for $\tan\beta = 10$ in the mSUGRA model, where
conventions are the same as in Figs.~\ref{fig:sugra45} and
\ref{fig:sugradirect45}.  

In the above calculations, we have adopted the following values of the
proton matrix elements for each of the three light quarks:
\begin{eqnarray}
 f_{T_u}= 0.019,\,\, f_{T_d}=0.041,\,\, f_{T_s}=0.14,
\label{eq:proton}
\end{eqnarray}
where $f_{T_q}\equiv\langle p|m_qq\bar{q}|p\rangle/m_p$.
For details about the calculation of the proton-$\chi$ cross section,
see Refs.~\cite{Drees_Nojiri,Jungman}.
%%%%%%%%%%%%%%%%%%%%%%%%%%%%%%%%%%%%%%%%%%%%%%%%%%%%%%%%%%%%%%%%%%%%%
%%%%%%%%%%  Omega vs SUGRA parameter, T_d vs m_{\chi} tanb=10%%%%%%%%
%%%%%%%%%%%%%%%%%%%%%%%%%%%%%%%%%%%%%%%%%%%%%%%%%%%%%%%%%%%%%%%%%%%%%
%%%%%%%%%%%%%%%%%%%%%%%%%%%%%%%%%%%%%%%%%%%%%%%%%%%%%%%%%%%%%%%%%%%%%%%%%%%%
%%%%%%%%%%%%%%%%%%%%%%%%%%%%%%%%%%%%%%%%%%%%%%%%%%%%%%%%%%%%%%%%%%%%%
%%%%%%%%%%  The Direct detection rates  tanb=10        %%%%%%%%%%%%%%
%%%%%%%%%%%%%%%%%%%%%%%%%%%%%%%%%%%%%%%%%%%%%%%%%%%%%%%%%%%%%%%%%%%%%
%%%%%%%%%%%%%%%%%%%%%%%%%%%%%%%%%%%%%%%%%%%%%%%%%%%%%%%%%%%%%%%%%%%%%%%%%%%%

{}From Figs.~\ref{fig:sugradirect45} and \ref{fig:sugradirect10},
we can see a clear difference between the thermal and non-thermal
scenarios. First of all, the detection rates in the non-thermal scenario 
are larger than those in the standard scenario, denoted by $\spadesuit$
points, by several times,
and most of the parameter space can be thoroughly surveyed by next generation 
detectors.~\footnote{In Fig.~\ref{fig:sugradirect45}, there are 
some points which lie fairly below the $\spadesuit$ points, where
the LSP is nearly pure Higgsino. At these points, we need to include 
higher order corrections to the neutralino-Higgs coupling constant
to obtain accurate detection rates~\cite{Drees_Higher}.}
Secondly, since the preferred parameter sets for AD baryogenesis predict
quite small relic densities 
in the thermal freeze-out scenario, the detection rates
for the corresponding points in the standard scenario become much
smaller than in the non-thermal scenario.
This may play a crucial role to reveal the true thermal history of our
Universe in the future.

%%%%%%%%%%%%%%%%%%%%%%%%%%%%%%%%%%%%%%%%%%%%%%%%%%%%%%%%%%%%%%%%%
\subsubsection{Parameter space and direct detection in the mAMSB}
%%%%%%%%%%%%%%%%%%%%%%%%%%%%%%%%%%%%%%%%%%%%%%%%%%%%%%%%%%%%%%%%%
Anomaly-mediated SUSY breaking (AMSB)~\cite{AMSB_models} is another interesting way to
mediate SUSY-breaking effects to the MSSM sector without conflicting
with the well known FCNC problem.
In the pure AMSB model, 
all the soft SUSY-breaking parameters are fully determined by 
$\beta$ functions of gauge and Yukawa coupling constants, and anomalous
dimensions of matter fields. 
Quite unfortunately, however, the pure AMSB model predicts negative
slepton masses, and hence is not capable of describing the real world.

Although many possible solutions have been proposed to this problem,
we adopt the simplest solution in the present paper.
We just assume the additional universal scalar mass $m_{0}$ at the GUT
scale, and then evolve the RG equations to obtain the low-energy
spectrum.
In this minimal framework (mAMSB), the entire parameter space is
specified by the following $4$ parameters:
\begin{eqnarray}
 m_{3/2},\,\, m_0,\,\, \tan\beta,\,\, {\rm sgn(\mu)}.
\end{eqnarray}
In this model, the gaugino masses are not modified and almost 
the same as those in the pure AMSB model 
except higher oder quantum corrections.
Their ratios at the weak scale
are approximately give by 
\begin{equation}
M_{1}\;:\;M_{2}\;:\; M_{3}\;\approx\; 2.8 \;:\;1\;:\;-8.3\;,
\end{equation}
and the Wino-like LSP is realized in almost the entire parameter space.
This fact has an important impact on the cosmology in the mAMSB model.
%%%%%%%%%%%%%%%%%%%%%%%%%%%%%%%%%%%%%%%%%%%%%%%%%%%%%%%%%%%%%%%%%%%%%
%%%% Omega vs SUGRA parameter, T_d vs m_{\chi} in AMSB %%%%%%%%%%%%%% 
%%%%%%%%%%%%%%%%%%%%%%%%%%%%%%%%%%%%%%%%%%%%%%%%%%%%%%%%%%%%%%%%%%%%%
\begin{figure}[ht]
 \begin{minipage}{0.48\linewidth}
\begin{center}(a)

 \includegraphics[width=0.9\linewidth]{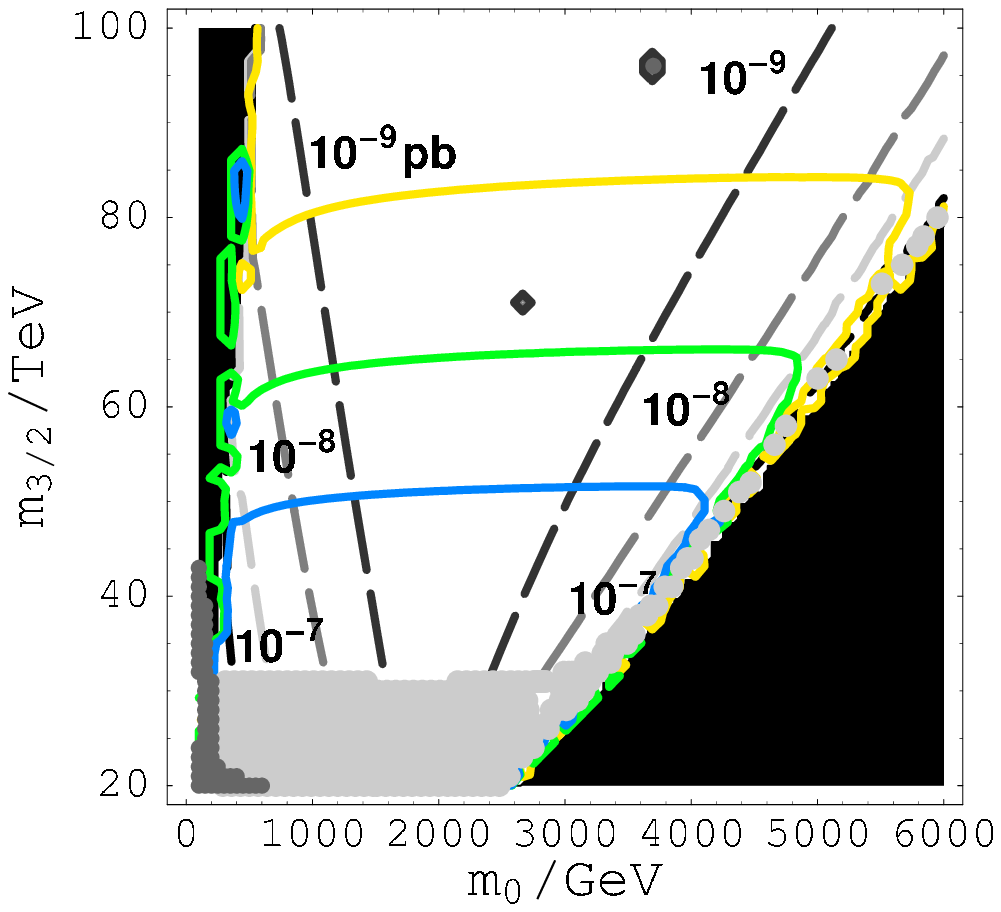} 
\end{center}
\end{minipage}
\begin{minipage}{0.48\linewidth}
\begin{center}(b)

  \includegraphics[width=0.9\linewidth]{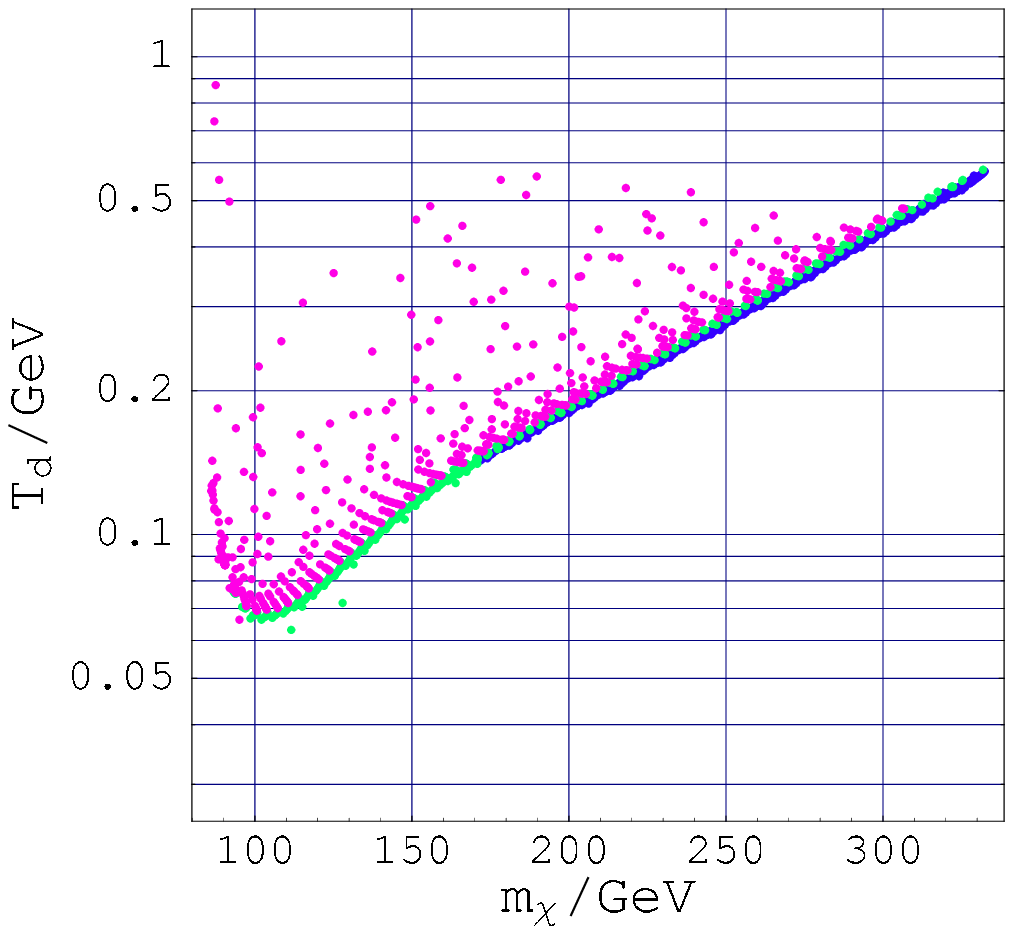} 
\end{center}
\end{minipage}
\caption{(a) The contours of $\Omega^{\rm th}_{\chi}h^2$ and 
$\sigma_{\chi-p}~[{\rm pb}]$. (b) The decay
 temperature of Q-ball which leads to the desired mass density of dark matter.
($\tan\beta=30$)}
\label{fig:anom30}
\end{figure}
%%%%%%%%%%%%%%%%%%%%%%%%%%%%%%%%%%%%%%%%%%%%%%%%%%%%%%%%%%%%%%%%%%%%%%%%%%%%

In Fig.~\ref{fig:anom30}-(a), we show contours of the relic neutralino
density in the thermal freeze-out scenario and the $\chi$-proton scalar
cross section for the case of  $\tan\beta = 30$ in the ($m_0$, $m_{3/2}$) plane. 
The horizontal lines correspond to the contours of the relic density,
with $\Omega^{\rm th}_{\chi} h^2= 10^{-3.2}$, $10^{-3.0}$ and
$10^{-2.8}$, from the bottom up, respectively. 
Here, we have taken sgn($\mu$) negative to avoid too large contribution to the
 branching ratio of $b\to s\gamma$.
The dashed lines are contours of the $\chi$-proton scalar cross section
$\sigma_{\chi-p}$ whose value is explicitly denoted in the figure. 
The light shaded region is excluded by the chargino mass 
limit~\cite{chargino_ALEPH}. 
The lightest Higgs boson is lighter than 114 GeV in the dark shaded
region.
The black shaded region denotes the region  where the electroweak symmetry breaking
cannot be implemented, or the lightest stau or sneutrino becomes the LSP.

As we can see from Fig.~\ref{fig:anom30}-(a), 
$\Omega^{\rm th}_{\chi}h^2$ is much 
smaller than the required abundance of the LSP because of the large  
annihilation cross section of the Wino-like LSP in most of the
parameter space. 
Therefore, in anyway, we need some non-standard thermal history to
explain the required mass density of dark matter in
this model, if we insist on the LSP as a primary component of the cold
dark matter. The most natural answer to this problem is given by 
Affleck--Dine baryogenesis. If we use this mechanism to explain the
observed baryon asymmetry, the associated late-time Q-ball decays 
with typical decay temperature $T_{d}\lsim {\cal{O}}(1)\GEV$ can
naturally generate the required abundance of the LSP at the 
same time~\cite{Fujii_Hamaguchi1,Hamaguchi_Fujii2}.
In fact, almost the entire parameter space of the mAMSB model 
is consistent with the Affleck--Dine baryo/DM-genesis 
scenario.~\footnote{In Ref.~\cite{Moroi_Randall}, the authors have proposed a
generation mechanism of Wino dark matter by late-time
decays of heavy moduli field. In this case, however,
we have to tune its coupling to the SM fields to obtain the
correct Wino abundance. Furthermore, in any way,
we have to rely on AD baryogenesis 
to produce enough baryon asymmetry in the presence of the huge entropy 
production from  the moduli decays.}

In Fig.~\ref{fig:anom30}-(b), we show the decay temperature of Q-ball in
the ($m_{\chi}$, $T_d$) plane, which leads to the required mass
density.  The dark shaded (blue) points correspond to the parameters with $Z_H\le
0.01$, the light shaded points (green) to those with $0.01\le Z_H \le
0.03$ and the medium shaded (purple) to those with $0.03\le Z_H$,
where $Z_H$  is a Higgsino fraction in the LSP.~\footnote{
The lightest neutralino $\chi_{1}^{0}$ is defined as:
\begin{eqnarray}
 \chi^0_1 = N_{11} \tilde{B} + N_{12} \tilde{W}^3 + N_{13}\tilde{H}^0_1 + N_{14}\tilde{H}^0_2,
\end{eqnarray}
where the coefficients $N_{1j}$ are obtained by diagonalizing the
neutralino mass matrix.
Here, we call $Z_H= |N_{13}|^2 + |N_{14}|^2$ the Higgsino fraction.
}
One can see that the required decay temperatures are about one magnitude
smaller than those in the mSUGRA model, which comes from a larger
annihilation cross section of the Wino-like LSP.
As in the case of the mSUGRA model, we have included only the $s$-wave 
contributions in the annihilation cross section of the neutralino to
calculate the decay temperature.
We have also neglected the possible co-annihilation effects with the
lightest charginos.
In the mAMSB model, however, the mass splitting between the lightest
chargino and neutralino is of the order of 
$100\MEV\sim 1\GEV$~\cite{wino_degeneracy},
which has comparable size as $T_d$.
Hence, the co-annihilation effects may slightly 
change the required decay temperature, which is at most a factor of few.
%%%%%%%%%%%%%%%%%%%%%%%%%%%%%%%%%%%%%%%%%%%%%%%%%%%%%%%%%%%%%%%%%%%%%
%%%% DIRECT in mAMSB                                   %%%%%%%%%%%%%% 
%%%%%%%%%%%%%%%%%%%%%%%%%%%%%%%%%%%%%%%%%%%%%%%%%%%%%%%%%%%%%%%%%%%%%
\begin{figure}[h!]
 \begin{minipage}{0.48\linewidth}
\begin{center}(a)

 \includegraphics[width=0.9\linewidth]{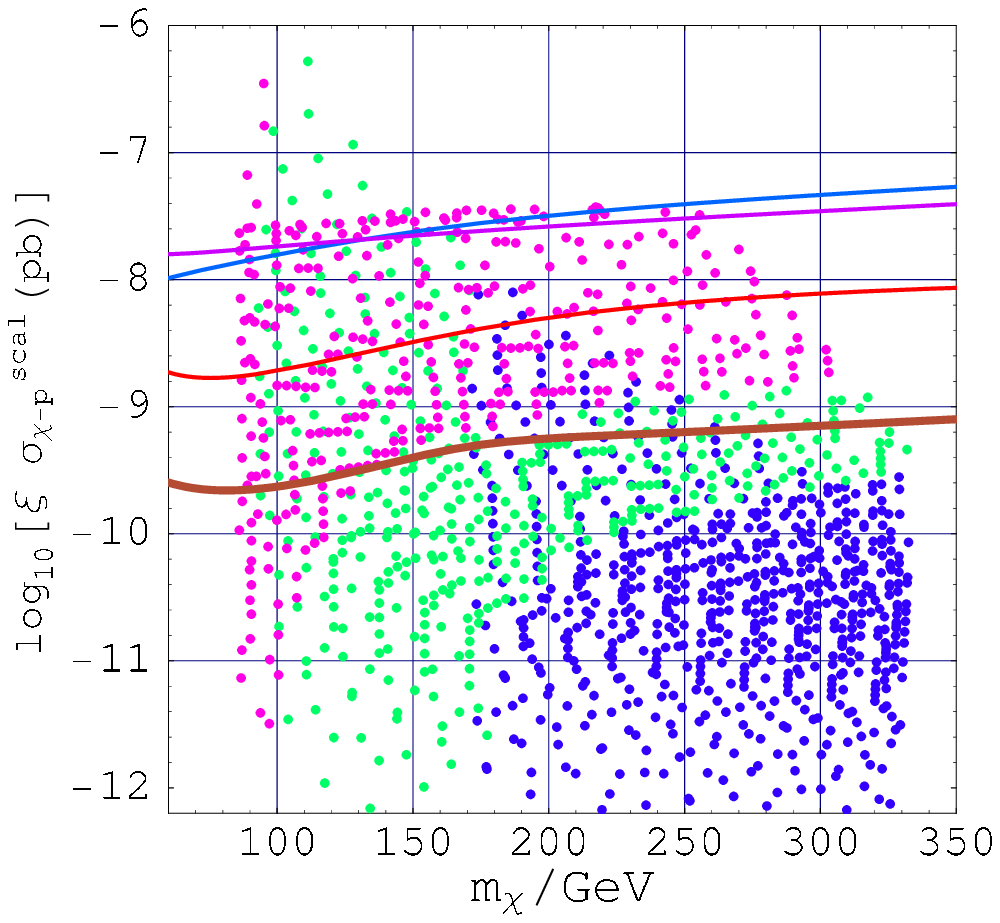} 
\end{center}
\end{minipage}
\begin{minipage}{0.48\linewidth}
\begin{center}(b)

 \includegraphics[width=0.9\linewidth]{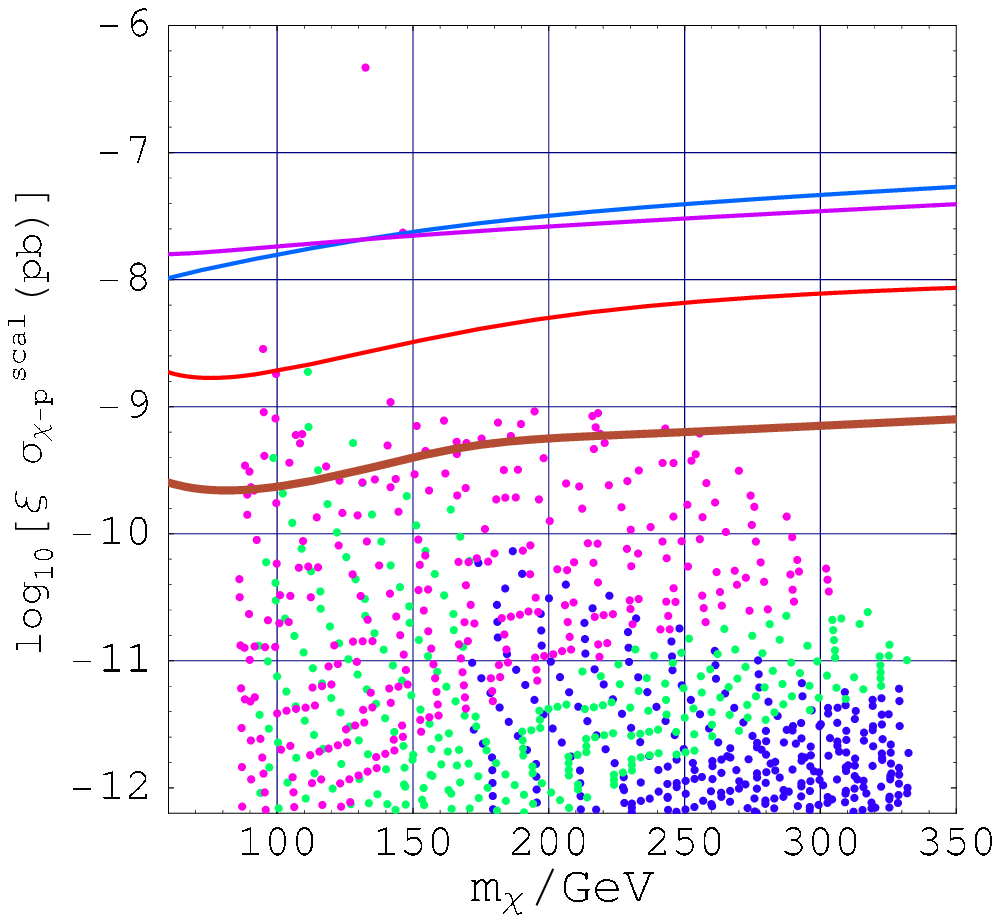} 
\end{center}
\end{minipage}\caption{The effective cross section of proton-$\chi$ interaction in the 
 mAMSB scenario. (a) The non-thermal scenario. (b)
The standard thermal freeze-out scenario. $(\tan\beta=30)$}
\label{fig:anomdirect30}
\end{figure}
%%%%%%%%%%%%%%%%%%%%%%%%%%%%%%%%%%%%%%%%%%%%%%%%%%%%%%%%%%%%%%%%%%%%%%%%%%%%
%%%%%%%%%%%%%%%%%%%%%%%%%%%%%%%%%%%%%%%%%%%%%%%%%%%%%%%%%%%%%%%%%%%%%%%%%%%
%%%%%%%%%%%%%%%%%%%%%%%%%%%%%%%%%%%%%%%%%%%%%%%%%%%%%%%%%%%%%%%%%%%%%%%%%%%

Now, let us discuss the direct detection rates in the mAMSB model.
In Fig.~\ref{fig:anom30}-(a), there are two important
factors to determine the $\chi$-proton cross section in this model.
When the LSP is nearly pure Wino $(m_{0}\lsim 2\TEV)$, 
the $\chi$-proton cross section is determined dominantly 
by the heavy Higgs exchange because of a large $\tan \beta$
enhancement. In this region, the shape of contours are controlled 
by the heavy Higgs boson mass $m_{H}$, and the $\chi$-proton cross
section scales as $\sigma_{\chi-p}\propto 1/m_{H}^{4}$.
As the parameter sets come close to the focus point region, 
the Higgsino component in the LSP becomes significant.
In this region, the $\chi$-proton cross section is 
primarily determined  by the light Higgs exchange, and the contours of
$\sigma_{\chi-p}$ are controlled by the Higgsino fraction
and have the same behavior seen in the mSUGRA model.

In Fig.~\ref{fig:anomdirect30}, we show the effective $\chi$-proton 
cross section in the ($m_{\chi}$, $\xi\sigma_{\chi -p}$) plane 
for both the non-thermal (a) and the thermal freeze-out  scenarios (b).
The four lines are the sensitivities of several direct detection
experiments explained in the previous section.
Conventions of the shading (coloring) of the plotted points are the  same 
as those in Fig.~\ref{fig:anom30}-(b): it shows the Higgsino fraction in the
LSP. As in the mSUGRA model, we have set $\xi=1$ in the non-thermal
scenario and $\xi=(\Omega_{\chi}^{\rm th}h^2/\Omega_{DM}h^2)$ in the 
thermal scenario to obtain 
the effective $\chi$-proton cross section.

\begin{figure}[ht]
 \begin{minipage}{0.49\linewidth}
\begin{center}(a)

\vspace{.1cm}
 \includegraphics[width=0.9\linewidth]{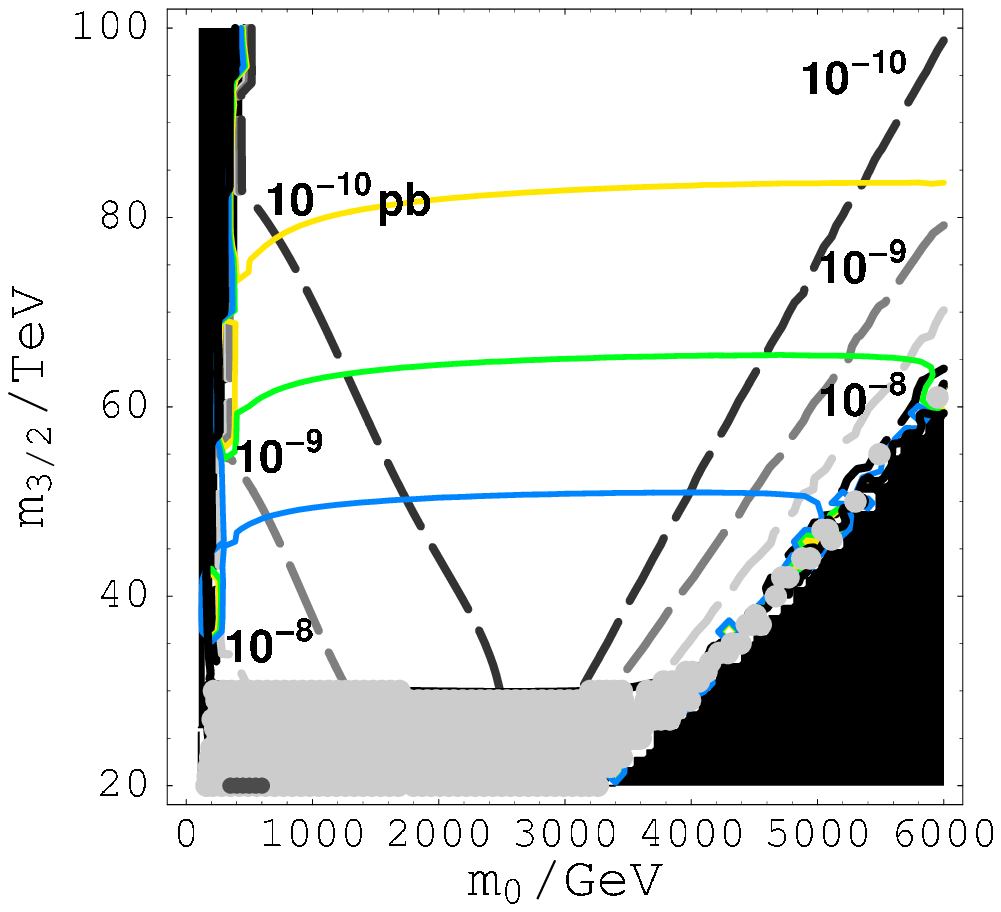} 
\end{center}
\end{minipage}
\begin{minipage}{0.48\linewidth}
\begin{center}(b)

  \includegraphics[width=0.9\linewidth]{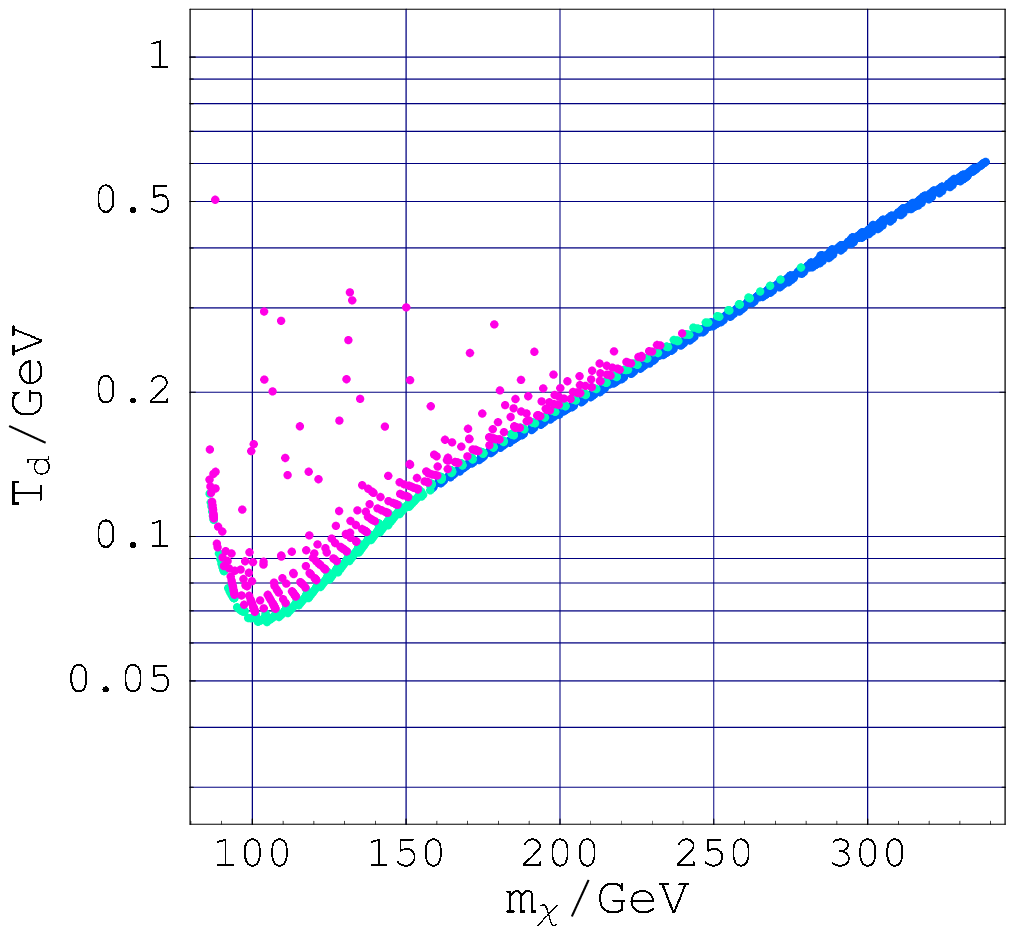} 
\end{center}
\end{minipage}
\caption{(a) The contours of $\Omega^{\rm th}_{\chi}h^2$ and 
$\sigma_{\chi-p}~[{\rm pb}]$. (b) The decay
 temperature of Q-ball which leads to the desired mass density of dark matter.
($\tan\beta=10$)}
\label{fig:anom10}
\end{figure}

\begin{figure}[h!]
 \begin{minipage}{0.48\linewidth}
\begin{center}(a)

 \includegraphics[width=0.9\linewidth]{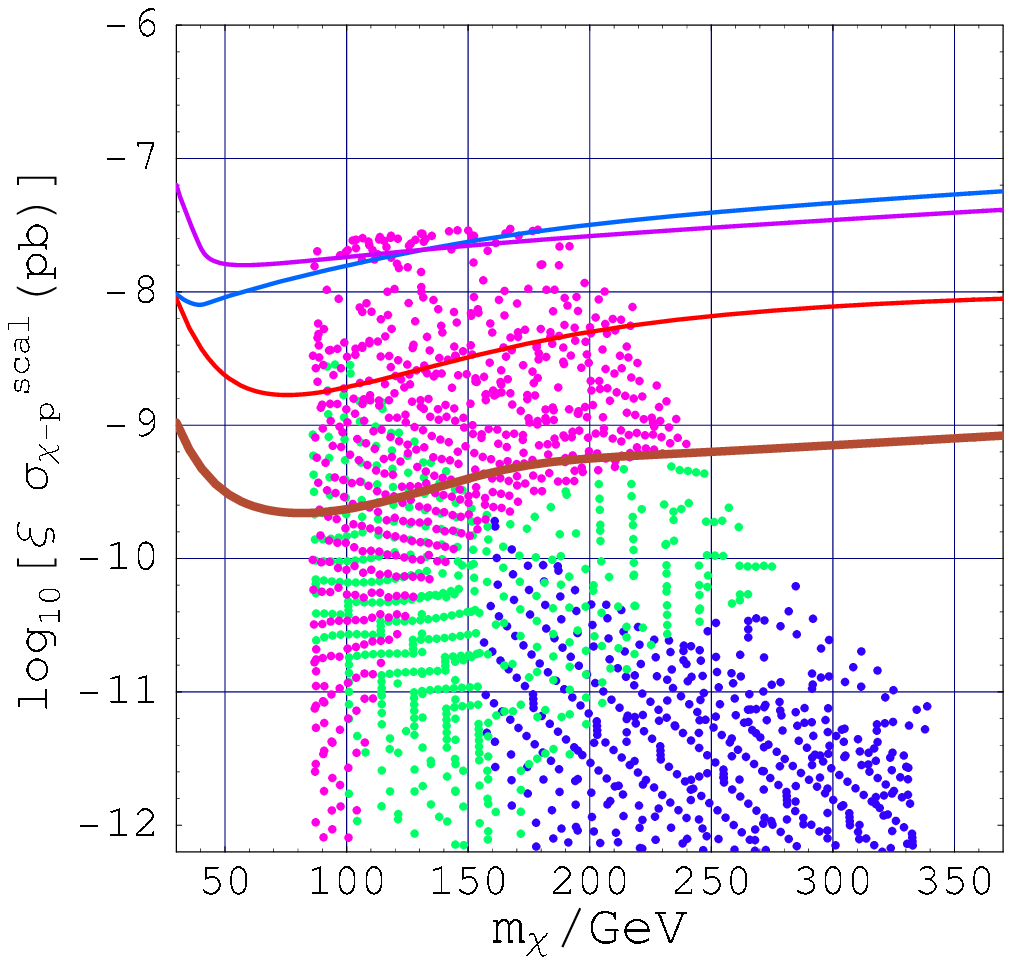} 
\end{center}
\end{minipage}
\begin{minipage}{0.48\linewidth}
\begin{center}(b)

 \includegraphics[width=0.9\linewidth]{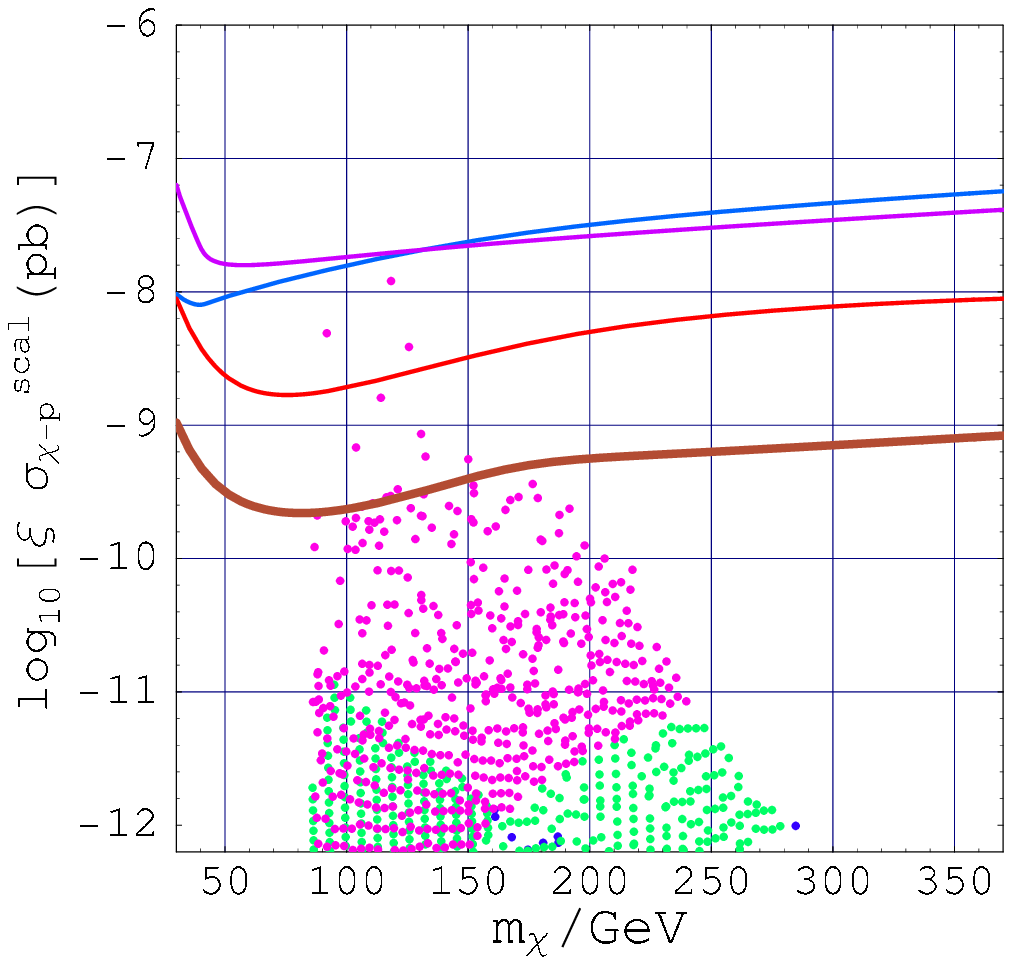} 
\end{center}
\end{minipage}\caption{The effective cross section of proton-$\chi$ interaction in the 
 mAMSB scenario. (a) The non-thermal scenario. (b)
The standard thermal freeze-out scenario. $(\tan\beta=10)$}
\label{fig:anomdirect10}
\end{figure}

>From the figures, in the non-thermal scenario, 
one can see that a large portion of the 
focus point region and the small $m_{H}$ region $(m_{0}\lsim 1\TEV)$
are within the reach of next generation experiments.
The bulk of the parameter space, where the Higgsino component of the 
LSP is very small and $m_{H}$ is large, is difficult to survey.
In the case of the thermal freeze-out scenario, 
there is almost no hope to detect the signal of SUSY dark matter 
in the entire parameter space
because of the smallness of the relic density of the LSP $(\xi\ll1)$.
In Figs.~\ref{fig:anom10} and \ref{fig:anomdirect10}, we 
also show the corresponding figures for $\tan\beta = 10$, 
where the conventions are the same as those in
Figs.~\ref{fig:anom30} and \ref{fig:anomdirect30}. 

%%%%%%%%%%%%%%%%%%%%%%%%%%%%%%%%%%%%%%%%%%%%%%%%%%%%%%%%%%%%%%%%%%%%%
%%%% Omega vs SUGRA parameter, T_d vs m_{\chi} in AMSB %%%%%%%%%%%%%% 
%%%%%%%%%%%%%%%%%%%%%%%%%%%%%%%%%%%%%%%%%%%%%%%%%%%%%%%%%%%%%%%%%%%%%
%%%%%%%%%%%%%%%%%%%%%%%%%%%%%%%%%%%%%%%%%%%%%%%%%%%%%%%%%%%%%%%%%%%%%%%%%%%%

%%%%%%%%%%%%%%%%%%%%%%%%%%%%%%%%%%%%%%%%%%%%%%%%%%%%%%%%%%%%%%%%%%%%%
%%%% DIRECT in mAMSB                                   %%%%%%%%%%%%%% 
%%%%%%%%%%%%%%%%%%%%%%%%%%%%%%%%%%%%%%%%%%%%%%%%%%%%%%%%%%%%%%%%%%%%%
%%%%%%%%%%%%%%%%%%%%%%%%%%%%%%%%%%%%%%%%%%%%%%%%%%%%%%%%%%%%%%%%%%%%%%%%%%%%%
%%%%%%%%%%%%%%%%%%%%%%%%%%%%%%%%%%%%%%%%%%%%%%%%%%%%%%%%%%%%%%%%%%%%%%%%%%%%%

%%%%%%%%%%%%%%%%%%%%%%%%%%%%%%%%%%%%%%%%%%%%%%%%%%%%%%%%%%%%%%%%%%%
\subsection{\Large Indirect detection observing  neutrino flux 
from the Sun}%%%%
%%%%%%%%%%%%%%%%%%%%%%%%%%%%%%%%%%%%%%%%%%%%%%%%%%%%%%%%%%%%%%%%%%%
In this section, we discuss one of the most promising methods 
of indirect detection for the neutralino dark matter, which observes 
energetic neutrinos from annihilation of the LSP in the Sun.
If the halo dark matter consists of the LSPs, the LSP has a finite
possibility to be captured by the Sun by an elastic scattering with
a nucleus therein. Once captured, the LSPs accumulate around the center
of the Sun through additional scatterings with nuclei.
Those LSPs which have accumulated in this way can annihilate with another 
LSP producing various decay products. Although most of them are immediately
absorbed through interactions with surrounding matter to leave no
evidence of their existence, produced 
neutrinos can escape out of the Sun and  reach 
terrestrial detectors. 

Especially, an energetic muon  neutrino
which escapes from the Sun and reaches the Earth, 
can be converted into a muon through a charged current interaction 
during passing through the rock below the detector.
These muons induced from the energetic neutrinos can be detected by 
various astrophysical neutrino detectors installed deep under ground,
sea water, or Antarctic ice.  
Since competing backgrounds are relatively well understood and also  
the nearby local halo density is
constrained better than the entire halo profile, which is still highly 
controversial, we can make more definite predictions on the expected
signal than they are in the cosmic-ray searches.
In the rest of this section, we investigate the consequences of the 
Affleck--Dine baryo/DM-genesis scenario in this indirect detection
method in the mSUGRA and mAMSB models, in turn.

\subsubsection{Neutrino-induced muon flux from the Sun in the mSUGRA}
First, let us discuss the prospects of the indirect detection of 
the neutrino-induced muon from the Sun in the mSUGRA model.
The readers who are interested in  
a full detail of the required calculations, please consult the 
excellent review given in  Ref.~\cite{Jungman}.
%%%%%%%%%%%%%%%%%%%%%%%%%%%%%%%%%%%%%%%%%%%%%%%%%%%%%%%%%%%%%%%%%%%%%
%%%%%%%%%%  neutrino flux contour sugra        %%%%%%%%%%%%%%
%%%%%%%%%%%%%%%%%%%%%%%%%%%%%%%%%%%%%%%%%%%%%%%%%%%%%%%%%%%%%%%%%%%%%
\begin{figure}[t]
 \begin{minipage}{0.48\linewidth}
\begin{center}(a)

 \includegraphics[width=0.9\linewidth]{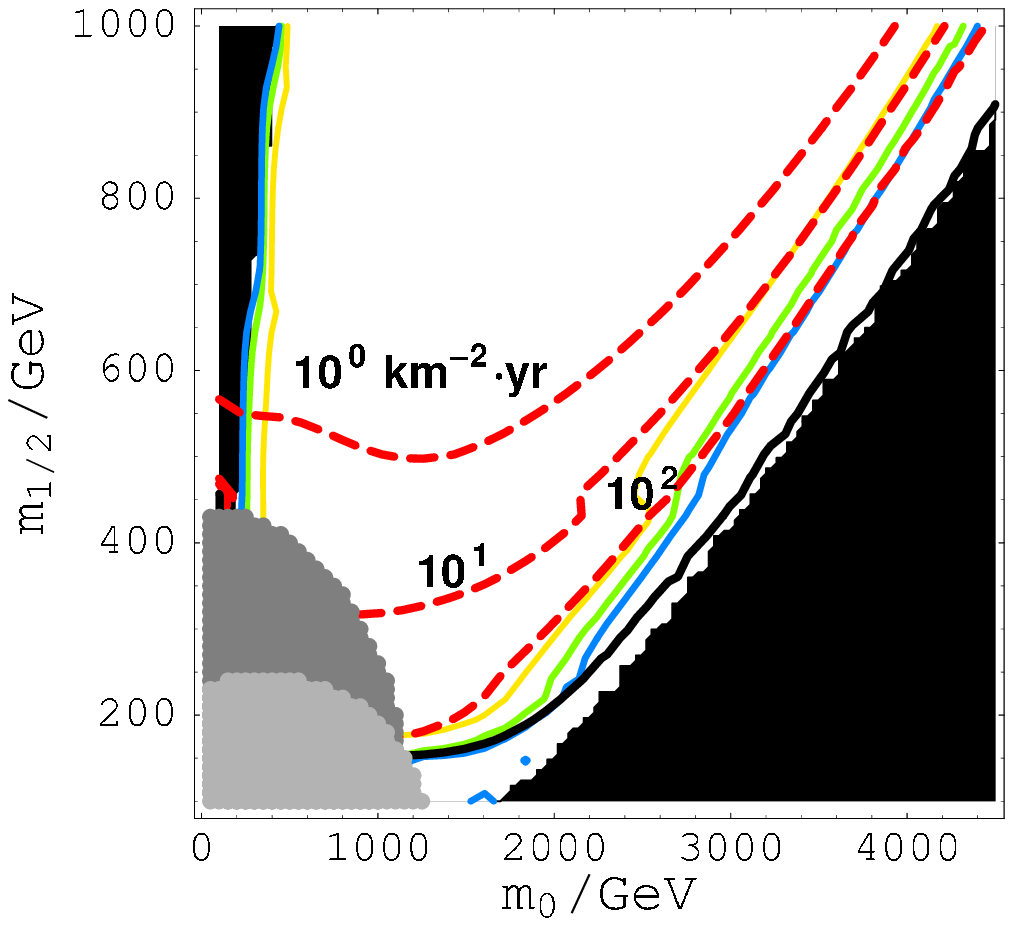} 
\end{center}
\end{minipage}
\begin{minipage}{0.48\linewidth}
\begin{center}(b)

  \includegraphics[width=0.9\linewidth]{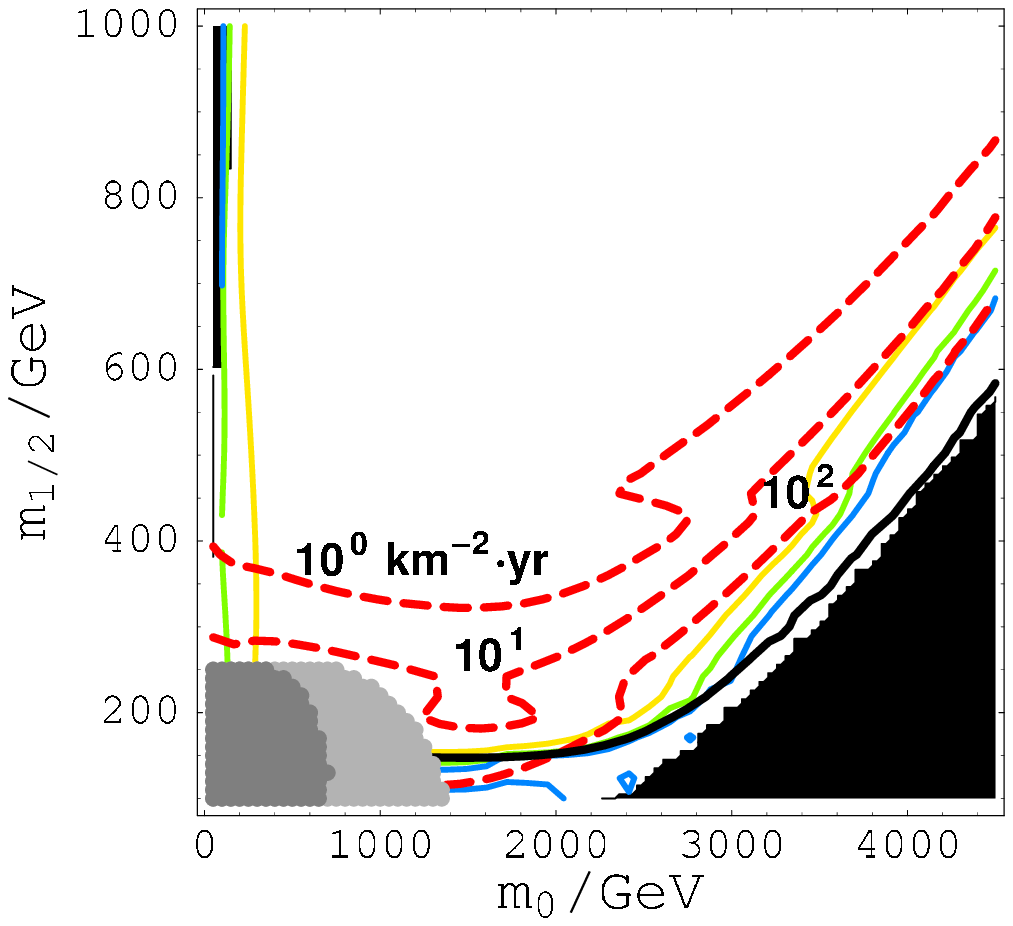} 
\end{center}
\end{minipage}
 
\caption{ The $\mu$ flux from the Sun in the  mSUGRA scenario. 
(a) $\tan\beta=45$.
(b) $\tan\beta=10$. Here, the local neutralino density is fixed as
 $\rho_{\chi}=0.3\GEV/{\rm cm}^3$.
}
\label{fig:sugrasun}
\end{figure}
%%%%%%%%%%%%%%%%%%%%%%%%%%%%%%%%%%%%%%%%%%%%%%%%%%%%%%%%%%%%%%%%%%%%%%%%%%%%
%%%%%%%%%%%%%%%%%%%%%%%%%%%%%%%%%%%%%%%%%%%%%%%%%%%%%%%%%%%%%%
In Fig.~\ref{fig:sugrasun}, we show the contours of the induced muon flux 
expected to be observed in a detector, where the dashed lines denote 
the expected flux and the other conventions are the same as those in 
Figs.~\ref{fig:sugra45} and \ref{fig:sugra10}. 
Note that, in this figure, 
the local neutralino density is
fixed as $\rho_{\chi}=0.3\GEV/{\rm cm}^3$ irrespective of the relic
density, and hence, the actual
detection rate in the thermal freeze-out scenario must be modified
according to the value of $\Omega_{\chi}^{\rm th}h^2$.

Once we fix the local neutralino density, the size of the neutrino flux 
is primarily controlled by the neutralino capture rate of the Sun. 
Hence, the elastic
scattering cross section between the LSP and nucleus in the Sun,
not the annihilation cross section of the LSP, determines the 
resultant neutrino flux.
Since the matter of the Sun largely consists of hydrogens, the
spin-dependent interaction through the Z-boson exchange is the most
important ingredient to determine the scattering cross section.
The coupling to the Z-gauge boson is proportional to
$N_{1,3(4)}^2$, and thus the flux becomes 
larger as Higgsino component in the LSP increases. 

A large Higgsino fraction in the LSP has another advantage in 
this detection method. The detection probability for an energetic
neutrino by observing the neutrino-induced upward muon is proportional to 
the second moment of the neutrino energy.
This is because that both of the charged-current cross section and the 
range of the produced muon are roughly proportional to its energy.
If the LSP has a significant fraction of Higgsino component,
it can annihilate into a pair of W- or Z-gauge bosons with a large 
branching ratio. Since the subsequent decays of these gauge bosons produce the 
most energetic neutrinos, a large Higgsino component is very
advantageous for the neutrino detection.

%%%%%%%%%%%%%%%%%%%%%%%%%%%%%%%%%%%%%%%%%%%%%%%%%%%%%%%%%%%%%%%%%%%%%
%%%%%%%%%%  The indirect detection rates  tanb=45        %%%%%%%%%%%%%%
%%%%%%%%%%%%%%%%%%%%%%%%%%%%%%%%%%%%%%%%%%%%%%%%%%%%%%%%%%%%%%%%%%%%%
\begin{figure}[ht]
 \begin{minipage}{0.48\linewidth}
\begin{center}(a)

 \includegraphics[width=0.9\linewidth]{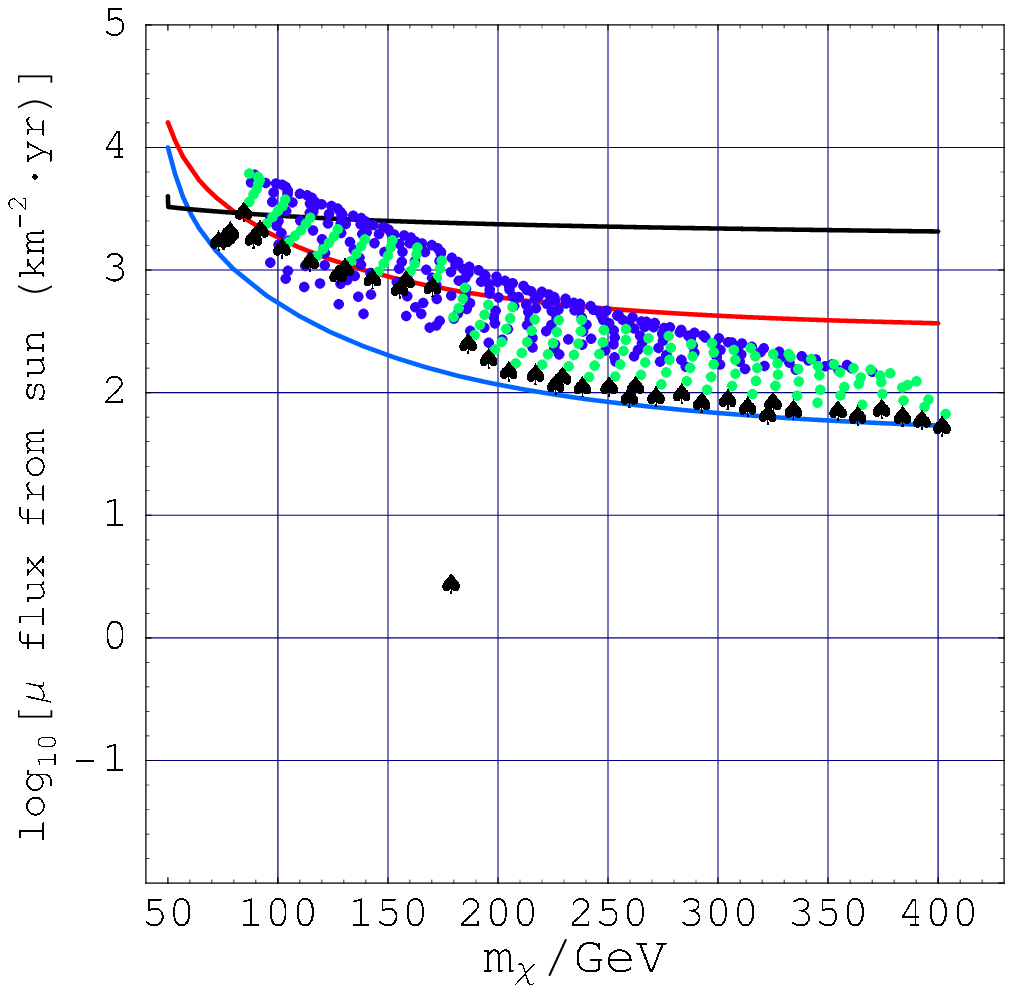} 
\end{center}
\begin{picture}(0,0)
\put(170,128){ICECUBE}
\put(170,161){ANTARES}
\put(190,180){SK}
\end{picture}
\end{minipage}
\begin{minipage}{0.48\linewidth}
\vspace{-1cm}
\begin{center}(b)

  \includegraphics[width=0.9\linewidth]{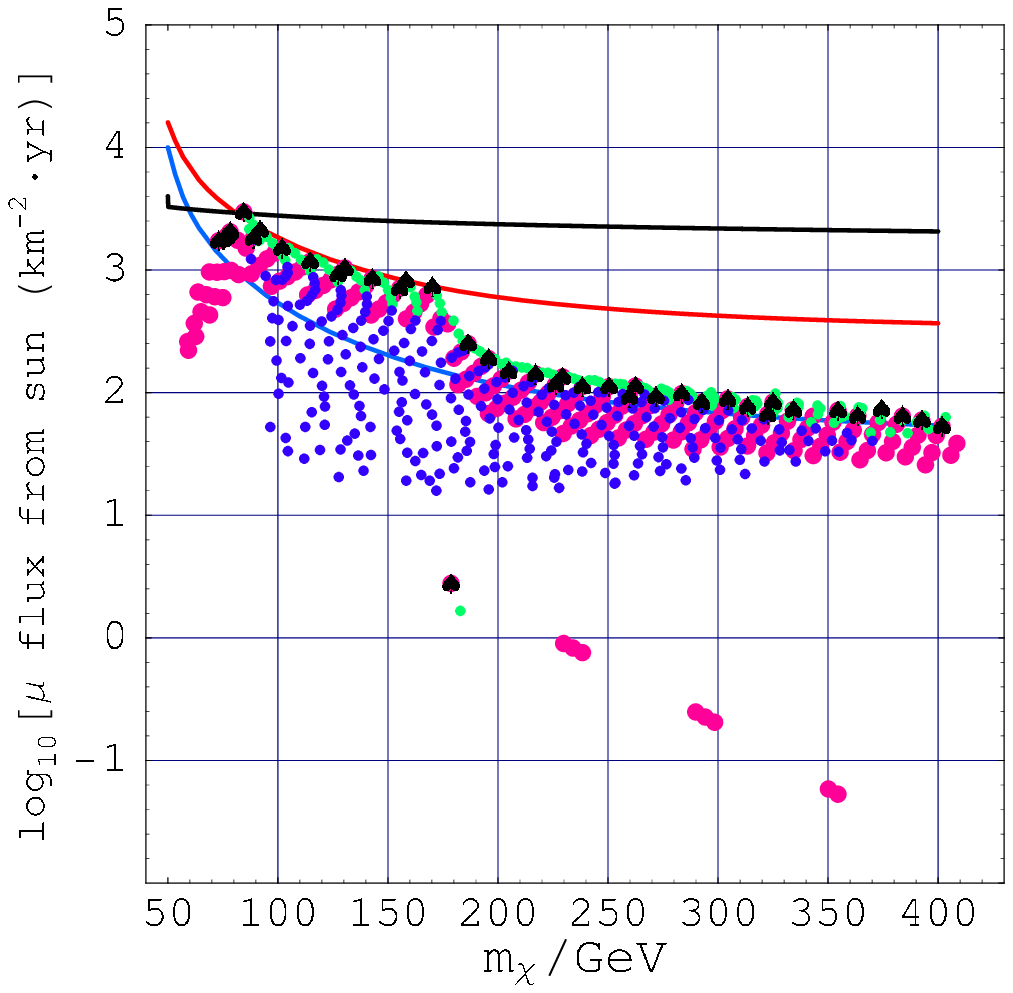} 
\end{center}
\end{minipage}
 
\caption{ The $\mu$ flux from the Sun in the  mSUGRA scenario. 
(a) The non-thermal scenario.
(b) The standard thermal freeze-out scenario. ($\tan\beta=45$)
}
\label{fig:sugrasun45}
\end{figure}
%%%%%%%%%%%%%%%%%%%%%%%%%%%%%%%%%%%%%%%%%%%%%%%%%%%%%%%%%%%%%%%%%%%%%%%%%%%%
%%%%%%%%%%%%%%%%%%%%%%%%%%%%%%%%%%%%%%%%%%%%%%%%%%%%%%%%%%%%%%

In Fig.~\ref{fig:sugrasun45}, we show the expected $\mu$ flux from the
Sun in the mSUGRA model with $\tan\beta=45$, in the non-thermal (a) 
and in the thermal freeze-out scenarios (b).
Shading (coloring) conventions are the same as those in
Fig.~\ref{fig:sugradirect45}: dark (blue) points 
for $\Omega_{\chi}^{\rm th}h^2\leq 0.03$, light (green) points for 
$0.03\leq \Omega_{\chi}^{\rm th}h^2\leq 0.1$ and
medium (purple) points for $0.1\leq \Omega_{\chi}^{\rm th}h^2\leq 0.3$.
For the non-thermal scenario, we have calculated the expected muon flux 
with a fixed local neutralino density $\rho_{\chi}=0.3\GEV/{\rm cm}^3$.
As for the thermal freeze-out scenario, we have taken smallness of the
local neutralino density into account by multiplying a factor 
$\xi=(\Omega_{\chi}^{\rm th}h^2/\Omega_{DM}h^2)$ as before.~\footnote{
Although this is not the exact treatment, it gives an excellent approximation,
since the equilibrium state between
capture and annihilation is well realized in almost 
the entire relevant parameter space.}
The black solid lines denote the present bound on the muon flux from 
Super Kamiokande~\cite{SK}. The other two solid lines represent expected
sensitivities for the muon flux in the near future experiments:
ICECUBE~\cite{ICECUBE} and ANTARES (3 years)~\cite{ANTARES} 
from the bottom up, respectively. 

\begin{figure}[h!]
 \begin{minipage}{0.48\linewidth}
\begin{center}(a)

 \includegraphics[width=0.9\linewidth]{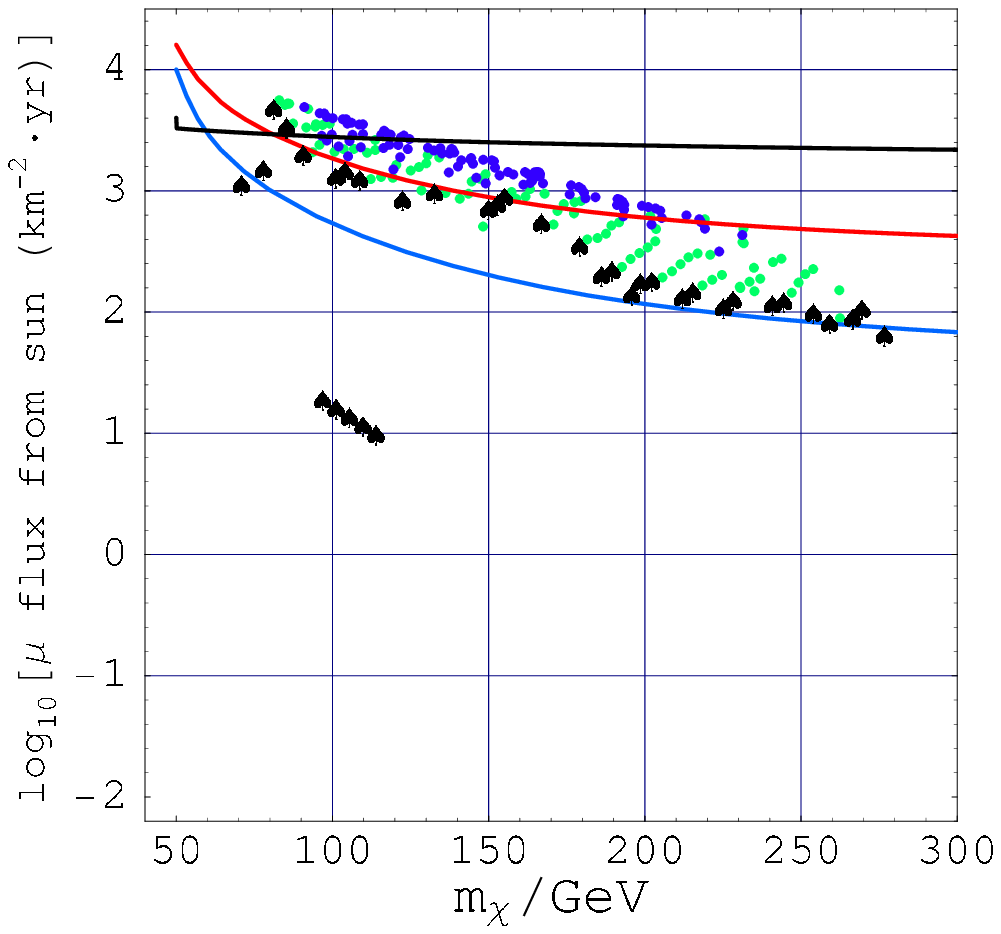} 
\end{center}
\end{minipage}
\begin{minipage}{0.48\linewidth}
\begin{center}(b)

  \includegraphics[width=0.9\linewidth]{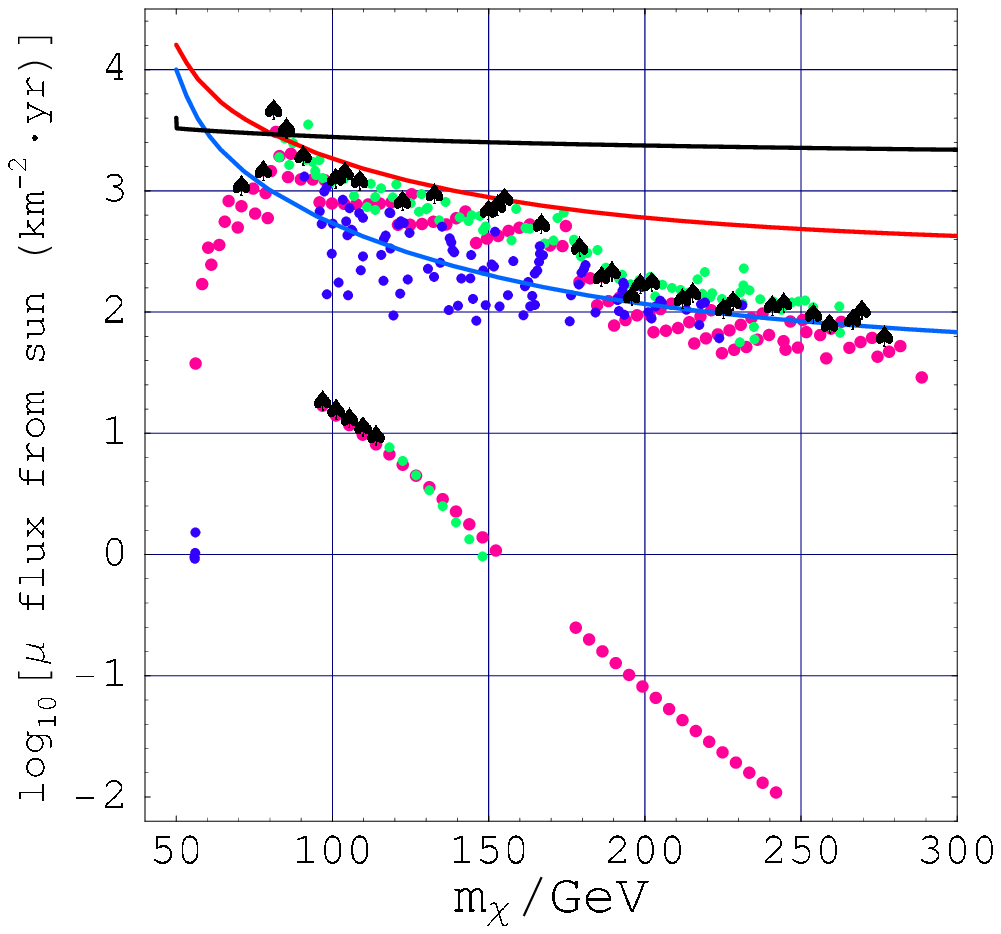} 
\end{center}
\end{minipage}
 \caption{ The $\mu$ flux from the Sun in the  mSUGRA scenario. 
(a) The non-thermal scenario.
(b) The standard thermal freeze-out scenario. ($\tan\beta=10$)
}
\label{fig:sugrasun2}
\end{figure}

Fig.~\ref{fig:sugrasun45} clearly shows the advantage of the non-thermal
scenario in the neutrino-induced muon detection. In the mSUGRA model, 
the late-time Q-ball decay requires a quite large annihilation cross
section, which in turn, requires a significant fraction of Higgsino
component in the LSP. This promises us a significant possibility to 
discover high-energy neutrino signals in the near future.
Especially, for relative light neutralinos $m_{\chi}\lsim m_{t}$, 
there is a big
possibility even for ANTARES, which is now in the last phase of 
its construction, to find the signals. 
Furthermore, after the deployment of the ICECUBE detectors, we can
survey the whole parameter space of the non-thermal scenario.
Similar features can be seen in
Fig.~\ref{fig:sugrasun2}, which is a corresponding figure for $\tan\beta=10$.
%%%%%%%%%%%%%%%%%%%%%%%%%%%%%%%%%%%%%%%%%%%%%%%%%%%%%%%%%%%%%%%%%%%%%
%%%%%%%%%%  The indirect detection rates  tanb=45        %%%%%%%%%%%%%%
%%%%%%%%%%%%%%%%%%%%%%%%%%%%%%%%%%%%%%%%%%%%%%%%%%%%%%%%%%%%%%%%%%%%%
%%%%%%%%%%%%%%%%%%%%%%%%%%%%%%%%%%%%%%%%%%%%%%%%%%%%%%%%%%%%%%%
%%%%%%%%%%%%%%%%%%%%%%%%%%%%%%%%%%%%%%%%%%%%%%%%%%%%%%%%%%%%%%%
%%%%%%%%%%%%%%%%%%%%%%%%%%%%%%%%%%%%%%%%%%%%%%%%%%%%%%%%%%%%%%%
\newpage
%%%%%%%%%%%%%%%%%%%%%%%%%%%%%%%%%%%%%%%%%%%%%%%%%%%%%%%%%%%%%%%%%%%%%%%
\subsubsection{Neutrino-induced muon flux from the Sun in the mAMSB}
%%%%%%%%%%%%%%%%%%%%%%%%%%%%%%%%%%%%%%%%%%%%%%%%%%%%%%%%%%%%%%%%%%%%%%
Now, let turn our attention to the mAMSB model.
Since the LSP in the mAMSB model is mostly composed 
of Wino, the spin-dependent (and also scalar) scattering cross section of
the LSP with  matter in the Sun is relatively small. This reduces the 
expected muon flux compared to the Higgsino-like dark matter.

%%%%%%%%%%%%%%%%%%%%%%%%%%%%%%%%%%%%%%%%%%%%%%%%%%%%%%%%%%%%%%%%%%%%%
%%%%%%%%%%  neutrino flux contour anom        %%%%%%%%%%%%%%
%%%%%%%%%%%%%%%%%%%%%%%%%%%%%%%%%%%%%%%%%%%%%%%%%%%%%%%%%%%%%%%%%%%%%
\begin{figure}[ht!]
 \begin{minipage}{0.48\linewidth}
\begin{center}(a)

 \includegraphics[width=0.9\linewidth]{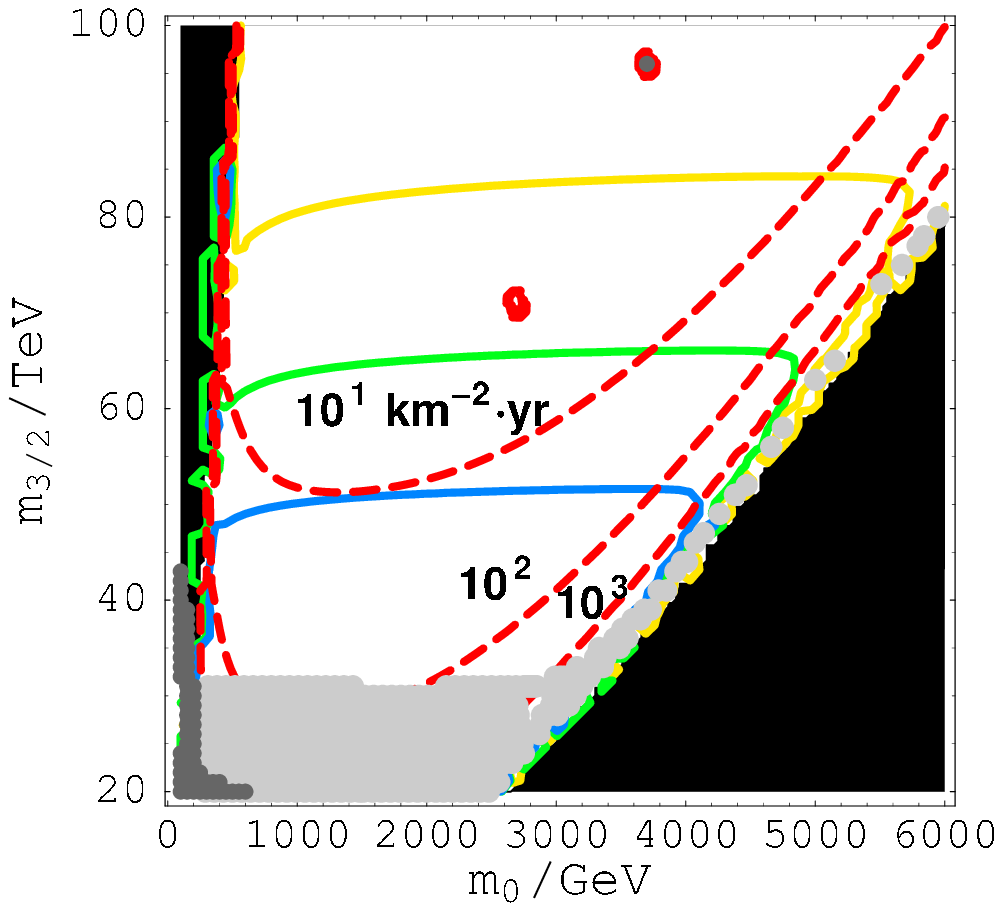} 
\end{center}
\end{minipage}
\begin{minipage}{0.48\linewidth}
\begin{center}(b)

  \includegraphics[width=0.9\linewidth]{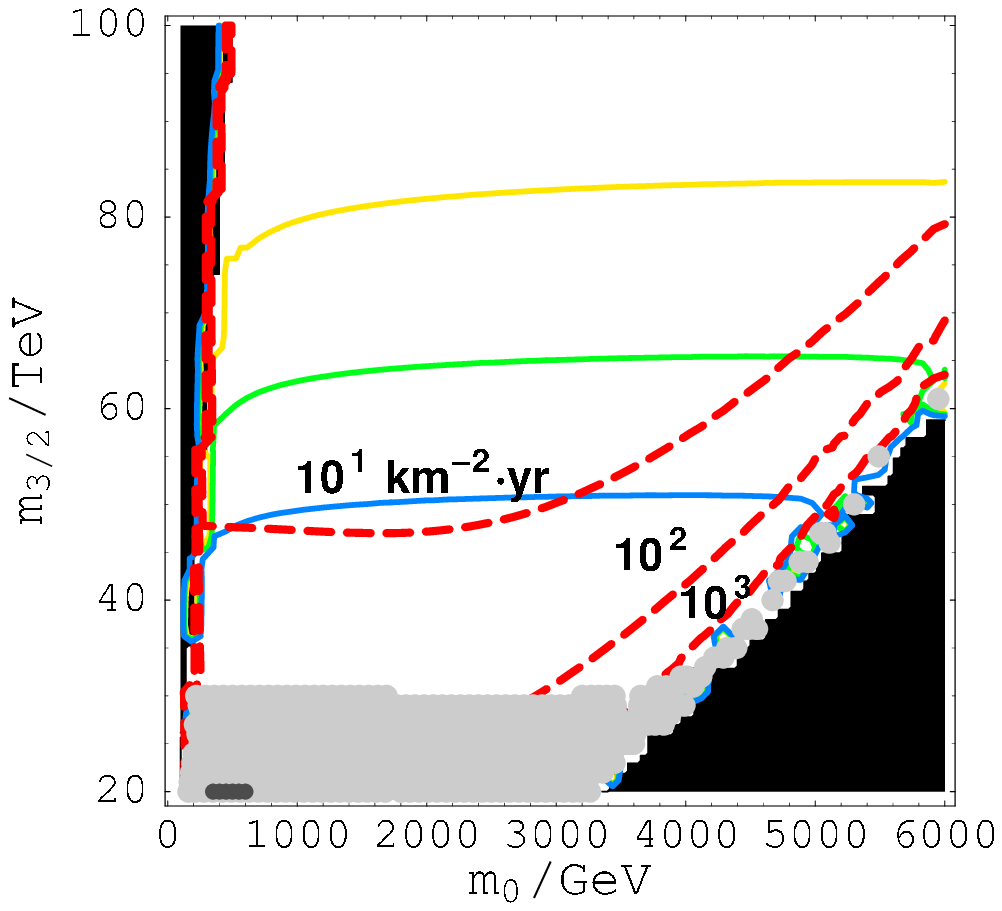} 
\end{center}
\end{minipage}
 
\caption{ The $\mu$ flux from the Sun in the  mAMSB scenario. 
(a) $\tan \beta=30$.
(b) $\tan \beta=10$.
}
\label{fig:anomsun30_10}
\end{figure}
%%%%%%%%%%%%%%%%%%%%%%%%%%%%%%%%%%%%%%%%%%%%%%%%%%%%%%%%%%%%%%%%%%%%%%%%%%%%
%%%%%%%%%%%%%%%%%%%%%%%%%%%%%%%%%%%%%%%%%%%%%%%%%%%%%%%%%%%%%%

In Fig.~\ref{fig:anomsun30_10}, we show the contours of the expected muon 
flux in the mAMSB model with $\tan\beta=30$ (a) and $\tan\beta=10$ (b).
The dashed lines denote the expected muon flux, whose value is 
explicitly presented in the figure. The other conventions are the same 
as those in Fig.~\ref{fig:anom30}-(a). As in the previous section,
the local neutralino density is fixed as $\rho=0.3\GEV/{\rm cm}^3$, and
hence we need an adjustment in the thermal freeze-out scenario to obtain 
correct predictions.
The muon flux increases as the parameter sets approach to the focus point 
region, where the LSP has a significant 
Higgsino fraction  as in the mSUGRA
case.

%%%%%%%%%%%%%%%%%%%%%%%%%%%%%%%%%%%%%%%%%%%%%%%%%%%%%%%%%%%%%%%%%%%%%
%%%%%%%%%%  neutrino flux anom        %%%%%%%%%%%%%%
%%%%%%%%%%%%%%%%%%%%%%%%%%%%%%%%%%%%%%%%%%%%%%%%%%%%%%%%%%%%%%%%%%%%%
\begin{figure}[h!]
 \begin{minipage}{0.48\linewidth}
\begin{center}(a)

 \includegraphics[width=0.9\linewidth]{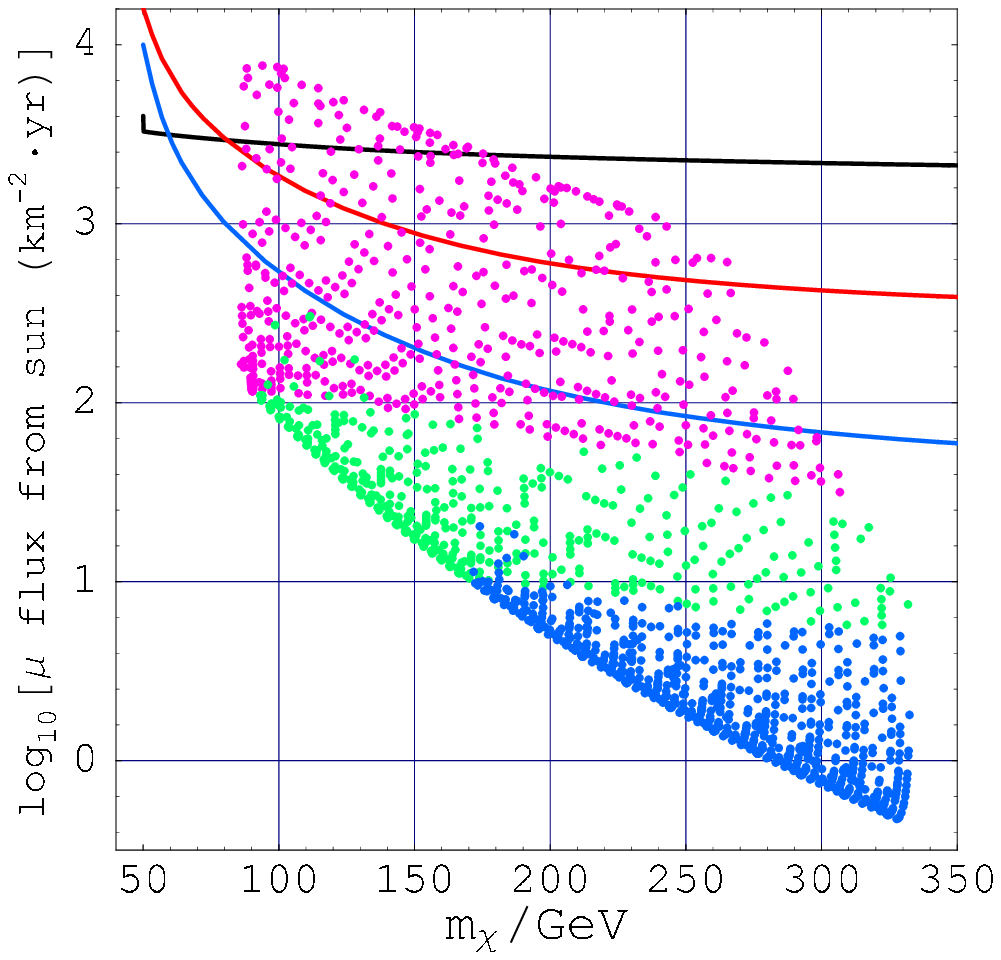} 
\end{center}
\end{minipage}
\begin{minipage}{0.48\linewidth}
\begin{center}(b)

  \includegraphics[width=0.9\linewidth]{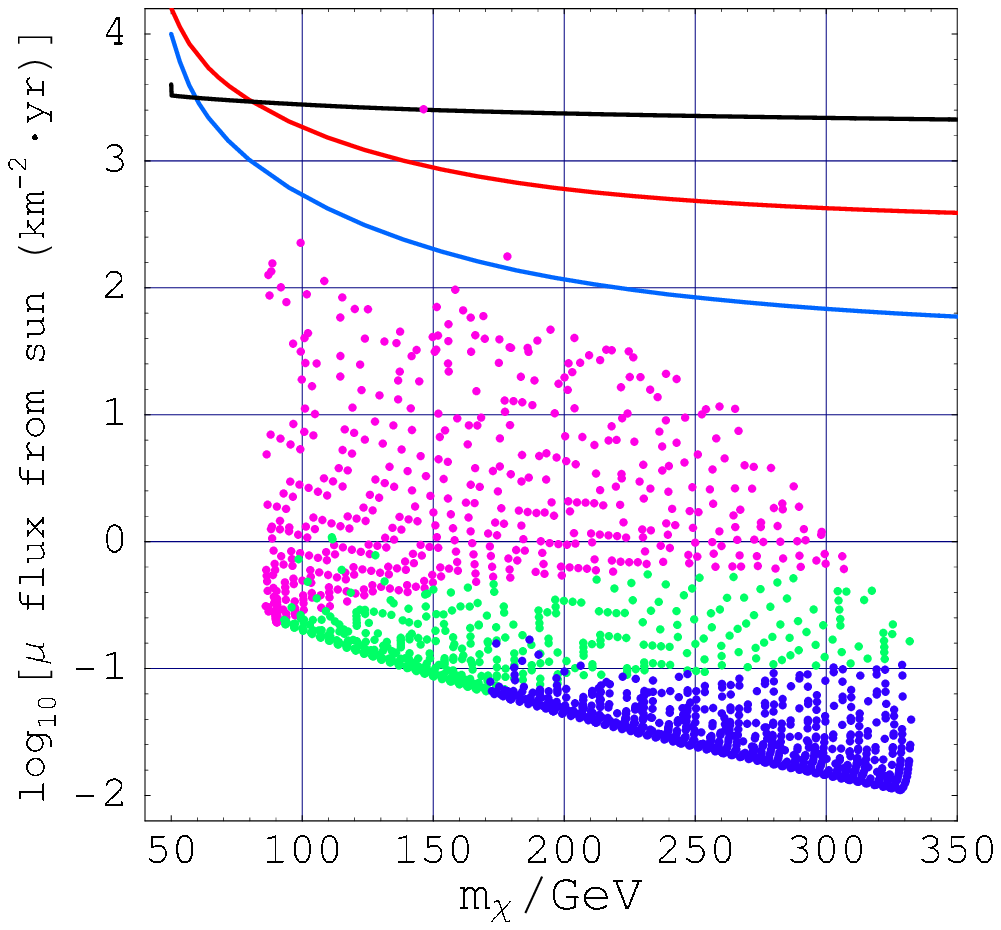} 
\end{center}
\end{minipage}
 
\caption{ The $\mu$ flux from the Sun in the  mAMSB scenario. 
(a) The non-thermal scenario.
(b) The standard thermal freeze-out scenario. ($\tan\beta=30$)
}
\label{fig:anomsunscale30}
\end{figure}
%%%%%%%%%%%%%%%%%%%%%%%%%%%%%%%%%%%%%%%%%%%%%%%%%%%%%%%%%%%%%%%%%%%%%%%%%%%%
%%%%%%%%%%%%%%%%%%%%%%%%%%%%%%%%%%%%%%%%%%%%%%%%%%%%%%%%%%%%%%

%%%%%%%%%%%%%%%%%%%%%%%%%%%%%%%%%%%%%%%%%%%%%%%%%%%%%%%%%%%%%%%%%%%%%
%%%%%%%%%%  neutrino flux contour anom        %%%%%%%%%%%%%%
%%%%%%%%%%%%%%%%%%%%%%%%%%%%%%%%%%%%%%%%%%%%%%%%%%%%%%%%%%%%%%%%%%%%%
\begin{figure}[h!]
 \begin{minipage}{0.48\linewidth}
\begin{center}(a)

 \includegraphics[width=0.9\linewidth]{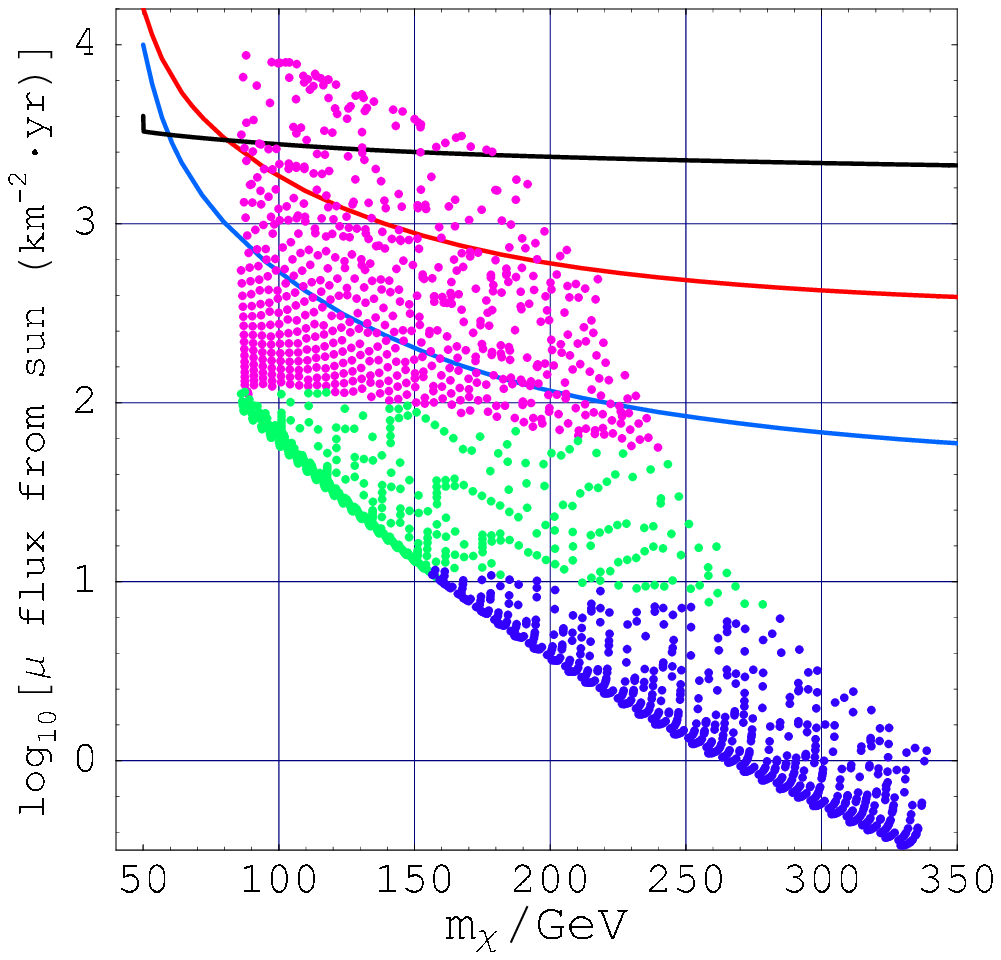} 
\end{center}
\end{minipage}
\begin{minipage}{0.48\linewidth}
\begin{center}(b)

  \includegraphics[width=0.9\linewidth]{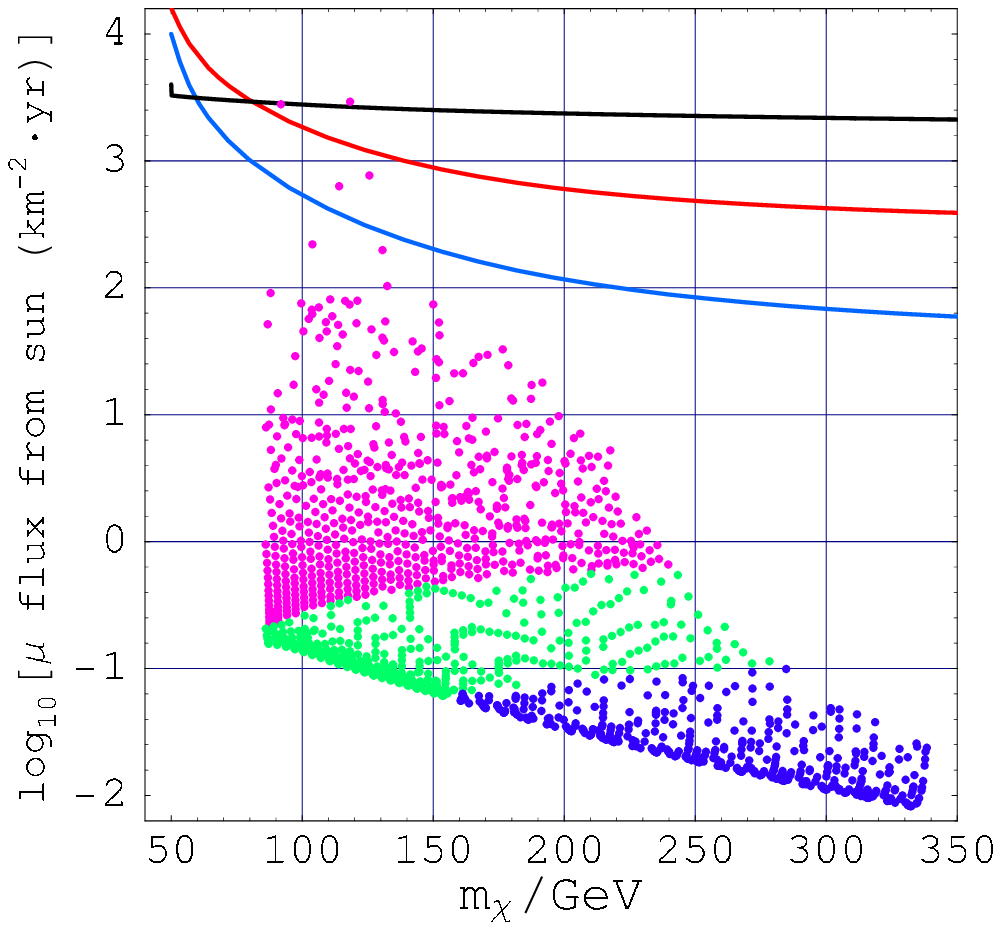} 
\end{center}
\end{minipage}
 
\caption{ The $\mu$ flux from the Sun in the  mAMSB scenario. 
(a) The non-thermal scenario.
(b) The standard thermal freeze-out scenario. ($\tan\beta=10$)
}
\label{fig:anomsunscale10}
\end{figure}
%%%%%%%%%%%%%%%%%%%%%%%%%%%%%%%%%%%%%%%%%%%%%%%%%%%%%%%%%%%%%%%%%%%%%%%%%%%%
%%%%%%%%%%%%%%%%%%%%%%%%%%%%%%%%%%%%%%%%%%%%%%%%%%%%%%%%%%%%%%

Fig.~\ref{fig:anomsunscale30} shows the expected muon flux for
$(\tan\beta=30)$ in the non-thermal (a)
and in the thermal freeze-out scenarios (b).
For the non-thermal case, we have set the local neutralino density as 
$\rho_{\chi}=0.3\GEV/{\rm cm}^3$, and for the thermal case, we have rescaled
the flux by multiplying a factor $\xi=(\Omega_{\chi}^{\rm
th}h^2/\Omega_{DM}h^2)$ to take the smallness of the local neutralino
density into account.~\footnote{The equilibrium  
between capture and annihilation of the LSP is realized also in the
mAMSB model. We found that the error larger than ${\cal O}(1\%)$ due to this 
rescaling method only appears in the region where the
 muon flux is quite small, $\Phi_{\mu}\leq {\cal {O}}(1) {\rm
km^{-2}\cdot yr}$, in the non-thermal scenario.} 
>From this figure,  we can see that there is almost no chance 
to find energetic neutrino signals in the thermal freeze-out scenario, 
even after the completion of ICECUBE project. 
In other words, in the mAMSB model, it strongly indicates the 
existence of the non-thermal dark matter if we find energetic
neutrino signals in the future experiments. 
We present the corresponding figure for $\tan\beta=10$ in Fig.~\ref{fig:anomsunscale10}.

%%%%%%%%%%%%%%%%%%%%%%%%%%%%%%%%%%%%%%%%%%%%%%%%%%%%%%%%%%%%%%%%%%%%%%%%%%%%%%
\subsection{\Large Indirect detection observing hard positron flux from the
halo}%%%%%%%%%%%%%%%%%%%%%%%%%%%%%%%%%%%%%%%%%%%%%%%%%%%%
%%%%%%%%%%%%%%%%%%%%%%%%%%%%%%%%%%%%%%%%%%%%%%%%%%%%%%%%%

Finally, in this section, 
we discuss another promising way to indirectly detect 
the existence of neutralino dark matter: search for an excess of positron
flux in cosmic rays in space based or balloon-borne experiments.
At low energies, the expected positron flux has large uncertainty 
for lack of precise knowledge about competing backgrounds.
Although the positron background is most 
likely to be composed of secondaries produced in the interactions of
cosmic ray nuclei with interstellar gas, which is expected to fall 
as $\sim E_{e^{+}}^{-3.1}$, this background is suffering from large 
ambiguity coming from the solar wind at energies 
below $\sim 10\GEV$~\cite{posi1,posi2}.
It is also affected by the orbit path of the experiment.
Fortunately, at high energies, these effects are strongly suppressed,
and we can hope to have a meaningful signal.
In addition, at high energies, the positrons lose their energy through 
various processes, and it is known that they can reach the detectors 
only when they are produced within a few $\rm kpc$~\cite{posi1,posi2}.
Therefore, as for the hard positron spectrum, the result is rather 
insensitive relative to the controversial halo profile near the 
galactic center.~\footnote{But it is affected by the ``clumpiness'' of
the local dark matter density.}

The dominant source of the most energetic positron flux is the
annihilation of two neutralinos into a $W^{\pm}$ or $Z^{0}$ pair,
which is followed by the direct decay of the $W^{+}$ into an $e^{+}$
and $\nu_{e}$ or decay of the $Z^{0}$ into an $e^{\pm}$ pair.~\footnote{
The positron line signal from the direct annihilation into an $e^{\pm}$
pair is helicity suppressed, and we will not consider it in this paper.}
The positrons produced as this way have an average energy of half the 
parent neutralino mass, and their spectrum has a peak around this
energy, where the signal to background ratio is maximized.

We use the following result of the differential positron flux given in 
Ref.~\cite{posi2}:
\begin{eqnarray}
E^2\frac{d \Phi_{e^{+}}}{d\Omega dE}&=&2.7\times 10^{-6} {\rm cm}^{-2}
{\rm s}^{-1}{\rm sr}^{-1}\GEV\nonumber\\
&&\times \left(\frac{\rho_{\chi}}{0.3 \GEV/{\rm cm}^3}\right)^2
\left(\frac{100\GEV}{m_{\chi}}\right)^2\sum_{i}\frac{\sigma_{i}v}{{\rm pb}
\cdot \beta_{i}}B_{e^{+}}^{i}\int^{z^{i}_{+}}_{z^{i}_{-}}dz\;
g(z,E/m_{\chi})\;,
\end{eqnarray}
where $i$ denotes an annihilation channel of the neutralinos into gauge
bosons. The other required expressions  are available in 
Refs.~\cite{posi2,Matchev}.
Although we adopt the modified isothermal distribution 
with halo size $4{\rm kpc}$ as the halo profile, other choices 
do not change the main results for the reason explained before. We use 
$E^2 d\Phi_{e^{+}}/d\Omega dE=1.16\times 10^{-3}E^{-1.23}$
as a fit of the positron background~\cite{Matchev}.

\subsubsection{Hard positron flux in the mSUGRA}
First, let us discuss the hard positron flux in the mSUGRA model.
As discussed above, the hard positron flux is determined by the 
neutralino annihilation cross section into a pair of gauge bosons.
%%%%%%%%%%%%%%%%%%%%%%%%%%%%%%%%%%%%%%%%%%%%%%%%%%%%%%%%%%%%%%%%%%%%%
%%%%%%%%%%  The indirect detection rates  tanb=45        %%%%%%%%%%%%%%
%%%%%%%%%%%%%%%%%%%%%%%%%%%%%%%%%%%%%%%%%%%%%%%%%%%%%%%%%%%%%%%%%%%%%
\begin{figure}[htbp]
\begin{minipage}{0.48\linewidth}
\begin{center}(a)
 
\includegraphics[width=0.9\linewidth]{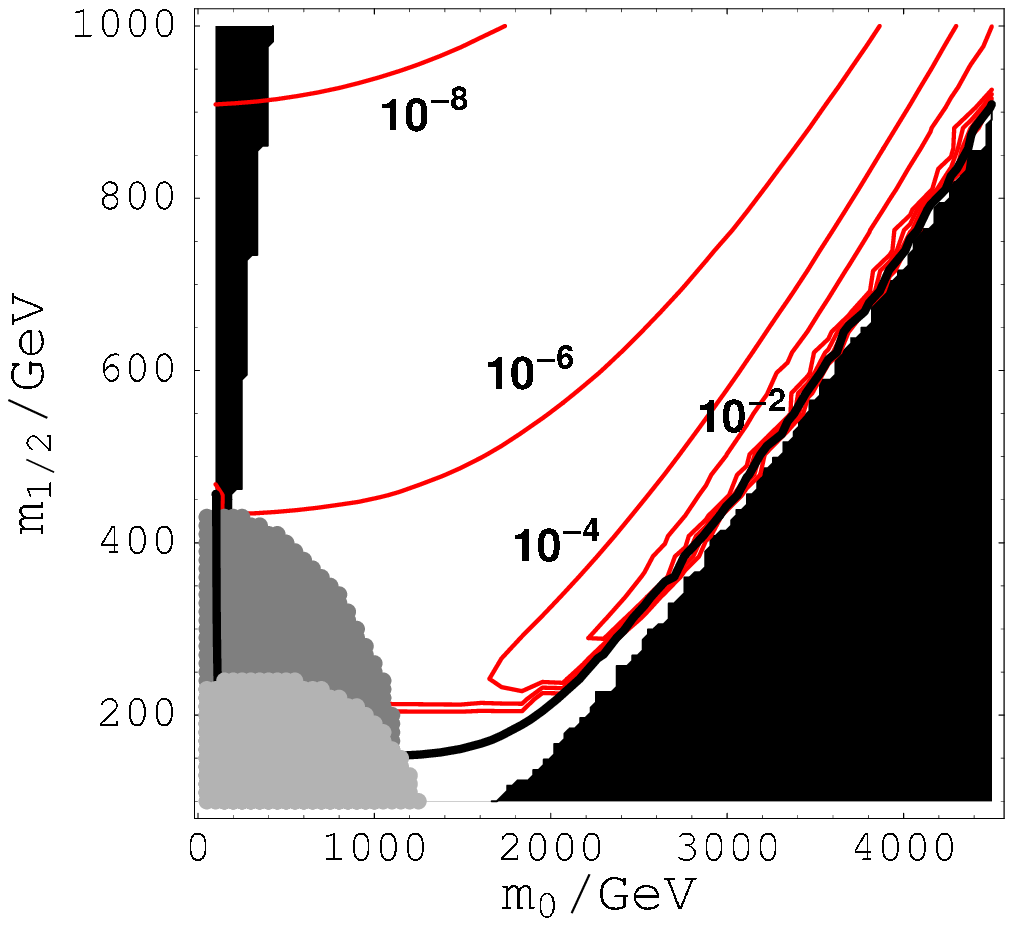} 
\end{center}

\end{minipage}
\begin{minipage}{0.48\linewidth}
\begin{center}(b)

 \includegraphics[width=0.9\linewidth]{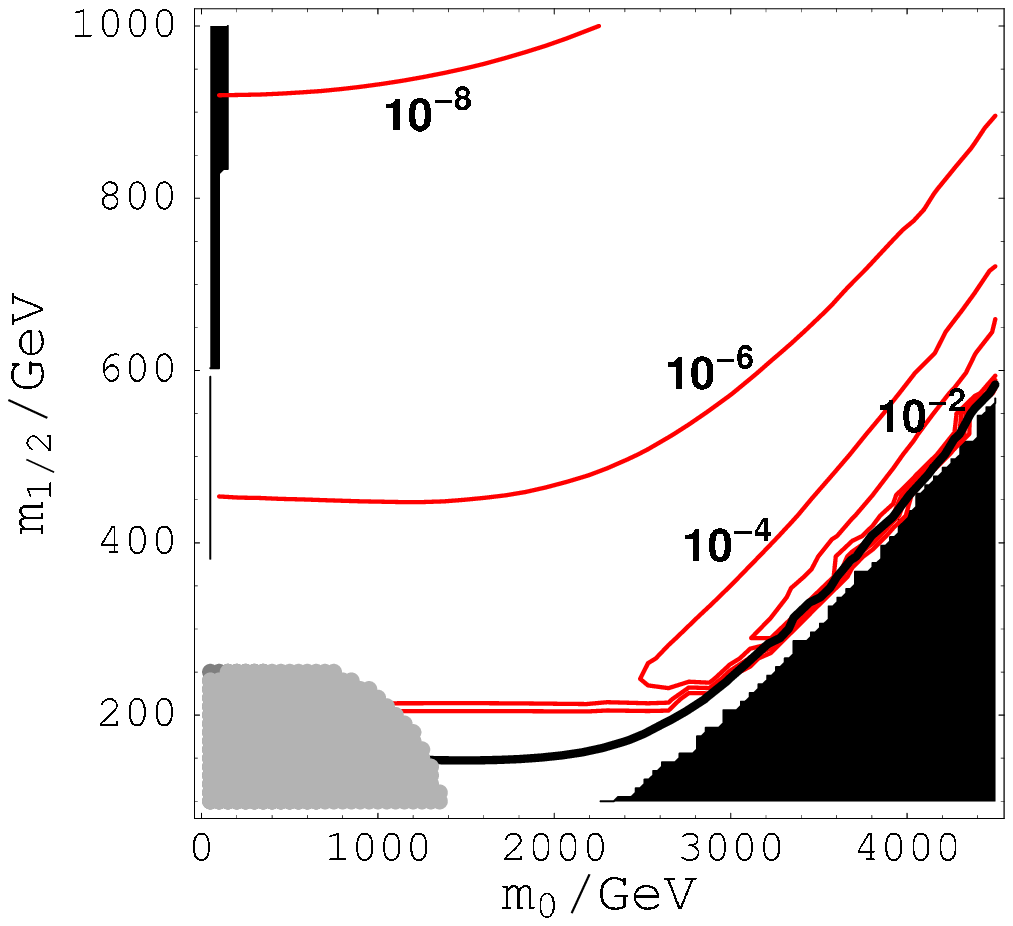} 
\end{center}
\end{minipage}
 
\caption{ The positron signal to background ratio $S/B$ at $E_{\rm opt}$ 
 in the mSUGRA scenario. 
(a) $\tan \beta=45$.
(b) $\tan \beta=10$.
}
\label{fig:POSImsugracont}
 \end{figure}
%%%%%%%%%%%%%%%%%%%%%%%%%%%%%%%%%%%%%%%%%%%%%%%%%%%%%%%%%%%%%%%%%%%%%%%%%
%%%%%%%%%%%%%%%%%%%%%%%%%%%%%%%%%%%%%%%%%%%%%%%%%%%%%%%%%%%%%%%%%%%%%%%%%
Therefore, we can expect that the flux increases as Higgsino component in
the LSP increases, which makes the current non-thermal scenario much more
advantageous than the standard thermal freeze-out scenario.

In Fig.~\ref{fig:POSImsugracont}, we show the contour plot of the
positron signal to background ratio $S/B$ at $E_{\rm opt}$, where
the $S/B$ is maximized. The solid (red) lines are the contours of $S/B$ 
whose value is denoted in the figure. Other conventions are the same as
before. Note that, in this calculation, 
we have fixed the local neutralino density as 
$\rho_{\chi}=0.3\GEV/{\rm cm}^{3}$ irrespective of the thermal relic
abundance.
%%%%%%%%%%%%%%%%%%%%%%%%%%%%%%%%%%%%%%%%%%%%%%%%%%%%%%%%%%%%%%%%%%%%%
%%%%%%%%%%%%%%%%%%%%%%%%%%%%%%%%%%%%%%%%%%%%%%%%%%%%%%%%%%%%%%%%%%%%%
%%%%%%%%%%%%%%%%%%%%%%%%%%%%%%%%%%%%%%%%%%%%%%%%%%%%%%%%%%%%%%%%%%%%%%%
\begin{figure}[h!]
 \begin{minipage}{0.48\linewidth}
\begin{center}(a)

 \includegraphics[width=0.9\linewidth]{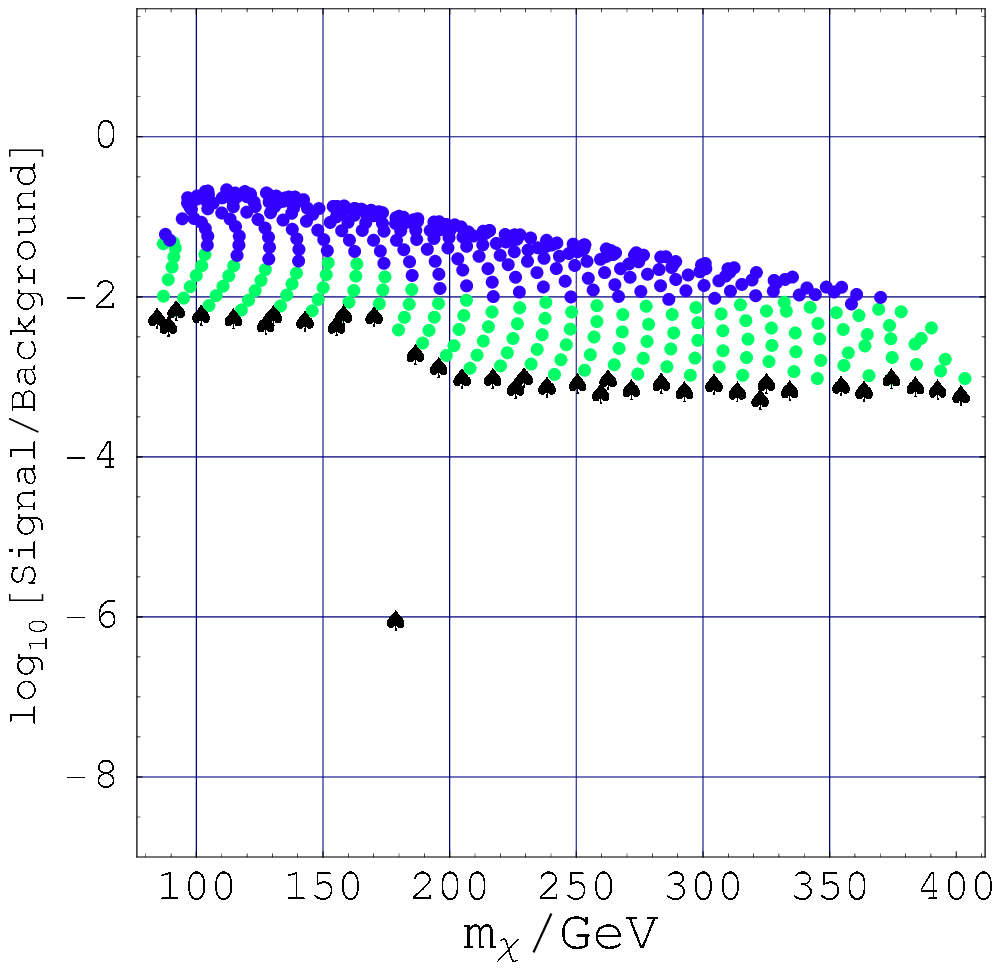} 
\end{center}
\end{minipage}
\begin{minipage}{0.48\linewidth}
\begin{center}(b)

  \includegraphics[width=0.9\linewidth]{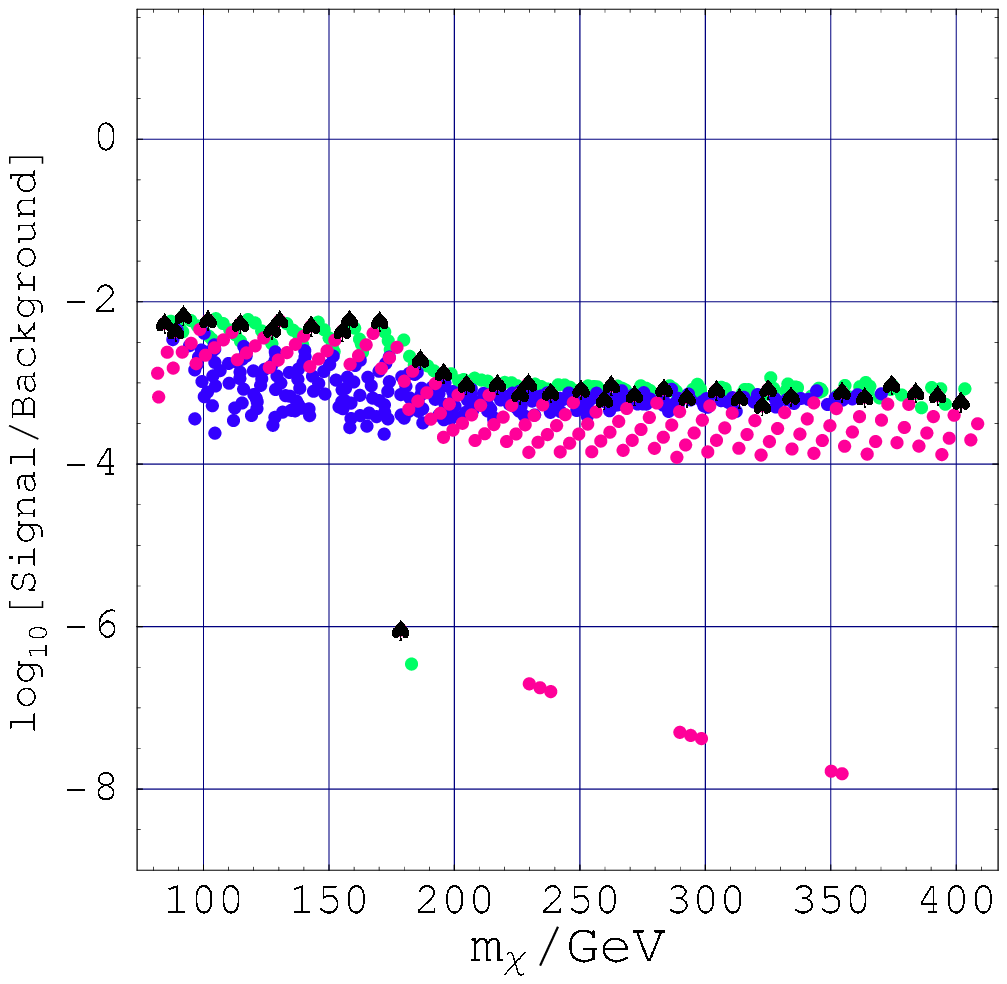} 
\end{center}
\end{minipage}
 
\caption{ The positron signal to background ratio $S/B$ at $E_{\rm opt}$ 
 in the mSUGRA scenario. 
(a) The non-thermal scenario.
(b) The thermal freeze-out scenario. $(\tan\beta=45)$
}
\label{fig:msugraposit45}
\end{figure}
%%%%%%%%%%%%%%%%%%%%%%%%%%%%%%%%%%%%%%%%%%%%%%%%%%%%%%%%%%%%%%%%%%%%%%%%%%%%
%%%%%%%%%%%%%%%%%%%%%%%%%%%%%%%%%%%%%%%%%%%%%%%%%%%%%%%%%%%%%%

In Fig.~\ref{fig:msugraposit45}, we show the positron $S/B$ ratio 
in the non-thermal (a) and thermal freeze-out  scenarios (b)  in the
mSUGRA model with $\tan \beta=45$.
In the case of the non-thermal scenario, we have set the local
neutralino density as $\rho_{\chi}=0.3\GEV/{\rm cm}^3$, which means
$\xi=1$. As for the thermal freeze-out scenario, we have rescaled the 
$S/B$ ratio by multiplying a factor $\xi^2=
(\Omega_{\chi}^{\rm th}h^2/\Omega_{DM}h^2)^2$ where $\Omega_{\chi}^{\rm th}h^2
<\Omega_{DM}h^2$ so that we can take the size of the
neutralino relic abundance into account.
The shading (coloring) conventions are the same as those in 
Fig.~\ref{fig:sugradirect45}: dark (blue) points denote the parameter
sets that  would lead to $\Omega_{\chi}^{\rm th}h^2\leq 0.03$ in the 
thermal scenario.

\begin{figure}[h!]
 \begin{minipage}{0.48\linewidth}
\begin{center}(a)

 \includegraphics[width=0.9\linewidth]{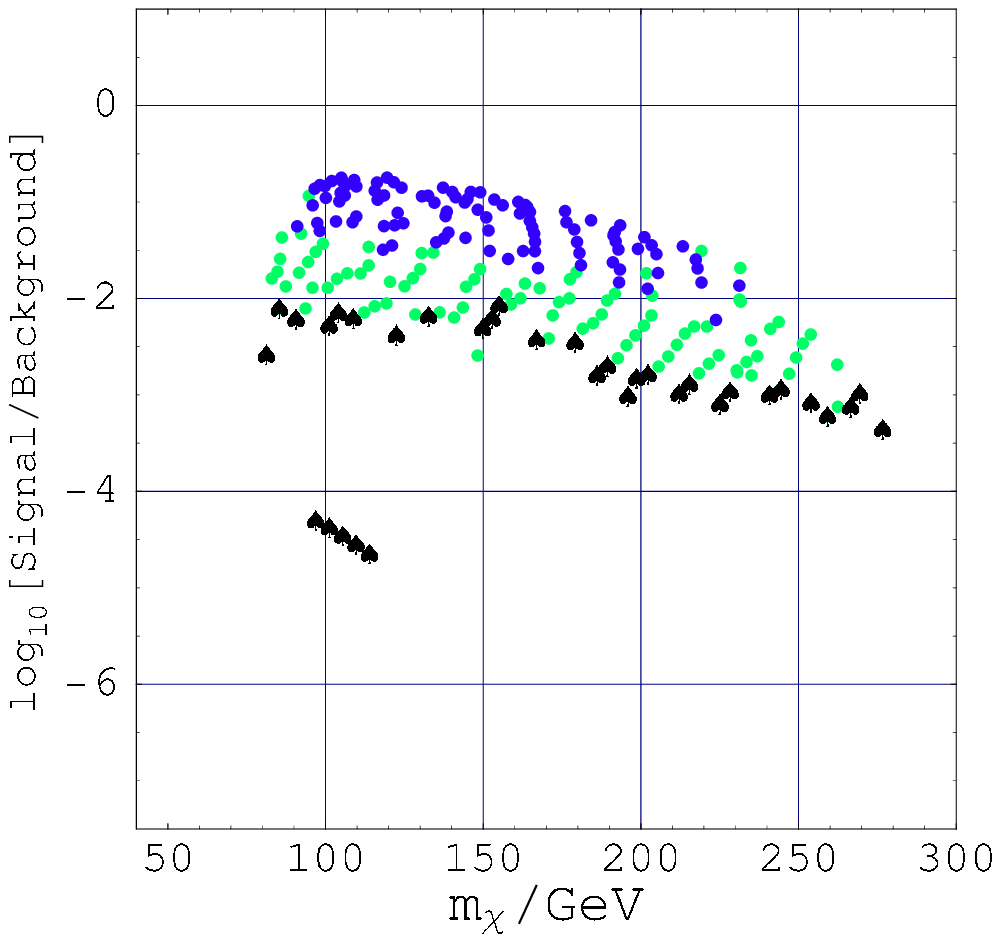} 
\end{center}
\end{minipage}
\begin{minipage}{0.48\linewidth}
\begin{center}(b)

  \includegraphics[width=0.9\linewidth]{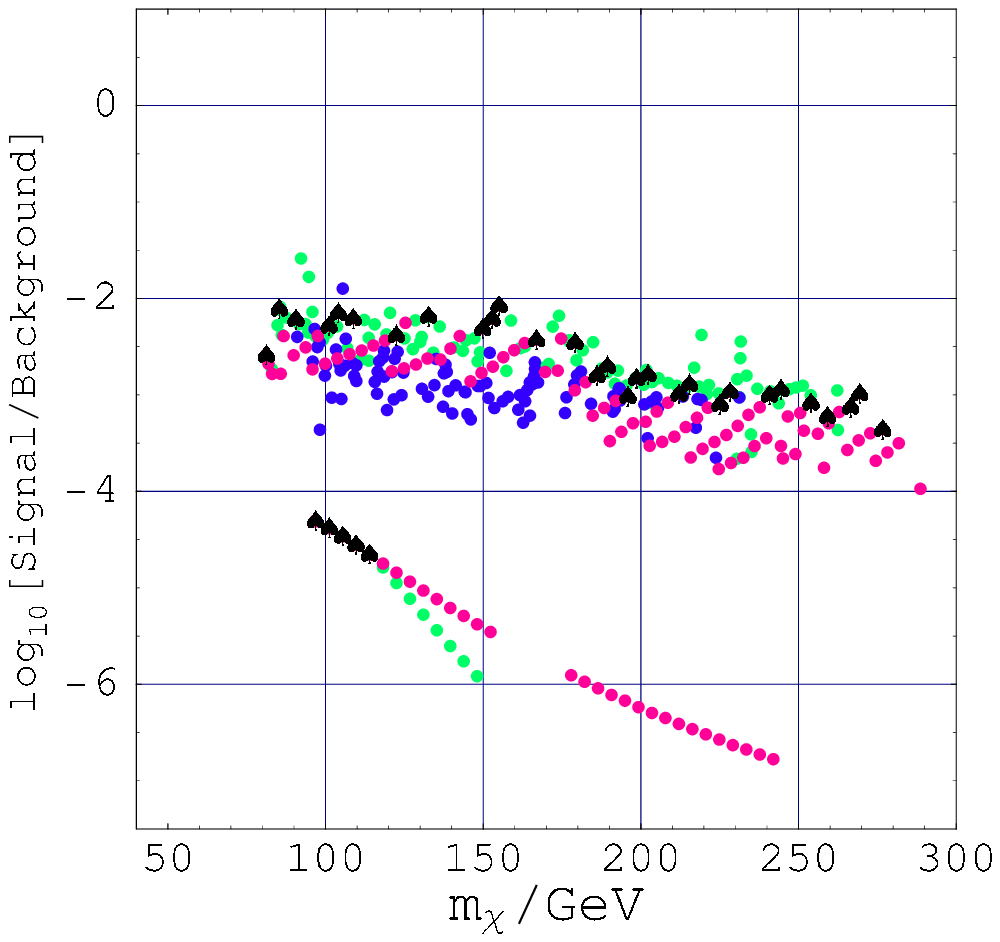} 
\end{center}
\end{minipage}
  
\caption{  The positron signal to background ratio $S/B$ at $E_{\rm opt}$ in the 
mSUGRA scenario. 
(a) The non-thermal scenario.
(b) The thermal freeze-out scenario. $(\tan\beta=10)$
}

\label{fig:msugraposit10}
\end{figure}

In the figure, the advantage of the non-thermal scenario is really 
distinctive. The preferred region for AD baryogenesis
($\Omega_{\chi}^{\rm th} h^2\leq 0.03$) provides quite large $S/B$
ratio, especially at the region $m_{\chi}\lsim m_{\rm top}$, where even 
$S/B \sim 10\%$ is possible. On the other hand, the thermal
freeze-out scenario predicts very small $S/B$ ratio $<1\%$, 
particularly at the co-annihilation region.
The expected sensitivity 
of the future space based 
experiments, such as AMS-$02$~\cite{AMS}, is roughly $\sim 1\%$.
Therefore, although the estimation of the positron flux is suffering from various
uncertainties, such as ``clumpiness'' of the local neutralino density,
we can expect a good possibility to find a kind of ``smoking-gun''
signals of the non-thermal dark matter in the near future.
In Fig.~\ref{fig:msugraposit10}, we show the corresponding calculations 
for  $\tan\beta=10$.
%%%%%%%%%%%%%%%%%%%%%%%%%%%%%%%%%%%%%%%%%%%%%%%%%%%%%%%%%%%%%%%%%%%%%
%%%%%%%%%%  The indirect detection rates  tanb=10       %%%%%%%%%%%%%%
%%%%%%%%%%%%%%%%%%%%%%%%%%%%%%%%%%%%%%%%%%%%%%%%%%%%%%%%%%%%%%%%%%%%%
%%%%%%%%%%%%%%%%%%%%%%%%%%%%%%%%%%%%%%%%%%%%%%%%%%%%%%%%%%%%%%%%%%%%%%%%%%%%
%%%%%%%%%%%%%%%%%%%%%%%%%%%%%%%%%%%%%%%%%%%%%%%%%%%%%%%%%%%%%%%%%%%%%%%%%%%%

%%%%%%%%%%%%%%%%%%%%%%%%%%%%%%%%%%%%%%%%%%%%%%%%%%%%%%%%%%%%%%%%%%%%%%%%%%
\subsubsection{Hard positron flux in the mAMSB}%%%%%%%%%%%%%%%%%%%%%%%%%%%
\begin{figure}[h!]
\begin{minipage}{0.48\linewidth}
\begin{center}(a)

 \includegraphics[width=0.9\linewidth]{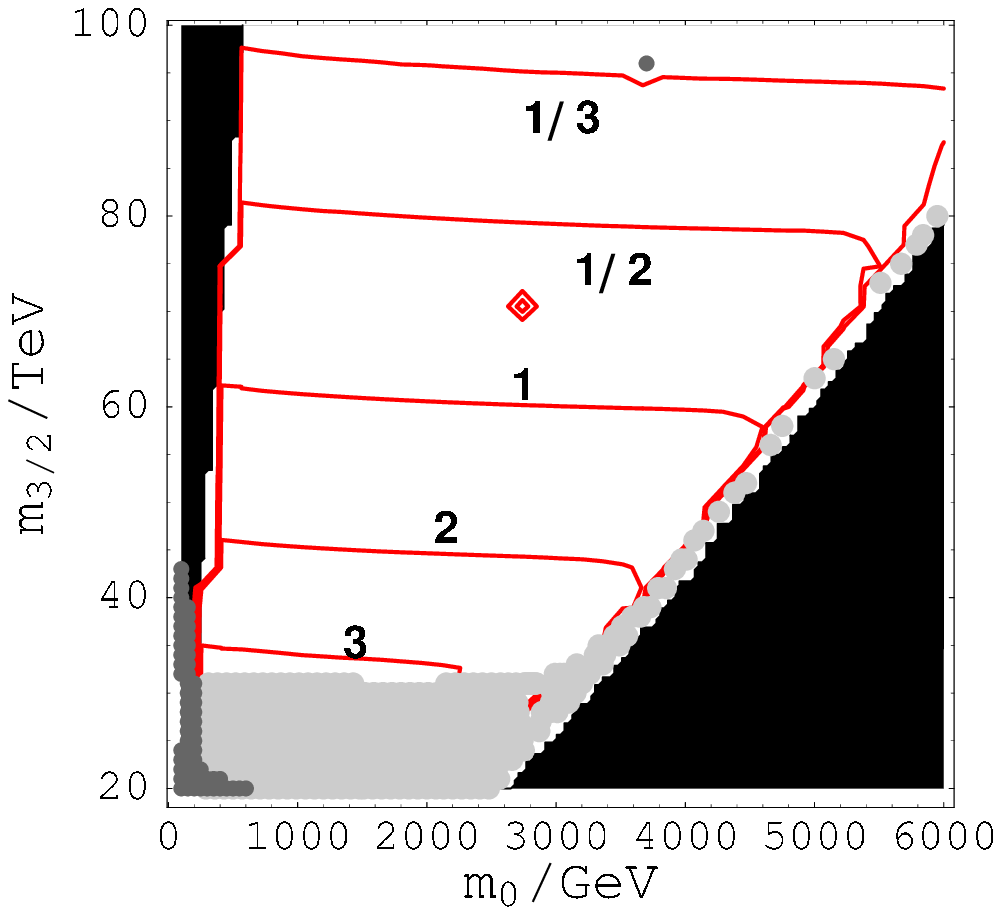} 
\end{center}
\end{minipage}
\begin{minipage}{0.48\linewidth}
\begin{center}(b)

 \includegraphics[width=0.9\linewidth]{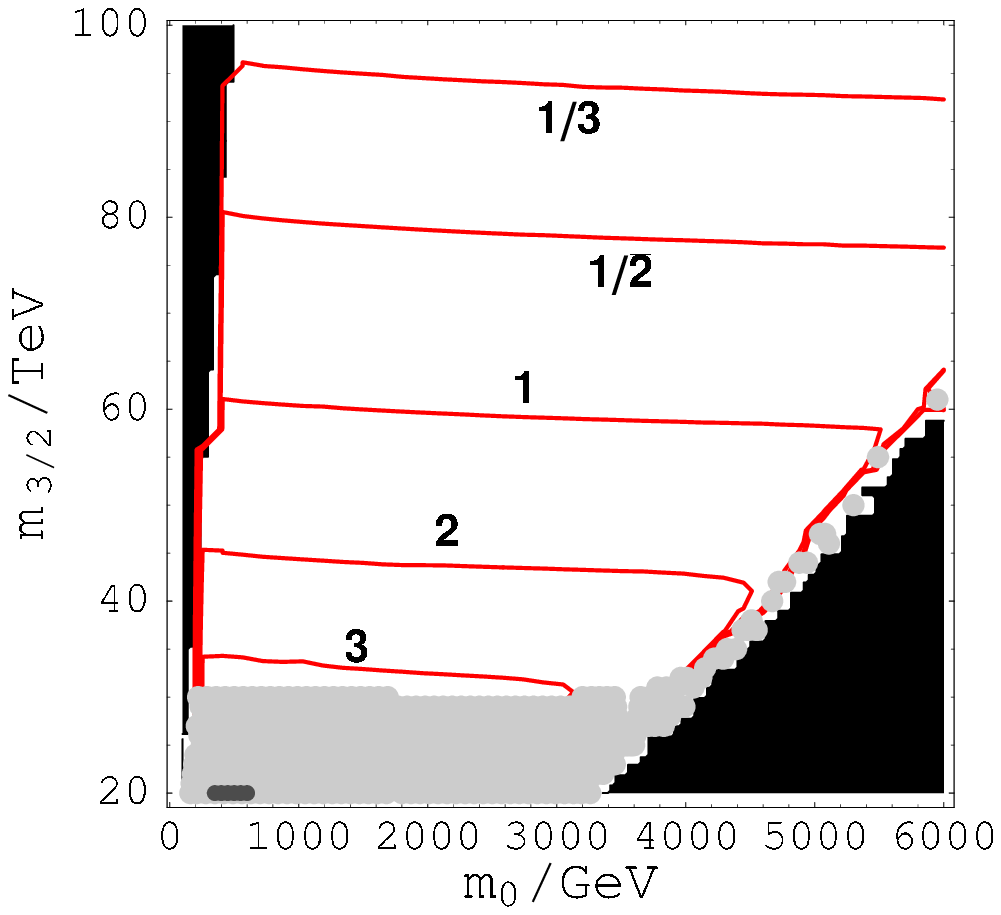} 
\end{center}
\end{minipage}

\caption{  The positron signal to background ratio $S/B$ at $E_{\rm opt}$
in the mAMSB scenario. 
(a) $\tan \beta=30$.
(b) $\tan \beta=10$.
}
\label{fig:anompositcont}
 \end{figure}

%%%%%%%%%%%%%%%%%%%%%%%%%%%%%%%%%%%%%%%%%%%%%%%%%%%%%%%%%%%%%%%%%%%%%%%%%%
Now, let us discuss the expected positron flux in the mAMSB scenario.
In this model, Wino-like LSP is realized in most of the parameter space,
which has a larger annihilation cross section into gauge bosons 
than Higgsino-like LSP by roughly one order of magnitude. 
This fact allows us to have much more
distinctive signals than those in the mSUGRA model, if the non-thermal 
DM-genesis had taken place in the early Universe.
In Fig.~\ref{fig:anompositcont}, we show the contour plot of the positron 
signal to background ratio $S/B$ at $E_{\rm opt}$ with fixed local
neutralino density as  $\rho_{\chi}=0.3\GEV/{\rm cm}^3$.
In most of the allowed parameter region, $S/B\sim 1$ is expected.
%%%%%%%%%%%%%%%%%%%%%%%%%%%%%%%%%%%%%%%%%%%%%%%%%%%%%%%%%%%%%%%%%%%%%
%%%%%%%%%%  The indirect detection rates  tanb=30       %%%%%%%%%%%%%%
%%%%%%%%%%%%%%%%%%%%%%%%%%%%%%%%%%%%%%%%%%%%%%%%%%%%%%%%%%%%%%%%%%%%%
%%%%%%%%%%%%%%%%%%%%%%%%%%%%%%%%%%%%%%%%%%%%%%%%%%%%%%%%%%%%%%%%%%%%%%%
%%%%%%%%%%%%%%%%%%%%%%%%%%%%%%%%%%%%%%%%%%%%%%%%%%%%%%%%%%%%%%%%%%%%%%%

%%%%%%%%%%%%%%%%%%%%%%%%%%%%%%%%%%%%%%%%%%%%%%%%%%%%%%%%%%%%%%%%%%%%%
%%%%%%%%%%%%%%%%%%%%%%%%%%%%%%%%%%%%%%%%%%%%%%%%%%%%%%%%%%%%%%%%%%%%%
\begin{figure}[h!]
 \begin{minipage}{0.48\linewidth}
\begin{center}(a)

 \includegraphics[width=0.9\linewidth]{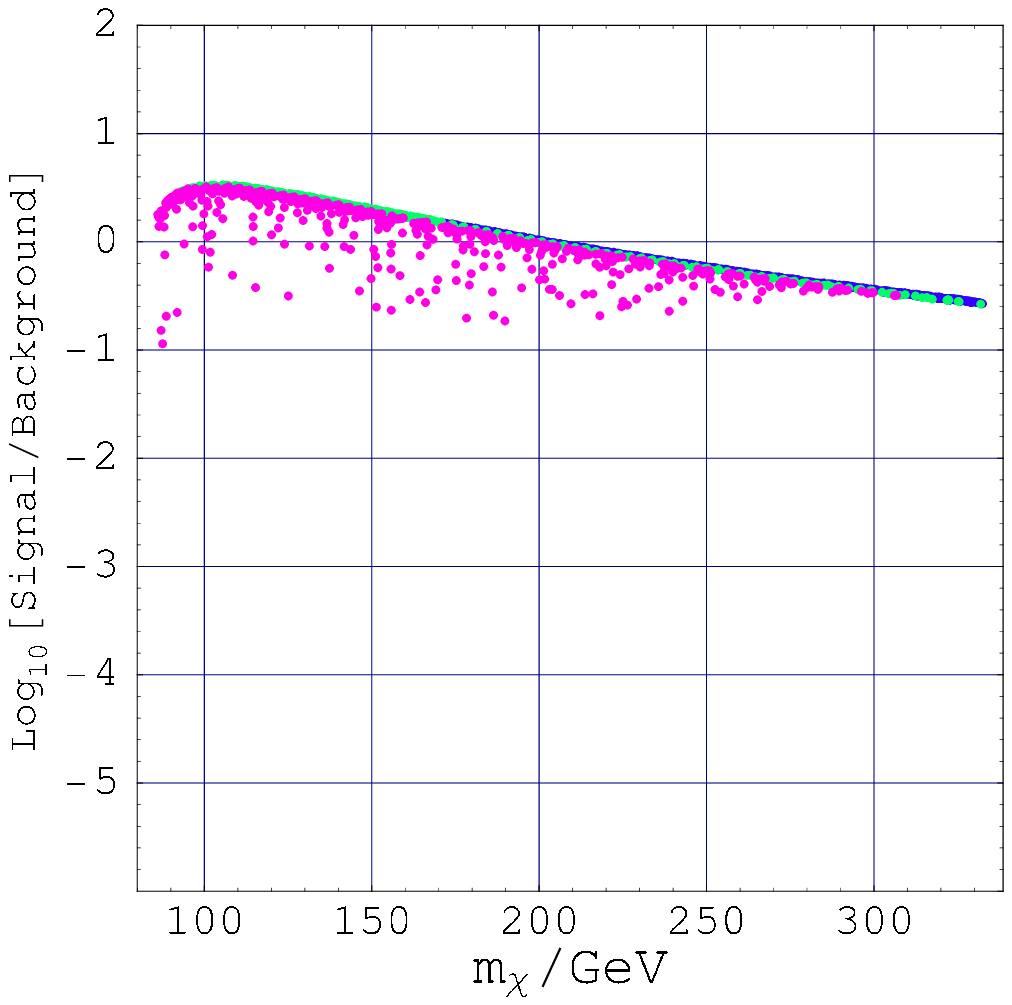} 
\end{center}
\end{minipage}
\begin{minipage}{0.48\linewidth}
\begin{center}(b)

  \includegraphics[width=0.9\linewidth]{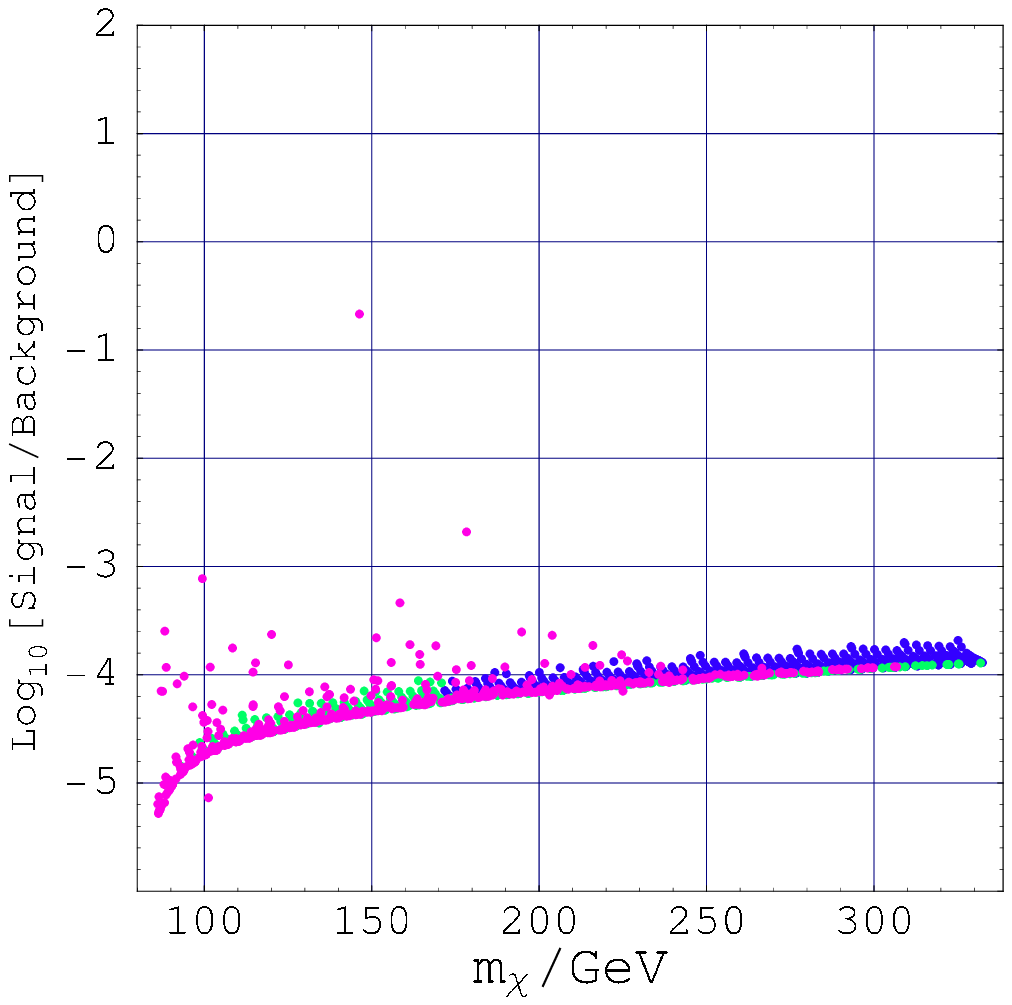} 
\end{center}
\end{minipage}

\caption{  The positron signal to background ratio $S/B$ at $E_{\rm opt}$ in 
the mAMSB scenario. 
(a) The non-thermal scenario.
(b) The thermal freeze-out scenario. $(\tan \beta=30)$
} 

\label{fig:anomposit30}
\end{figure}
%%%%%%%%%%%%%%%%%%%%%%%%%%%%%%%%%%%%%%%%%%%%%%%%%%%%%%%%%%%%%%%%%%%%%%%%%%%%%
%%%%%%%%%%%%%%%%%%%%%%%%%%%%%%%%%%%%%%%%%%%%%%%%%%%%%%%%%%%%%%%%%%%%%
%%%%%%%%%%%%%%%%%%%%%%%%%%%%%%%%%%%%%%%%%%%%%%%%%%%%%%%%%%%%%%%%%%%%%
\begin{figure}[h!]
 \begin{minipage}{0.48\linewidth}
\begin{center}(a)

 \includegraphics[width=0.9\linewidth]{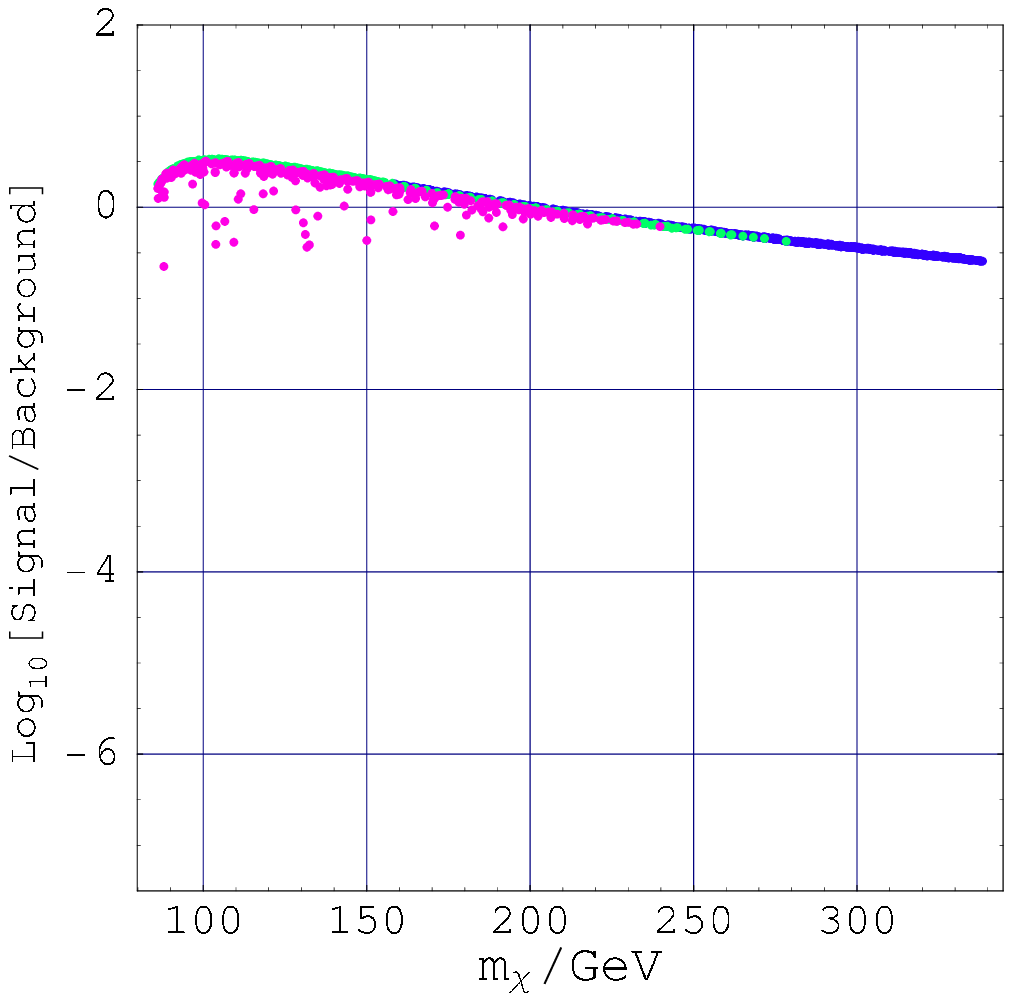} 
\end{center}
\end{minipage}
\begin{minipage}{0.48\linewidth}
\begin{center}(b)

  \includegraphics[width=0.9\linewidth]{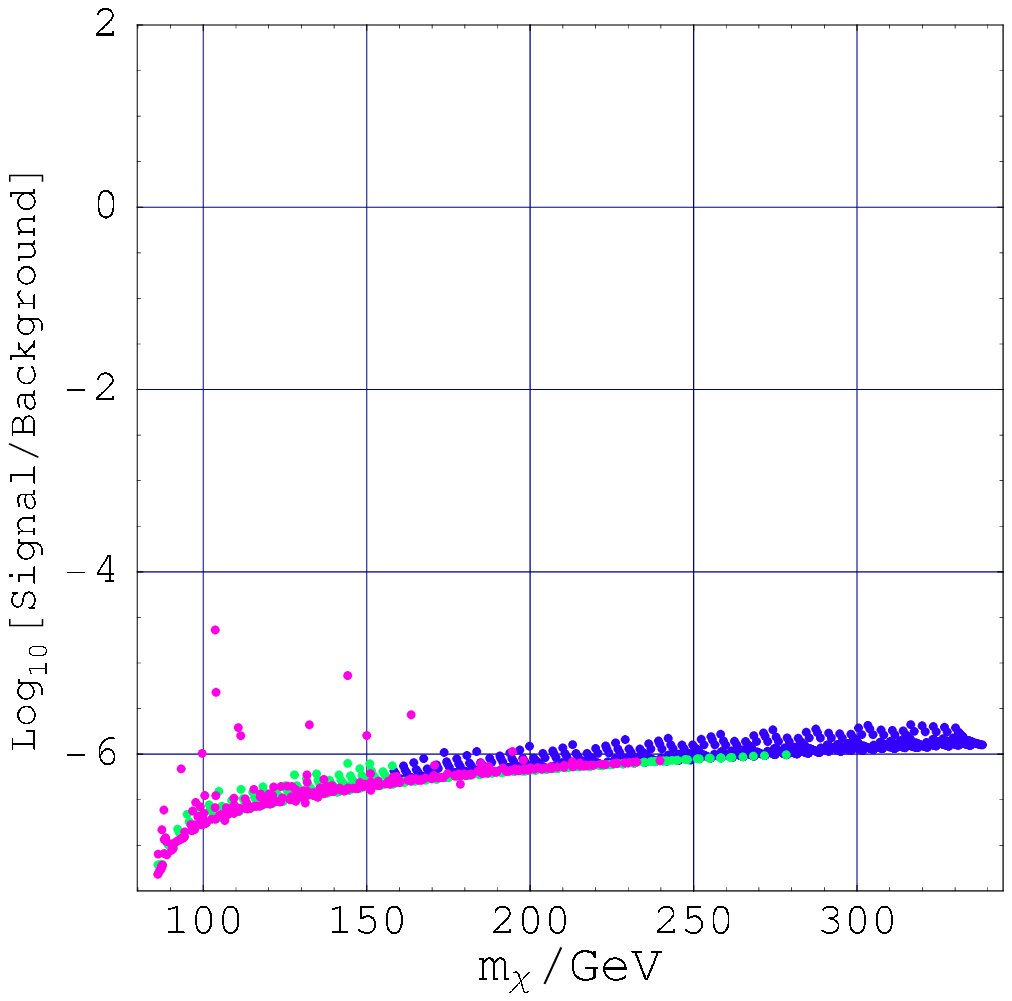} 
\end{center}
\end{minipage}

\caption{  The positron signal to background ratio $S/B$ at $E_{\rm opt}$ in the 
mAMSB scenario. 
(a) The non-thermal scenario.
(b) The thermal freeze-out scenario. $(\tan \beta=10)$
} 

\label{fig:anomposit10}
\end{figure}
%%%%%%%%%%%%%%%%%%%%%%%%%%%%%%%%%%%%%%%%%%%%%%%%%%%%%%%%%%%%%%%%%%%%%%%%%%%%%

In Fig.~\ref{fig:anomposit30}, we present the signal to background ratio $S/B$
in the non-thermal (a) and the thermal freeze-out  scenarios (b) for
$\tan \beta=30$. The shading (coloring ) conventions are the same as
those in Fig.~\ref{fig:anomdirect30}, which denote the Higgsino fraction in
the LSP. As in the case of the mSUGRA model, we have set $\xi=1$ in the 
non-thermal scenario, and rescaled the $S/B$ ratio by multiplying 
$\xi^2=(\Omega_{\chi}^{\rm th}h^2/\Omega_{DM}h^2)^2$ in the thermal 
freeze-out scenario. As one can see, 
if the AD baryo/DM-genesis had really taken place
in the early Universe, we will have 
really distinctive signals in the near future. On the other hand, in the 
thermal scenario, we cannot expect any observable signal because of
smallness of the relic Wino density. Because  we can 
survey only a limited parameter space in the mAMSB model 
by the direct detection and 
indirect dark matter search observing energetic neutrinos,
the observation of the hard positron flux will play a crucial role 
to reveal the nature of dark matter in this model.
In Fig.~\ref{fig:anomposit10}, we present a corresponding figure 
for  $\tan \beta=10$.

%%%%%%%%%%%%%%%%%%%%%%%%%%%%%%%%%%%%%%%%%%%%%%%%%%%%%%%%%%%%%%
\section{Conclusions and discussion}%%%%%%%%%%%%%%%%%%%%%%%%%
%%%%%%%%%%%%%%%%%%%%%%%%%%%%%%%%%%%%%%%%%%%%%%%%%%%%%%%%%%%%%%
In this paper, we have discussed the implications of the 
Affleck--Dine baryo/DM-genesis scenario in several ways of 
dark matter search. We have investigated two promising ways of
indirect detection: one is the observation of the muon flux
induced by the energetic neutrinos from the center of the Sun,
and the other is to observe the hard positron flux from the halo.
We have also updated the previous analysis of the direct 
detection by implementing the recent WMAP result to constrain the 
allowed parameter space, which allowed us to 
have more definitive predictions of the
non-thermal scenario.

We have adopted the mSUGRA and mAMSB scenarios to demonstrate the 
predictions of the non-thermal scenario. In the mSUGRA model, 
Affleck--Dine baryogenesis prefers the ``focus point'' region to
avoid overclosing the Universe,
where the LSP contains non-negligible component of Higgsino.
A large Higgsino fraction in the LSP increases all of the above 
mentioned detection rates, and we can survey the whole parameter space
in the future experiments. Especially, for relatively light neutralinos
$m_{\chi}\lsim m_{t}$, we have an intriguing possibility to discover 
dark matter signals in near future experiments, such as 
CDMS (Soudan)~\cite{CDMS}, EDELWEISS II~\cite{EDELWEISS} and 
ANTARES~\cite{ANTARES}.
In the case of the mAMSB model, the thermal relic density is very small
$\Omega_{\chi}^{\rm th}h^2\sim 10^{-3}$ in the entire parameter space,
which makes the AD baryogenesis consistent in all over the region.
Unfortunately,  we can survey only a limited parameter space by 
the direct detection and indirect detection observing neutrino flux 
from the Sun.
However, the quite large annihilation cross section into W-bosons 
promises us distinctive signals of the hard positron flux (and also of
the mono-energetic photon from the direct annihilation channel:
$\chi\chi\rightarrow \gamma\gamma/Z$) in the future experiments~\cite{ullio1}.
Although we have adopted the mSUGRA and mAMSB models for demonstration,
since discussed detection rates are primarily determined 
by the Higgsino or Wino fraction in the LSP,
we hope that the main predictions are not changed in other
SUSY-breaking models.

Very encouragingly, though we have to wait father confirmations, 
there already exist 
some interesting experimental signals which can be naturally 
explained by Higgsino- or Wino-like non-thermal dark matter.
The recent HEAT balloon experiment~\cite{HEAT} has reported a significant 
excess of positrons in cosmic rays. The  authors of Refs.~\cite{KANE,KANE2}
have argued that a Higgsino or Wino LSP with mass $m_{W}<m_{\chi}\lsim
200\GEV$ could yield a consistent positron flux provided the relic
abundance is from a non-thermal source.
The EGRET~\cite{EGRET} telescope has also identified a gamma-ray source at 
the galactic center. The authors of Ref.~\cite{ullio2} have argued 
that the spectrum features of this source are compatible 
with the gamma-ray flux induced by pair annihilations of 
dark matter neutralinos. They have shown that the discrimination 
between this interesting interpretation and other viable explanations 
will be possible with GLAST~\cite{GLAST}, the next major gamma-ray telescope in space.

Finally, let us comment on the generality of the non-thermal dark matter. 
In the present work, we have assumed Affleck--Dine
baryogenesis as the origin of the non-thermal source of the LSP.
However, we think that the existence of the non-thermal dark matter is a much
more generic prediction of the MSSM, or any kind of SUSY standard models. 
Once we assume MSSM-like models, we inevitably have many flat directions.
Because we believe the existence of an inflationary era in the very
beginning of the Universe, we have a good reason to expect that there are couplings
between inflaton and SM fields in the K\"ahler potential in the
following form:
\begin{eqnarray}
\delta K=\frac{b}{M_{*}^2}I^{\dag}I\Phi^{\dag}\Phi\;,
\nonumber
\end{eqnarray}
where $I$ is the inflaton superfield and $\Phi$ denotes any kind of 
SM field. In order to ensure for all the flat directions to have positive
Hubble-order mass term, we have to assume $b< 1$ for arbitrary
combinations of ${\Phi}$ along flat directions.~\footnote{$b\gsim
1 + 0.3$ is enough to drive $\Phi$ away from the
origin.} 

This seems a rather
strong assumption. We think that it is much more natural, or at least
comparably natural, that some flat direction has a negative Hubble-order
mass term and develops a large expectation value during the inflationary 
stage. This generally leads to the same non-thermal DM-genesis as discussed in
this paper. There is no need for the flat direction to carry non-zero
baryon number. If the flat direction is lifted by some
non-renormalizable operator in the superpotential, the flat-direction
field is likely not to dominate the energy density of the
Universe, and hence it does not lead to an additional entropy
production.  In this case, we can make use of the standard leptogenesis to produce 
the observed baryon asymmetry.
Note that the decay temperature of Q-balls is mainly determined by the 
initial amplitude of the flat-direction field, and hence, $T_{d}\lsim {\cal O}(1)\GEV$
is a quite generic prediction.~\footnote{Detailed discussion on each flat 
direction will be published elsewhere.} 
These observations lead us to consider the Higgsino- or Wino-like
non-thermal dark matter as a quite natural consequence of the MSSM or 
other SUSY standard models. We hope that this work
will encourage serious research on the non-thermal 
dark matter by many researchers. 

%%%%%%%%%%%%%%%%%%%%%%%%%%%%%%%%%%%%%%%%%%%%%%%%%%%%%%%%%%%%%%
\section*{Acknowledgments}%%%%%%%%%%%%%%%%%%%%%%%%%%%%%%%%%%%
%%%%%%%%%%%%%%%%%%%%%%%%%%%%%%%%%%%%%%%%%%%%%%%%%%%%%%%%%%%%%%
M.F. thanks the Japan Society for the Promotion of Science for financial support.

%%%%%%%%%%%%%%%%%%%%%%%%%%%%%%%%%%%%

%%%%%%%%%%%%%%%%%%%%%%%%%%%%%%%%%%%%%%%%%%%%%%%%

%%%%%%%%%%%%%%%%%%%%%%%%%%%%%%%%%%%%%%%%%%%%%%%%%%%%%%%%%%%%
%%%%%%%%%%%%%%%
\end{document}